\documentclass[longauth]{aa}

\usepackage{graphicx}
\usepackage{txfonts}
\begin{document}

\title{The GAPS Programme at TNG\thanks{Based on observations made with the Italian Telescopio Nazionale Galileo (TNG) operated by the Fundaci\'{o}n Galileo Galilei (FGG) of the Istituto Nazionale di Astrofisica (INAF) at the Observatorio del Roque
de los Muchachos (La Palma, Canary Islands, Spain).}}
\subtitle{XXXVI. Measurement of the Rossiter-McLaughlin effect and revising the physical and orbital parameters of the HAT-P-15, HAT-P-17, HAT-P-21, HAT-P-26, HAT-P-29 eccentric planetary systems}

\author{
L. Mancini\inst{1,2,3}
\and
M. Esposito\inst{4}
\and
E. Covino\inst{5}
\and
J. Southworth\inst{6}
\and
E.\ Poretti\inst{7,8}
\and
G.\ Andreuzzi\inst{8,9}
\and
D.\ Barbato\inst{10,3}
\and
K.\ Biazzo\inst{9}
\and
L.\ Borsato\inst{11,12}
\and
I.\ Bruni\inst{13}
\and
M.\ Damasso\inst{3}
\and
L. Di Fabrizio\inst{8}
\and
D.~F.\ Evans\inst{6}
\and
V.\ Granata\inst{12,11}
\and
A.~F.\ Lanza\inst{14}
\and
L.\ Naponiello\inst{1,15}
\and
V.\ Nascimbeni\inst{11}
\and
M.\ Pinamonti\inst{3}
\and
A.\ Sozzetti\inst{3}
\and
J.\ Tregloan-Reed\inst{16}
\and
M.\ Basilicata\inst{1}
\and
A.\ Bignamini\inst{17}
\and
A.~S.\ Bonomo\inst{3}
\and
R.\ Claudi\inst{11}
\and
R.\ Cosentino\inst{8}
\and
S.\ Desidera\inst{11}
\and
A.~F.~M.\ Fiorenzano\inst{8}
\and
P.\ Giacobbe\inst{3}
\and
A.\ Harutyunyan\inst{8}
\and
Th.\ Henning\inst{2}
\and
C.\ Knapic\inst{17}
\and
A.\ Maggio\inst{18}
\and
G.\ Micela\inst{18}
\and
E.\ Molinari\inst{19}
\and
I.\ Pagano\inst{14}
\and
M.\ Pedani\inst{8}
\and
G.\ Piotto\inst{12}
}
\institute{
Department of Physics, University of Rome ``Tor Vergata'', Via della Ricerca Scientifica 1, 00133 -- Rome, Italy \\
\email{lmancini@roma2.infn.it}
\and
Max Planck Institute for Astronomy, K\"{o}nigstuhl 17, 69117 -- Heidelberg, Germany
\and
INAF -- Osservatorio Astrofisico di Torino, via Osservatorio 20, 10025 -- Pino Torinese, Italy
\and
Th\"{u}ringer Landessternwarte, Tautenburg Sternwarte 5, 07778 -- Tautenburg, Germany   \and
INAF -- Osservatorio Astronomico di Capodimonte, Salita Moiariello 16, 80131 -- Naples, Italy
\and
Astrophysics Group, Keele University, Keele ST5 5BG, UK
\and
INAF -- Osservatorio Astronomico di Brera, Via E. Bianchi 46, 23807 -- Merate (LC), Italy
\and
Fundaci\'{o}n Galileo Galilei - INAF, Rambla Jos\'{e} Ana Fernandez
P\'{e}rez 7, 38712 Bre\~{n}a Baja, TF, Spain
\and
INAF -- Osservatorio Astronomico di Roma, Via Frascati 33, 00078 -- Monte Porzio Catone (Roma), Italy
\and
Observatoire de Gen\`{e}ve, Universit\'{e} de Gen\`{e}ve, 51 Chemin Pegasi, 1290 -- Sauverny, Switzerland
\and
INAF -- Osservatorio Astronomico di Padova, Vicolo dell'Osservatorio 5, 35122 -- Padova, Italy
\and
Dipartimento di Fisica e Astronomia ``Galileo Galilei'' – Universit\`{a} di Padova, Vicolo dell’Osservatorio 2, 35122 -- Padova, Italy
\and
INAF -- OAS, Osservatorio di Astrofisica e Scienza dello Spazio di Bologna, Via P. Gobetti 93/3, 40129 -- Bologna, Italy
\and
INAF -- Osservatorio Astrofisico di Catania, via S. Sofia 78, 95123 -- Catania, Italy
\and
Department of Physics and Astronomy, University of Florence, Largo Enrico Fermi 5, 50125 Firenze
\and
Instituto de Investigaci\'{o}n en Astronomia y Ciencias Planetarias, Universidad de Atacama, Copiap\'{o}, Atacama, Chile
\and
INAF -- Osservatorio Astronomico di Trieste, via Tiepolo 11, 34143 -- Trieste, Italy
\and
INAF -- Osservatorio Astronomico di Palermo, Piazza del Parlamento, 1, 90134 -- Palermo, Italy
\and
INAF -- Osservatorio Astronomico di Cagliari, via della Scienza 5, 09047 -- Selargius (CA), Italy
}

  \date{Received ; accepted }
 
  \abstract
{The measurement of the spin-orbit alignment of hot Jupiters, with a range of orbital and physical properties, can provide information about the evolution of the orbits of this special class of giant planets.}
{We aim to refine the orbital and physical parameters and determine the sky-projected planet orbital obliquity, $\lambda$, of five eccentric ($e\cong0.1-0.3$) transiting planetary systems: HAT-P-15, HAT-P-17, HAT-P-21, HAT-P-26, and HAT-P-29, whose parent stars have an effective temperature between $5100\,{\rm K}<T_{\rm eff}<6200\,{\rm K}$. Each of the systems hosts a hot Jupiter, except for HAT-P-26 that hosts a Neptune-mass planet.}
{We observed transit events of these planets with the HARPS-N spectrograph, obtaining high-precision radial velocity measurements that allow us to measure the Rossiter-McLaughlin effect for each of the target systems. We used these new HARPS-N spectra and archival data, including those from \textit{Gaia}, to better characterise the stellar atmospheric parameters. The photometric parameters for four of the hot Jupiters were recalculated using 17 new transit light curves, obtained with an array of medium-class telescopes, and data from the TESS space telescope. HATNet time-series photometric data were checked for the signatures of rotation periods of the target stars and their spin axis inclination.}
{From the analysis of the Rossiter-McLaughlin effect, we derived a sky-projected obliquity of $\lambda=13^{\circ}\pm6^{\circ}$, $\lambda=-26.3^{\circ}\pm6.7^{\circ}$, $\lambda=-0.7^{\circ}\pm12.5^{\circ}$, $\lambda=-26^{\circ}\pm16^{\circ}$, for HAT-P-15\,b, HAT-P-17\,b, HAT-P-21\,b and  HAT-P-29\,b, respectively. 
Based on theoretical considerations, these small values of $\lambda$ should be of primordial origin, with the possible exception of HAT-P-21.
Due to the quality of the data, we were not able to well constrain $\lambda$ for HAT-P-26\,b, although a prograde orbit is favoured ($\lambda=18^{\circ}\pm49^{\circ}$). The stellar activity of HAT-P-21 indicates a rotation period of $15.88\pm0.02$\,days, which allowed us to determine its true misalignment angle $\psi= 25^{\circ}\pm16^{\circ}$. Our new analysis of the physical parameters of the five exoplanetary systems returned values compatible with those existing in the literature. Using TESS and the available transit light curves, we reviewed the orbital ephemeris for the five systems and confirmed that the HAT-P-26 system shows transit timing variations, which may tentatively be attributed to the presence of a third body.
}
   {}
   \keywords{Extrasolar planets -- Stars: late-type, fundamental parameters -- Techniques: radial velocities, photometric -- Stars: individual: HAT-P-15, HAT-P-17, HAT-P-21, HAT-P-26, HAT-P-29}
\titlerunning{RM effect in 5 HATNet exoplanets}
\authorrunning{L. Mancini et al.}
\maketitle
%
\section{Introduction}
In more than 20 years of exciting research, exoplanetary science has given many surprises for viewpoints based on our own Solar system. One major surprise came early in the development of this field: the small orbital distances and sometimes large orbital eccentricities of the first known exoplanets (e.g. \citealp{mayor:1995,cochran:1997}). Since the two giant planets and the two ice planets of the Solar system are much further from the Sun than the four rocky planets are, we would expect a similar situation for exoplanetary systems. This is because the formation of a giant planet requires a core of solid material to grow above a certain critical size of $\approx 10\,M_{\oplus}$ in order to start accreting hydrogen and helium (see, e.g., \citealp{raymond:2022}). 
This phenomenon happens in proto-planetary disks beyond the {\em snow line}, where there is a steep increase of the density of solid materials (especially volatile compounds with freezing points more than 100\,K) that can aggregate to form protoplanets with large cores, which then become giant gaseous planets. 

Furthermore, the orbits of these giant planets should be roughly circular. This is because the effective frictional force, which exists for planets that are orbiting within a massive disk of gas and dust, tend to circularize any orbits, even though they are initially eccentric (see, e.g, \citealp{raymond:2022}). 

Instead, observational results show there are a lot of giant planets with eccentric orbits and located very close to their stars. These planets, which are known as hot Jupiters, are very easy to detect via transit and radial-velocity (RV) methods, and their origin has been one of the main topics of discussion in planetary science for the last 20 years. They most likely formed at large distances from their star, as did Jupiter in the Solar system, but then, through some physical mechanism, have migrated towards the innermost regions. Different mechanisms have been proposed that are able to shrink the orbit of a giant planet. Among them, the main ones are based on ($i$) dynamical interactions, through planet-planet scattering \citep{rasio:1996,davies:2014} or the Kozai mechanism (see, e.g., \citealt{wu:2003}), and ($ii$) disc-planet interaction \citep{lin:1996,ward:1997}.

To determine which of the two theories is the right one, or at least the most efficient, one can examine how several parameters are distributed, like the metallicity of the parent stars (see, e.g., \citealt{dawson:2013}) or the eccentricity and the orientation of the planetary orbits of hot Jupiters (see, e.g., \citealt{matsumura:2010}). For example, it is expected that scattering encounters among planets should randomize the alignments of the orbital planes. Instead, if the existence of hot Jupiters is to be primarily ascribed to disc-planet interactions, we should see many flat architectures, as this mechanism keeps the planetary orbits coplanar throughout the entire migration process. Numerical simulations support these predictions (e.g., \citealt{Chatterjee:08,Marzari:09}).

Whenever a giant planet undergoes orbital migration, the responsible mechanism is expected also to affect the eccentricity, $e$, and/or the angle, $\psi$, between the planet's orbital axis and its host star's spin. Several authors have dealt with this question based on the available data (e.g., \citealt{bonomo:2017,wang:2021,rice:2021,rice:2022}), but it is still not clear what the degree of correlation between these two parameters is and how much tidal interactions intervene to complicate the interpretation, as recently pointed out by \citet{albrecht:2022}. Therefore, new measurements of the spin-orbit alignments are useful for enlarging the sample of exoplanetary systems in order to perform robust statistical analyses and shed new light on what the real cause of the giant-planet migration process is. 

While $\psi$ is, unfortunately, not easy to determine, its sky-projected value, $\lambda$, can be easily measured for systems containing transiting hot Jupiters, via the observation of the Rossiter-McLaughlin (RM) effect or from star-spot tracking in consecutive transit light curves. Once $\lambda$ is known, one can derive the true alignment of the projected rotational velocity of the parent star, $v\sin i$. If its radius is also known, its rotation period can be measured as well.

The long-term observational programme GAPS (Global Architecture of Planetary Systems) utilises the HARPS-N spectrograph, at the 3.5\,m Telescopio Nazionale Galileo (TNG), to execute a study of the spin-orbit alignment of a sample of known transiting exoplanetary systems for measuring the RM effect \citep{covino:2013}. We took advantage of the high spectral resolution of HARPS-N for targeting faint stars (up to $V < 14$\,mag), for which the RM effect is harder to measure at the required level of accuracy \citep{esposito:2014,esposito:2017,mancini:2015,mancini:2018}. 

Photometric follow-up observations with an array of medium-class telescopes support our programme. The aim is to obtain high-quality light curves of planetary-transit events to refine the whole set of physical and orbital parameters of the planetary systems in our target list. High-quality light curves also allow us to detect possible features of the transit light curve, which can be associated with stellar activity (e.g., \citealt{mancini:2017}), as well as transit timing variations (TTVs).

In this work, we present new measurements of the RM effect for five exoplanetary systems. They have been selected considering the effective temperature of their parent stars, which have $5100\,{\rm K}<T_{\rm eff}<6200\,{\rm K}$ and the eccentricity of the planetary orbits, which are  $0.1< e < 0.3$. 
We can divide these systems in two groups. Each of the stars of the first group hosts a hot Jupiter (HAT-P-15\,b, HAT-P-17\,b, HAT-P-21\,b and HAT-P-29\,b), whose orbital eccentricity is larger than zero at $3\, \sigma$ confidence level, as estimated by \citet{bonomo:2017}. The second group includes only one system, in which there is a Neptune-mass planet, HAT-P-26\,b, whose eccentricity was estimated larger than zero at $2\, \sigma$ confidence level by \citet{hartman:2011}.
For the planets in the first group, we were able to measure the orbital obliquity, but we do not detected any TTVs. For HAT-P-26, we were not able to deduce a precise value for $\lambda$, but we  detected a TTV.

The paper is organised as follows. In Sect.~\ref{sec:Literature} we describe the systems that are the subject of this study. Spectroscopic and photometric observations and data reduction procedures are presented in Sect.~\ref{Sec:ObsRed}. Sect.~\ref{sec:Light-curve_analysis} is devoted to the light-curve analysis. The results of the analysis of the HATNet time-series photometric data are reported in Sect.~\ref{Frequency_analysis_time-series}. An analysis of the stellar parameters, based on archival data, is performed in Sect.~\ref{sec:EXOFAST}. The stellar atmospheric properties and activity of the stars, based on HARPS-N data, and the measurements of the spin-orbit alignment angle for each of the systems are presented in Sect.~\ref{sec:HARPS-N_spectra_analysis}. The results of the analysis aimed at refining the physical parameters of the systems are reported in Sect.~\ref{sec:Final_Physical_parameters}. Finally, Sect.~\ref{discussion} contains the discussion, while the main results of this work are summarised in Sect.~\ref{summary}.

\section{Targets properties}
\label{sec:Literature}
In this section we summarize the main properties of the five planetary systems that are the subject of this study. The values of the parameters were taken from the Transiting Extrasolar Planet Catalogue (TEPCat\footnote{TEPCat is available at {\tt http://www.astro.keele.ac.uk/jkt/} {\tt tepcat/} \citep{southworth:2011}.}) and are also summarised in Tables \ref{tab:hatp15_final_parameters}, \ref{tab:hatp17_final_parameters}, \ref{tab:hatp21_final_parameters}, \ref{tab:hatp26_final_parameters} and \ref{tab:hatp29_final_parameters}.

\subsection{HAT-P-15}
\label{subsec:Lit_HATP15}
\citet{kovacs:2010} reported the discovery of HAT-P-15\,b, a giant planet ($M_{\rm p} =1.946 \pm 0.066\, M_{\rm Jup}$; $R_{\rm p} =1.072 \pm 0.043\, R_{\rm Jup}$) on an eccentric orbit ($e =0.190 \pm 0.019$) of period $P_{\rm orb} \sim 10.9$\,days, transiting a G5\,V dwarf star ($V=12.16$\,mag, $M_{\star} = 1.013 \pm 0.043 \, M_{\sun}$, $R_{\star} =1.080 \pm 0.039\, R_{\sun}$, $T_{\rm eff} = 5568 \pm 90$~K).
\begin{itemize}
\item[$\bullet$] \citet{kovacs:2010} gathered 24 high-precision RV measurements with the HIRES@KECK I, over the time interval August 2007 to December 2009. \\[-8pt]
\item[$\bullet$] \citet{knutson:2014} collected seven additional RV data points with HIRES (from August 2010 to September 2012) and found no evidence of long-term RV trends, which might indicate the presence of outer companions. \\[-8pt]
\item[$\bullet$] \citet{ngo:2015} used NIRC2 on Keck\,II to acquire $K$-band adaptive-optic (AO) images of HAT-P-15 in two epochs, finding no bound companions.  \\[-8pt]
\item[$\bullet$] Lucky images, which were taken with the AstraLux Norte at the Calar Alto 2.2\,m telescope, show two faint objects at separation of 6.2 and 7.1~arcsec, respectively \citep{wollert:2015}. Further observations will be needed to confirm whether these objects are physically associated with HAT-P-15.  \\[-8pt]
\item[$\bullet$] The \textit{Gaia} EDR3 catalogue \citep{gaia:2016,gaia:2021} also reports two objects at a separation of $\sim7$\,arcsec with $\Delta g \sim 7$\,mag; only one of them has a parallax entry that allows us to exclude it as a physical companion.  \\[-8pt]
\item[$\bullet$]  \citet{piskorz:2015} analysed NIRSPEC@Keck $K$-band spectra of HAT-P-15 and did not find evidence of a close redder stellar companion. Subsequent studies of this system \citep{bonomo:2017,stassun:2017} confirmed the original estimates of its main physical parameters as reported by \citet{kovacs:2010}.
\end{itemize}

\subsection{HAT-P-17}
\label{SubSec:Lit_HATP17}
HAT-P-17\,b is a giant planet ($M_{\rm p} =0.54 \pm 0.02 \, M_{\rm Jup}$; $R_{\rm p} = 1.04 \pm 0.02\, R_{\rm Jup}$) on an eccentric orbit ($e = 0.3417 \pm 0.0036$), with period $P_{\rm orb} \sim 10.3$\,d and hosted by a relatively bright early K\,V dwarf star ($V=10.54$\,mag, $M_{\star}= 0.88 \pm 0.04 \,M_{\sun}$, $R_{\star} = 0.84 \pm 0.01 \, R_{\sun}$, $T_{\rm eff} = 5322 \pm 55$\,K; \citealp{howard:2012}).
\begin{itemize}
\item[$\bullet$] In $\sim$50 HIRES-RV measurements spanning from October 2007 to August 2013 \citealp{fulton:2013,knutson:2014}), evidence was found for the presence of another long period planetary companion, HAT-P-17\,c, with $P_{\rm orb} \approx 10$ to 36\,yr, $m \sin{i}\approx 3.4 \,M_{\rm Jup}$, and $e \approx 0.4$. \citet{fulton:2013} also measured the projected obliquity of planet b, finding $\lambda=19^{+14}_{-16}$, a value which is consistent with zero. \\[-8pt]
\item[$\bullet$] \citet{bonomo:2017} reported an additional 25 RV measurements taken with HARPS-N from October 2012 to November 2015. With a combined fit of the HIRES and HARPS-N RV data sets, they put more stringent constraints on the HAT-P-17\,c parameters: $P_{\rm orb} = 3972_{-146}^{+185}$\,days; $e = 0.295 \pm 0.021$; $m \sin{i} = 2.88 \pm 0.10 \,M_{\rm Jup}$; $a = 4.67 \pm 0.14$\,au. \\[-8pt]
\item[$\bullet$] $K$-band AO images, taken with the NIRC2 at Keck II \citep{fulton:2013,ngo:2015}, ruled out the existence of companions with $\Delta K < 7$\,mag for separations beyond 0.7~arcsec (65~au according to the Gaia parallax). \\[-8pt]
\item[$\bullet$] \citet{wollertal:2015} observed HAT-P-17 with AstraLux Norte in the $i^{\prime}$ and $z^{\prime}$ passbands. They found no close companions, and reported $5\sigma$ detection limits of $\Delta z^{\prime}=3.84$, $4.95$, $6.29$, $7.09$ mag 
at $0.25$, $0.5$, $1.0$, $2.0$ arcsec, respectively. \\[-8pt]
\item[$\bullet$] Similar detection limits in the $K_{\rm s}$ band were reported by \citet{adams_2013}, who used ARIES at the MMT telescope. \\[-8pt]
\item[$\bullet$] \citet{piskorz:2015} used high-resolution $K$-band spectra, taken with NIRSPEC@KECK, to search for blended lines from cool stellar companions. They found that the spectral fit is significantly improved by the presence of a $3900_{-300}^{+200}$\,K companion in the HAT-P-17 system at a maximum separation of 36\,au.
\end{itemize}

\subsection{HAT-P-21}
\label{SubSec:Lit_HATP21}
The exoplanet HAT-P-21\,b was discovered by \citet{bakos:2011} to be a massive hot Jupiter ($M_{\rm p} = 4.063 \pm 0.161\, M_{\rm Jup}$; $R_{\rm p} = 1.024 \pm 0.092\, R_{\rm Jup}$) moving on a short-period and eccentric orbit ($P_{\rm orb} \sim 4.1$\,d; $e = 0.228 \pm 0.016$) around a G3\,V star ($V=11.69$\,mag, $M_{\star} =0.947 \pm 0.042\, M_{\sun}$, $R_{\star} =1.105 \pm 0.083\, R_{\sun}$, $T_{\rm eff} = 5588 \pm 80$\,K).
\begin{itemize}
\item[$\bullet$] The values of the main physical and orbital parameters of this system reported in the discovery paper are in a good agreement with those from subsequent studies \citep{torres:2012,bonomo:2017,stassun:2017}.  \\[-8pt]
\item[$\bullet$] For this star, 15 high-precision RV measurements were obtained with the HIRES@KECK I (from May 2009 to February 2010). $K$-band images, obtained with the NIRC2 at Keck II \citep{ngo:2016}, showed no evidence of bound companions. High-resolution Lucky-Imaging observations made with the AstraLux Norte camera also did not reveal the presence of any companions \citep{wollertal:2015}. \\[-8pt]
\item[$\bullet$] The Gaia EDR3 catalogue reports no objects close to HAT-P-21 within 10 arcsec.
\end{itemize}

\subsection{HAT-P-26}
\label{SubSec:Lit_HATP26}
The discovery of the HAT-P-26 planetary system was announced by \citet{hartman:2011}. It consists of a low-density Neptune-mass planet ($M_{\rm p} =0.059 \pm 0.007\, M_{\rm Jup}$; $R_{\rm p} =0.565^{+0.072}_{-0.032} \, R_{\rm Jup}$) transiting a K1\,V dwarf star ($V=11.74$\,mag, $M_{\star} = 0.816 \pm 0.033 \,M_{\sun}$, $R_{\star} = 0.788^{+0.098}_{-0.043} \, R_{\sun}$, $T_{\rm eff} = 5011 \pm 55$\,K) with a period of $P_{\rm orb} \sim 4.23$\,d. The orbit of this planet is also eccentric, with $e=0.124 \pm 0.060$. 
\begin{itemize}
\item[$\bullet$] 12 RV measurements were obtained for this star with the HIRES@KECK I between December 2009 and June 2010 \citep{hartman:2011}. A further 11 were obtained between December 2011 and June 2012 \citep{knutson:2014}. \\[-8pt]
\item[$\bullet$] High-resolution Lucky-Imaging observations performed with the AstraLux Norte camera did not reveal close-in bound companions \citep{wollertal:2015}. \\[-8pt]
\item[$\bullet$] Indications of TTVs in the system, with an amplitude of 4 min and a periodicity of 270 epochs, were observed by \citet{vonEssen:2019}. \\[-8pt]
\item[$\bullet$] Much more precise measurements of the physical parameters for this system have not been obtained by other authors \citep{torres:2012,mortier:2013,stassun:2017}. \\[-8pt]
\item[$\bullet$] Detailed studies of the atmosphere of HAT-P-26\,b have been conducted via transmission spectroscopy \citep{stevenson:2016,wakeford:2017,MacDonald:2019}. \\[-8pt]
\item[$\bullet$] The Gaia EDR3 catalogue reports no objects close to HAT-P-26 within 10 arcsec.
\end{itemize}

\subsection{HAT-P-29}
\label{SubSec:Lit_HATP29}
The HAT-P-29 planetary system is composed of a hot Jupiter ($M_{\rm p} =0.767^{+0.047}_{-0.045}\, M_{\rm Jup}$; $R_{\rm p} = 1.064^{+0.075}_{-0.068}\, R_{\rm Jup}$), orbiting an F8\,V star ($V=11.90$\,mag, $M_{\star} =1.199^{+0.063}_{-0.061} \,M_{\sun}$, $R_{\star} =1.237^{+0.077}_{-0.071} \, R_{\sun}$, $T_{\rm eff} =6115 \pm 86$\,K) every $\sim 5.72$\,d \citep{buchhave:2011}.
\begin{itemize}
\item[$\bullet$] 8 RV measurements were obtained for this star with the HIRES@KECK I between September and December 2010 \citep{buchhave:2011}. Joining these measurements with four others, which were taken with the same instrument between February and August 2012, \citet{knutson:2014} found a trend in the RV data and, hence, evidence for a companion in this system. \\[-8pt]
\item[$\bullet$] \citet{wollert:2015} found a stellar source $3.3^{\prime\prime}$ away from HAT-P-29, using the AstraLux Norte camera. \\[-8pt]
\item[$\bullet$] \citet{ngo:2016} confirmed the latter discovery with NIRC2 and found that this source is consistent with a bound stellar companion. \\[-8pt]
\item[$\bullet$] With 25 high-precision HARPS-N RVs, \citet{bonomo:2017} did not detect any significant trend that can be consistent with the RV drifts found by \citet{knutson:2014}, but they found a significant ($5.8\sigma$) small eccentricity ($e\approx 0.1$) for the orbit of HAT-P-29\,b. \\[-8pt]

\item[$\bullet$] The measurement of the orbital period of HAT-P-29\,b was recalculated thanks to new transit light curves, finding that it is $\approx 17.6$\,s longer than the previous measurement. No statistically significant TTVs were found \citep{wang:2018}. \\[-8pt]

\item[$\bullet$] The Gaia EDR3 catalogue \citep{gaia:2016,gaia:2021} reports a faint ($G=17.8$\,mag) object at a separation of 3.43\,arcsec from HAT-P-29. However, the parallax of this object indicates a distance roughly 30 times greater than that of HAT-P-29, excluding the possibility of it being a physical companion.
\end{itemize}

\section{Observations and data reduction}%
\label{Sec:ObsRed}
In this section we present new times-series spectroscopic data of HAT-P-15, HAT-P-17, HAT-P-21, HAT-P-26 and HAT-P-29, as well as new photometric follow-up observations with TESS and ground-based telescopes.

\subsection{HARPS-N spectroscopic observations}
\label{SubSec:HARPS-N_spectroscopic_observations}

All the spectra analysed in this work were acquired with the High Accuracy Radial velocity Planet Searcher -- North (HARPS-N; \citealt{cosentino:2012}) spectrograph at the Italian Telescopio Nazionale Galileo (TNG). The log of the HARPS-N observations is given in Table~\ref{tab:logspec}. 

\begin{figure}
\centering
\includegraphics[width=\hsize]{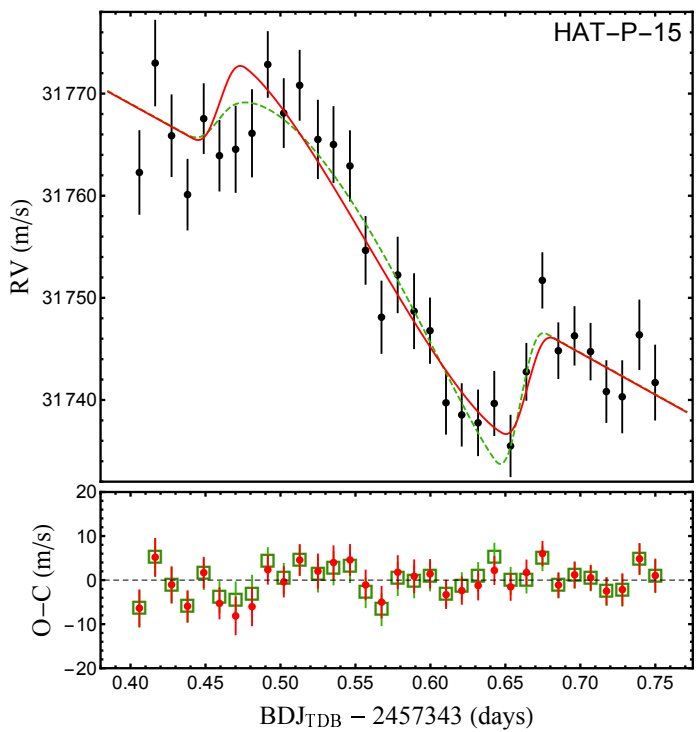}
\caption{RV data of the transit of HAT-P-15\,b observed with HARPS-N. Superimposed are the best-fitting RV-curve models (the red line does not include the modelling of the stellar convective blueshift (CB) effect, while the green-dashed line does). The corresponding residuals are plotted in the lower panel. For clarity, the error bars are displayed only for the model without the CB effect. See the discussion in Sect.~\ref{subsec:spin-orbit_alignment}}
\label{fig:RM_hatp15}
\end{figure}

\begin{figure*}
\centering
\includegraphics[width=18cm]{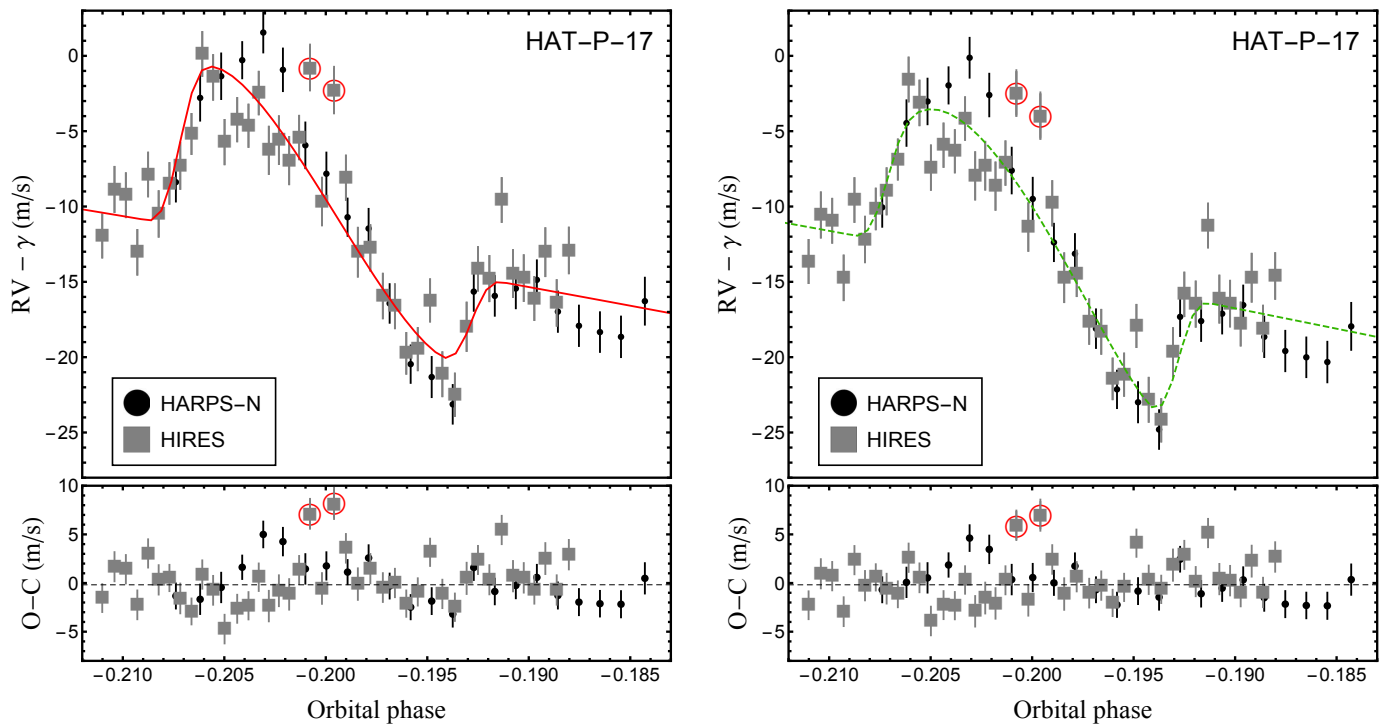}
\caption{A joint RV-data plot of two different transit events of HAT-P-17\,b, fitted with two different models: RV only ({\it left-hand panels}) and RM+CB ({\it right-hand panels}). {\it Left-hand panels}: grey squares are the HIRES RVs \citep{fulton:2013}. Note that the two circled data points were considered outliers (they showed residuals larger than $4\,\sigma$ and, therefore, were not considered in the modelling process. Black points are the HARPS-N RVs (this work). The best-fitting values of the systemic RVs were subtracted in order to compare the two data sets. The red line represents the best-fitting model of the RM effect; the RV residuals are plotted in the lower panel. {\it Right-hand panels}: same as left panels but with the RM effect and CB effect modelled simultaneously (green-dashed line).}
\label{fig:RM_hatp17}
\end{figure*}

\begin{figure}
\centering
\includegraphics[width=\hsize]{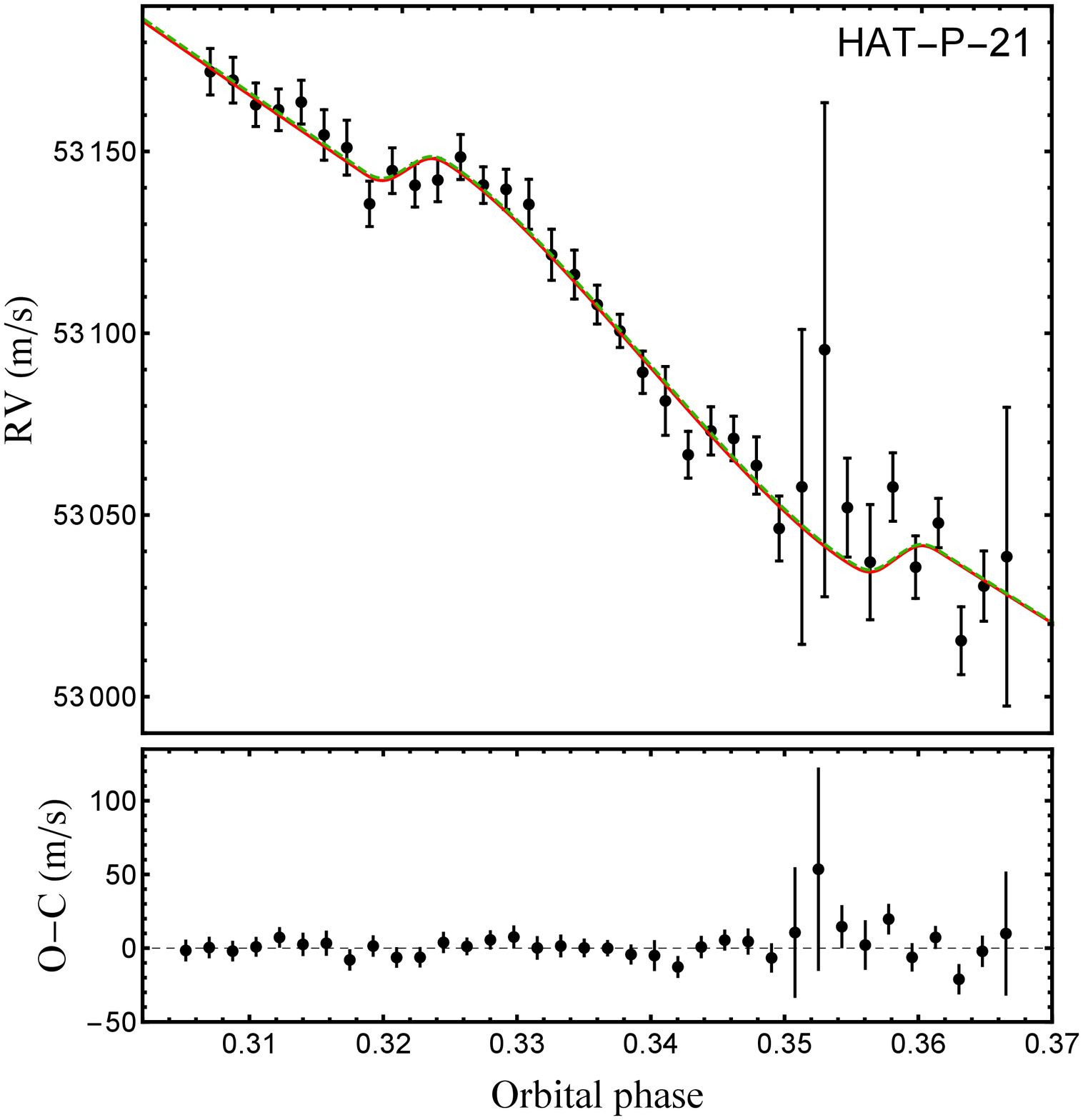}
\caption{RV data of the transit of HAT-P-21\,b observed with HARPS-N (this work). Superimposed are the best-fitting RV-curve models (red line is without, green-dashed line with the CB effect). The corresponding residuals are plotted in the lower panel. For clarity, the error bars are displayed only for the model without the CB effect.}
\label{fig:RM_hatp21}
\end{figure}

\begin{figure}
\centering
\includegraphics[width=\hsize]{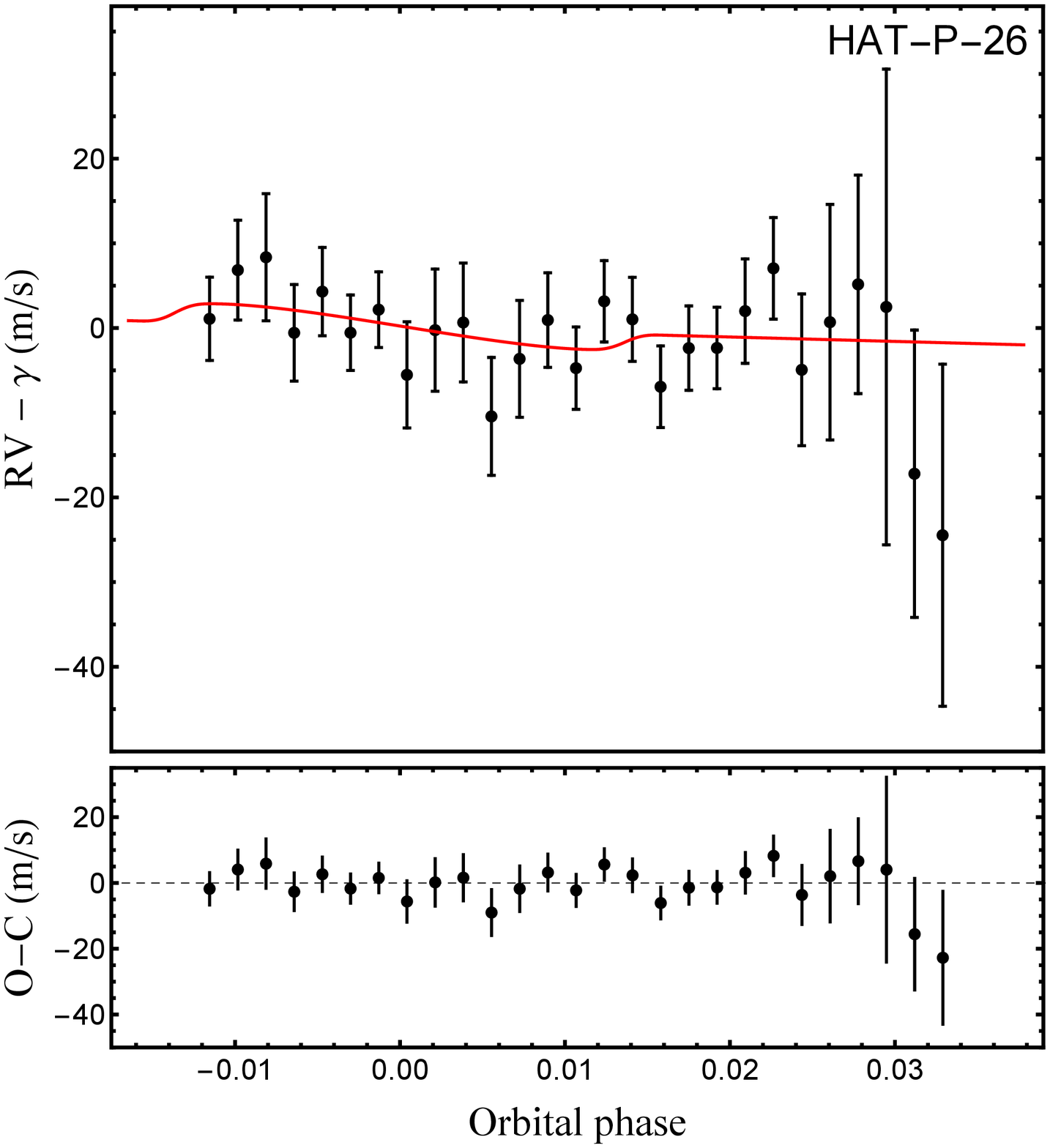}
\caption{RV data of the transit of HAT-P-26\,b observed with HARPS-N (this work). Superimposed is the best-fitting RV-curve model without the CB effect (red line). The corresponding residuals are plotted in the lower panel.}
\label{fig:RM_hatp26}
\end{figure}

\begin{figure*}
\centering
\includegraphics[width=18cm]{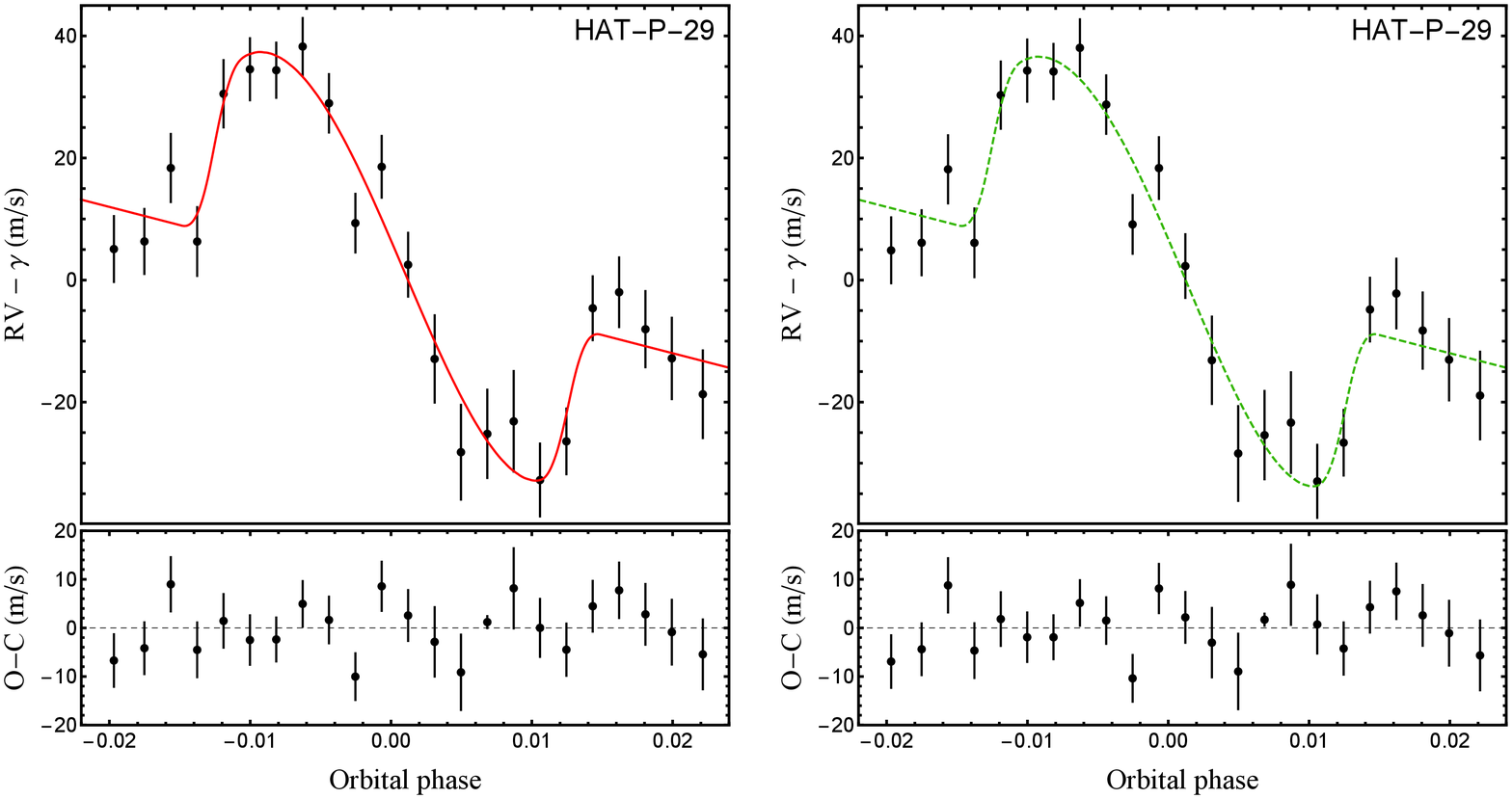}
\caption{{\it Left-hand panels}: RV data of the transit of HAT-P-29\,b observed with HARPS-N (this work). The red line represents the best-fitting model of the RM effect; the RV residuals are plotted in the lower panel. {\it Right-hand panels}: same as left panels but with the RM effect and CB effect modelled simultaneously (green-dashed line).}
\label{fig:RM_hatp29}
\end{figure*}

HARPS-N provides high resolution spectra ($R\sim115000$) spanning almost the entire optical range ($\lambda=383-690$\,nm), and was designed to deliver very precise stellar RVs \citep{cosentino:2014}. HARPS-N is equipped with its own Data Reduction Software (DRS) that, in addition to 1-D wavelength-calibrated spectra, provides RVs, which are calculated by cross-correlating the spectra with a numerical mask \citep{baranne:1996,pepe:2002,lovis:2007} and line bisectors. The DRS also measures the Mount Wilson $S$ index and, if the stellar $B-V$ colour index is lower than 1.2, also the log($R'_{\rm HK}$) chromospheric activity index \citep{lovis:2011}. Specifically, we utilised the latest version of the DRS available offline at the  Italian center for Astronomical Archives (IA2) \citep{smareglia:2014}.
\begin{table*}
\caption{Details of the spectroscopic observations of the planetary transits recorded with HARPS-N.}             
\label{tab:logspec}      
\centering          
\begin{tabular}{l c c c c c c c c}     
\hline\hline       
Object  & Date\tablefootmark{(a)} & UT Start & UT End  & $N_{\rm obs}$ & $T_{\rm exp}$[s] &   Airmass\tablefootmark{(b)}  &  Moon\tablefootmark{(c)}            & 2$^{\rm nd}$ fibre \\
\hline 
 HAT-P-15  & 2015-11-16   &  21:29     &  05:59   &  33   &   900  &  1.62$\rightarrow$1.02$\rightarrow$1.59  & NO                    &  Sky          \\
 HAT-P-17  & 2013-10-13   &  19:29     &  01:28   &  23   &   900  &  1.07$\rightarrow$1.00$\rightarrow$1.67  & 73\%/41$^{\circ}$     &  Sky          \\
 HAT-P-21  & 2014-03-07   &  22:24     &  04:37   &  36   &   600  &  1.28$\rightarrow$1.02$\rightarrow$1.28  & 46\%/85$^{\circ}$     &  Sky          \\
 HAT-P-26  & 2015-03-26   &  23:48     &  04:30   &  27   &   600  &  1.60$\rightarrow$1.10$\rightarrow$1.17  & NO                    &  Sky          \\
 HAT-P-29  & 2013-10-16   &  00:15     &  06:15   &  23   &   900  &  1.12$\rightarrow$1.09$\rightarrow$1.72  & 96\%/56$^{\circ}$     &  Sky          \\ 
\hline 
\end{tabular}
\tablefoot{ 
\tablefoottext{a}{Dates refer to the beginning of the night.}
\tablefoottext{b}{Values at first$\rightarrow$meridian$\rightarrow$last exposure.}
\tablefoottext{c}{Fraction of illumination and angular distance from the target.}
}
\end{table*}
\\ [6pt]
{\bf HAT-P-15} was observed on the night of 2015/11/16, during a planetary transit. A time series of 33 spectra covered the 5.5 hours long transit from 0.9 hours before ingress to 1.8 hours after egress (see Table \ref{tab:logspec}). With an exposure time of 900 s, the spectra have an average signal-to-noise ratio of 30 (S/N per pixel in 1-D spectra at 5500\,\AA). The RV measurements were obtained using the G2 numerical mask. They are reported in Table \ref{tab:RV_HAT-P-15} and plotted in Fig.~\ref{fig:RM_hatp15}.
\\ [6pt]
{\bf HAT-P-17} was observed on the night of 2013/10/13, during a planetary transit. The acquisition series started before nautical twilight, while the last spectrum was taken 1.9\,hr after the end of the transit, when the star was at an airmass of $\sim$1.7. During the night $\sim$73\% of the moon was illuminated and it was at an angular separation of $\sim${$41^{\circ}$} from the target; we checked that no significant light contamination was present by analysing the spectra and the CCFs of the second fibre which was pointed at the sky. The resulted spectra have an average S/N of 60 (per pixel in 1-D spectra at 5500\,\AA) and a G2 mask was used to measure the RVs. They are reported in Table \ref{tab:RV_HAT-P-17} and plotted in Fig.~\ref{fig:RM_hatp17}.
\\ [6pt]
{\bf HAT-P-21} was observed on the night of 2014/03/07. A time-series of 36 spectra, lasting over six hours, bracketed a planetary transit from $\sim$80 minutes before the ingress up to $\sim$40 minutes after the egress. With an exposure time of 600\,s, the spectra have a typical S/N of 20 (per pixel in 1-D spectra at 5500\,\AA). However the last two hours of observations were affected by passing clouds, so some spectra have much lower S/N and large RV uncertainties. The RV values were obtained using a G2 mask; they are reported in Table \ref{tab:RV_HAT-P-21} and plotted in Fig.~\ref{fig:RM_hatp21}.
\\ [6pt]
{\bf HAT-P-26} was observed on the night of 2015/03/26.
The first spectrum was acquired about 15 minutes after the transit ingress, when the rising target was at an airmass of 1.6. The time series of 27 spectra stretched to $\sim2$  hours after the transit egress. The 600 s long exposures have a typical  S/N of 15 (per pixel in 1-D spectra at 5500\,\AA). The last 6 spectra of the series were affected by deteriorating weather conditions. The RV values were obtained using a K5 mask; they are reported in Table \ref{tab:RV_HAT-P-26} and plotted in Fig.~\ref{fig:RM_hatp26}
\\ [6pt]
{\bf HAT-P-29} was observed on the night of 2013/10/16.
A series of 23 spectra spanned the time interval
from $\sim1$ hour before transit ingress up to
$\sim1$ hour after egress. With an exposure
time of 900 s the spectra have an average S/N of
25 (per pixel in 1-D spectra at 5500\,\AA). The sky spectra acquired with the second fibre
show no detectable sign of light contamination
from the full Moon. The RV values were obtained using a G2 mask. They are reported in Table \ref{tab:RV_HAT-P-29} and plotted in Fig.~\ref{fig:RM_hatp29}
%

\subsection{Photometric follow-up observations}
\label{SubSec:Photometric_follow-up_observations}
Except for HAT-P-15, the planetary systems studied in this work were monitored with an array of medium-class telescopes with the aim of obtaining high-quality transit light curves, which can be used for refining the physical parameters of both the star and the planet, as well as checking stellar activity. As in our previous works based on photometric follow-up observations of transiting exoplanets (e.g., \citealp{southworth:2012,ciceri:2013}), we autoguided the telescopes and adopted the \emph{defocussing} technique in all the observations to significantly improve the precision of the photometry. The photometric data thus obtained were reduced using a modified version of the DEFOT pipeline \citep{southworth:2014} and the light curves were extracted by performing standard aperture photometry.
For the same purpose as above, we also analysed the light curves obtained by the TESS space telescope \citep{ricker:2015}.
\\ [6pt]
{\bf HAT-P-15.}
Having an orbital period larger than 10\,d, complete transits of HAT-P-15\,b are difficult to catch with ground-based facilities. Unfortunately, HAT-P-15 light curves are also not available in both the TESS 2\,min and 20\,s cadence target list\footnote{https://tess.mit.edu/observations/target-lists/} for sector 19.
In order to check the 30\,min cadence TPF files, we went to TESScut\footnote{https://mast.stsci.edu/tesscut/} and downloaded the TPFs for the RA and DEC of HAT-P-15. Having inspected the TPFs, we realised that HAT-P-15 was not observed by TESS.
\\ [6pt]
{\bf HAT-P-17.} 
HAT-P-17\,b also has an orbital period larger than 10\,d, so the observation of a complete transit is difficult to achieve using ground-based telescopes. We observed one incomplete transit of HAT-P-17\,b on July 2012 through a Gunn-$i$ filter with the BFOSC (Bologna Faint Object Spectrograph \& Camera) imager, which is mounted on the Cassini\,1.52\,m Telescope at the Astrophysics and Space Science Observatory of Bologna in Loiano (Italy). Another incomplete transit of HAT-P-17\,b was observed on July 2014 with the Calar Alto (CA) Zeiss\,1.23\,m telescope and using a Cousins-$I$ filter. Details of the instruments and telescopes were already reported in our previous works (see, e.g., \citealt{mancini:2017}). 
Two transits were observed by TESS with the 2\,min cadence during the monitoring of Sector 15 of its primary mission. Continuous observations of the target star were obtained from 2019-08-23 to 2019-09-03, for a total of 892 measurements. All the transit light curves of HAT-P-17, which were analysed in this work, are plotted in Fig.~\ref{fig:hatp17_lc}.
\\ [6pt]
{\bf HAT-P-21.} 
Two complete transit events of HAT-P-21\,b were observed on March 2012 with the Cassini\,1.52\,m and the CA\,1.23\,m telescopes, through a Gunn-$r$ and a Cousins-$R$ filter, respectively. The first data set was severely affected by clouds (Fig.~\ref{fig:hatp21_lc}.). Five transits were observed by TESS with the 2\,min cadence during the monitoring of Sector 22 of its primary mission. Continuous observations of the target star were obtained from 2020-02-22 to 2020-03-14, for a total of 1777 measurements. All the transit light curves of HAT-P-21, which were analysed in this work, are plotted in Fig.~\ref{fig:hatp21_lc}.
\\ [6pt]
{\bf HAT-P-26.} 
Four complete and one partial transit events of HAT-P-26\,b were observed with the CA\,1.23\,m telescope, through a Cousins-$I$ filter, between March 2012 and February 2018. They are plotted in Fig.~\ref{fig:hatp26_lc}. Again, there are no TESS data for this target. It is scheduled to be observed in March 2022 in sector 50. It was missed last time due to being close to the ecliptic, but for the extended mission the orientation of TESS has been changed to observe sections of the sky that were missed.
\\ [6pt]
{\bf HAT-P-29.} 
Four transits of HAT-P-29\,b were observed with the CA\,1.23\,m telescope, two through a Cousins-$R$ filter and two through a Cousins-$I$ filter. The last two were only partially observed because they occurred much later than expected. Another two incomplete transits were observed with the {\sc Dolores}\footnote{{\sc Dolores} (Device Optimized for the LOw RESolution) is a low-resolution spectrograph and camera installed at the Nasmyth-B focus of the Telescopio Nazionale Galileo.} instrument, mounted on the TNG, and with the IAC\,80\,cm telescope, through a Johnson-$R$ and Cousins-$R$ filter, respectively. Two complete transits were observed with the INAF-OAPd Copernico 1.82\,m telescope, which is located at Cima Ekar-Asiago (Italy), using the Asiago Faint Object Spectrograph and Camera (AFOSC) and a Sloan-$r$ filter within the long-term monitoring program of the TASTE project \citep{nascimbeni:2011}. Details about the last two telescopes were already reported in our previous works (see, e.g., \citealt{covino:2013,mancini:2015}). Finally, four transits were observed by TESS. They are plotted in Fig.~\ref{fig:hatp29_lc}.

\section{Light-curve analysis}
\label{sec:Light-curve_analysis}
The light curves of the transit events of HAT-P-17\,b, HAT-P-21\,b, HAT-P-26\,b and HAT-P-29\,b, which were presented in the previous section, were individually modelled with the {\sc jktebop} code \citep{southworth:2013} to make a new determination of the transit parameters. This code considers stars and planets as spheres and makes use of the Levenberg-Marquardt optimisation algorithm in order to fit the parameters of the transit light curves. These are the orbital period and inclination ($P_{\rm orb}$ and $i$), the time of transit midpoint ($T_0$), the sum and ratio of the fractional radii, i.e. $r_{\star}=R_{\star}/a$ and $r_{\rm p}=R_{\rm p}/a$; $R_{\star}$ and $R_{\rm p}$ are the radii of the star and planet, respectively, while $a$ is the semi-major axis of the planetary orbit. For modelling the limb darkening (LD) of the star, we used a quadratic law and fitted the LD coefficients ($u_{\star}$ and $v_{\star}$), taking into account the differences between the atmospheric properties of the four stars as well as the filters that were used. We also took into account the eccentric orbit of the four planets, as it has a slight effect on the shape of the transit light curves \citep{kipping:2008}. {\sc jktebop} allows the inclusion of the eccentricity, $e$, and periastron longitude, $\omega$, as fitted parameters constrained by their known values and uncertainties \citep{southworth:2009}, which are summarised in Table~\ref{tab:eccentricities}.
\begin{table}
\caption{Summary of the values of two orbital elements of the five systems analysed in this work.}             
\label{tab:eccentricities}  
\resizebox{\hsize}{!}{    
\centering          
\begin{tabular}{l c c c} %
\hline\hline       
Planet & $e$ & $\omega$ (deg) & Reference \\%
\hline\\[-6pt]%
HAT-P-15\,b & $0.200_{-0.028}^{+0.026}$ & $262.5^{+2.4}_{-2.9}$ & \citet{bonomo:2017} \\ [2pt]
HAT-P-17\,b & $0.3417 \pm 0.0036$       & $200.5 \pm 1.3$ & \citet{bonomo:2017} \\ [2pt]
HAT-P-21\,b & $0.217 \pm 0.010$         & $305.8^{+2.1}_{-1.9}$ & \citet{bonomo:2017} \\ [2pt]
HAT-P-26\,b & $0.124 \pm 0.060$         & $54 \pm 165$ & \citet{hartman:2011} \\ [2pt]
HAT-P-29\,b & $0.104^{+0.021}_{-0.018}$ & $159^{+20}_{-16}$ & \citet{bonomo:2017} \\ [2pt]
\hline 
\end{tabular}
}
\end{table}

Finally, to mitigate the correlated (red) noise, which generally affects time-series photometry obtained by the {\sc aper} routine\footnote{{\sc aper} is part of the {\sc astrolib} subroutine library distributed by NASA.}, we inflated the error bars of the photometric measurements so that each transit light curve had a reduced chi-square of $\chi_{\nu}^2=1$ during the best-fitting process. The light curves and the corresponding {\sc jktebop} best-fitting models are reported in Figs.~\ref{fig:hatp17_lc}--\ref{fig:hatp29_lc}.

The uncertainties of the fitted parameters were estimated by running both a Monte Carlo and a residual-permutation algorithm. For each of the light curves, we ran 10\,000 simulations for the Monte Carlo algorithm and the maximum number of simulations (which is one less than the number of data points) for the residual-permutation algorithm. We took the largest of the two $1\,\sigma$ values as the uncertainty for each parameter. Finally, for each planetary system, the final values of each parameter were calculated by taking the weighted average of the values extracted from the fit of all the individual light curves; the relative uncertainties were used as weights. These values are shown in the Tables reported in Appendix\,\ref{appendix_Revised_physical_parameters} and are in good agreement with those available in the literature.

\section{Orbital period determination}
\label{subsec:Orbital_period_determination}
In the modelling of the transit light curves with {\sc jktebop}, we also estimated each transit mid-time and placed them on the BJD\,(TDB) time system. By joining these new measurements with those already published, it is possible to review and refine the orbital ephemerides for the HAT-P-17, HAT-P-21, HAT-P-26 and HAT-P-29 planetary systems, as well as search for possible TTVs due to variations in the planetary orbital period. The timings that we used for each of the four systems and their residuals for a constant period are shown in the Tables reported in Appendix~\ref{Times_of_mid-transit}.
\\ [6pt]
{\bf HAT-P-17.} 
The few transit timings recorded for HAT-P-17\,b do not allow us to perform any detailed investigations about possible TTVs. Besides the timing from the discovery paper \citep{howard:2012}, we only have two from partial transit observations and two more from TESS observations. Assuming that the orbital period is constant (linear model), we performed a weighted linear least-squares fit to the mid-transit times versus their cycle number, i.e.
\begin{equation}
T_{\rm mid} = T_0 + P_{\rm orb} E\; ,
\end{equation}
where $E$ is the number of orbital cycles after the reference epoch $T_0$. The fit returned 
\begin{equation}
T_{\rm mid} = {\rm BJD_{TDB}}\, 2\,454\,801.16943\,(15)+10.33853781\,(60) \, E,
\end{equation}
with a $\chi^{2}_{\nu}=0.48$ (the quantities in brackets represent the uncertainties in the preceding digits). 
The residuals of the timings of mid-transit are plotted in Fig.\,\ref{fig:OC_plots}.
\\ [6pt]
{\bf HAT-P-21.} 
Also for HAT-P-21\,b, few transit mid-times are available and we can not investigate possible TTVs. Of the two transit light curves we have obtained, one is of low quality with large uncertainties (Fig.~\ref{fig:hatp21_lc}) so was excluded from the analysis. This left us with only seven timings: one from the discovery paper \citep{bakos:2011}, one from our observational program and five from TESS. The linear fit gives 
\begin{equation}
T_{\rm mid} = {\rm BJD_{TDB}}\, 2\,454\,996.41243\,(60)+4.12449009\,(90)\, E,
\end{equation}
with a $\chi^{2}_{\nu}=1.05$. The residuals of the timings of mid-transit are plotted in Fig.\,\ref{fig:OC_plots}.
\\ [6pt]
{\bf HAT-P-26.} 
This system is a special case for orbital period determination. \citet{hartman:2011} found a variation in the systemic velocity with a detection significance of 2.1$\sigma$. \citet{stevenson:2016} noted that there appeared to be a curvature in the diagram of the residuals from fitting a linear ephemeris to the measured times of minimum light. This was followed up by \citet{vonEssen:2019}, who found clear evidence for a sinusoidal variation with a period of 270 epochs (1140\,d) and an amplitude of 4\,min. They tentatively attributed this to the presence of a third body, in agreement with the marginal detection of a variable systemic velocity.

We assembled the times of mid-transit from previous works (see the list compiled by \citealt{vonEssen:2019}) and augmented these with our own measurements. We then fitted four types of orbital ephemeris to them: linear, quadratic, cubic, and linear plus sinusoid. The last ephemeris was modelled with the equation
\begin{equation}
T_{\rm mid} = T_0 + P_{\rm orb} E + A\sin\left(\frac{2\pi E}{P_{\rm sine}}-\phi\right)\,,
\end{equation}
where $P_{\rm sine}$ is the period of the sine wave (in units of the orbital period), $A$ is the sine amplitude and $\phi$ is the phase offset with respect to $T_0$. We were initially unable to find a good fit, with a best value of $\chi^{2}_{\nu}=2.7$. A close inspection of the published timings showed that three of them have implausibly small errorbars of 0.000011 to 0.000016 d (1.0 to 1.4 s). These three are based on transmission spectroscopy with HST \citep{wakeford:2017} and contain large gaps due to the low-Earth orbit: all fully cover the egress but have no observations during ingress. 

We therefore increased the errorbars of all four timings from \citet{wakeford:2017} by a factor of ten and refitted the full set of timings (a reanalysis of the HST data to obtain improved timings and uncertainties is outside the scope of the current work but would be useful in future studies). The factor of ten was chosen iteratively and is the factor by which the errorbars from \citet{wakeford:2017} must be increased by to get the sum of the absolute values of the residuals of these four datapoints to be equal to $4\,\sigma$ (i.e. the multiplicative factor was determined from the scatter of the data around the best fit instead of the quoted errorbars).

\begin{figure}
\resizebox{\hsize}{!}{\includegraphics{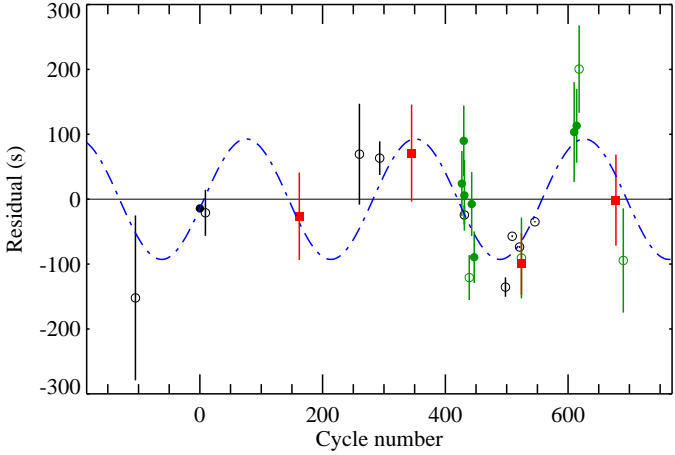}}
\caption{Plot of the residuals of the timings of mid-transit of HAT-P-26\,b versus the linear term in the sinusoidal ephemeris. The dotted blue line shows the sine curve and the points represent the residuals of the measured transit times versus the linear term in the ephemeris. Filled green circles are data from \citet{vonEssen:2019}, black circles are other timings from the literature that are also reported in \citet{vonEssen:2019}, and red squares are the new timings from this work. Empty circles refer to mid-transit times estimated from incomplete transit light curves. The errorbars of the timings coming from incomplete HST light curves were increased (see text).}
\label{Fig:OC_hatp26}
\end{figure}

With these revised error bars, we find decent fits for all four types of orbital ephemeris. The sinusoidal model fits the data best and is in good agreement with that from \citet{vonEssen:2019} and is also supported by our new transit timings. In Table\,\ref{tab:hatp26:ephemeris} we provide the fitted parameters, uncertainties, and Bayesian and Akaike Information Criterion (BIC and AIC) values for the four ephemerides. The sinusoidal variation is confirmed with an amplitude significant at a level of $6\sigma$ and much better BIC and AIC values. We therefore confirm that the HAT-P-26 system shows transit timing variations; see Fig.\,\ref{Fig:OC_hatp26}.

Interpreting the oscillating $O-C$ as a light time effect, we find a projected semi-major axis of the orbit of the star-planet system around  the centre of mass with the third body $a_{1} \sin i \sim 0.186 \pm 0.030$ au and a mass function in solar units  $(a_{1} \sin i)^{3}/P_{\rm cyc}^{2} = (6.3 \pm 4.3) \times 10^{-4}$ au$^{3}$~yr$^{-2}$, where $P_{\rm cyc} = 1167 \pm 39$~days is the $O-C$ modulation period. This corresponds to a minimum mass of the third body  $M_{\rm TB} \sim 0.07\,M_{\odot}$ assuming $M_{\rm TB} \ll M_{\rm s}$, where $M_{\rm s}$ is the mass of HAT-P-26. In the most favourable conditions, the angular separation of the third body from HAT-P-26 is only 14 mas, making its direct detection challenging, especially if it is a very faint  brown dwarf as expected if it has the same age as estimated for the star. Nevertheless, the third body hypothesis is apparently in conflict with the barycentre acceleration of the HAT-P-26 system as measured by \citet{hartman:2011} because our model predicts a mean minimum acceleration of $|\dot{\gamma}| \sim 2.2$ \,m\,s$^{-1}$\,day$^{-1}$ along the time interval of 182.69 days covered by their observations, while they gave $\dot{\gamma} =-0.028\pm0.014$\,m\,s$^{-1}$\,day$^{-1}$. An alternative hypothesis to explain the observed $O-C$ modulation is the perturbation of the orbit of HAT-P-26\,b by a distant third body as discussed in Sect.~4 of \citet{Agoletal05}. Their model requires an eccentric orbit of the third body which is associated with a non-sinusoidal shape of the $O-C$ modulation. We find a minimum eccentricity $e=0.67$ for the minimum third body mass as estimated from the mass function. Lower values of the eccentricity require a larger third body mass; for example, $e=0.33$ requires a mass four times  the minimum mass. Another kind of TTV model based on an exchange of angular momentum between the orbit and the rotation of the planet, as suggested by \citet{Lanza20}, is disfavoured by the small moment of inertia of the planet that does not allow to store enough angular momentum to account for the amplitude of the $O-C$ modulation.

HAT-P-26 is astrometrically well-behaved according to the Gaia EDR3 archive information \citep{gaia:2016,gaia:2021}: both the values of astrometric excess noise (0.10\,mas) and of renormalized unit weight error (${\rm RUWE}=1.04$) indicate that a single-star solution fits well the available astrometric data. Hot-Jupiter hosts harboring known or likely long-period, massive companions typically have ${\rm RUWE} \gtrsim 1.1$ (e.g., \citealt{belokurov:2020}), but these are not expected to be the ones responsible for the possible TTVs observed in the HAT-P-26 photometry.

We conclude that what seems to be a cyclic TTV in the HAT-P-26 system is worthy of further investigation by collecting more times of mid-transit. This will allow us to look for a non-sinusoidal shape of the modulation and to refine its period and amplitude before we can draw any sound conclusion on its origin.
\begin{table*}
\centering
\caption{\label{tab:hatp26:ephemeris} Parameters of the four ephemerides fitted to the measured times of mid-transit in the HAT-P-26 system. 
Quantities in brackets represent the uncertainties in the final digits of the preceding numbers.}
\begin{tabular}{lcccc}
\hline \hline
Quantity & Linear & Quadratic & Cubic & Sinusoidal \\
\hline
$T_0$ (BJD/TDB)          & 2455304.652182 (32) & 2455304.652181(32)              & 2455304.652180 (29)             & 2455304.65234 (35) \\
Linear term (d)          & 4.23450158 (18)     & 4.2345025 (15)                  & 4.2345179 (43)                  & 4.23450213 (76)    \\
Quadratic term (d)       &                     & $(-1.8 \pm 2.7) \times 10^{-9}$ & $(-6.7 \pm 1.7) \times 10^{-8}$ &                    \\
Cubic term (d)           &                     &                                 & $(6.7 \pm 1.8) \times 10^{-11}$ &                    \\
Sine period (epochs)     &                     &                                 &                                 & $275.5 \pm 9.1$    \\
Sine amplitude (s)       &                     &                                 &                                 & $93 \pm 15$        \\
Sine phase (BJD/TDB)     &                     &                                 &                                 & $0.15 \pm 0.33$    \\
\hline                                         
$\chi^2_\nu$             & 1.67                & 1.66                            & 1.40                            & 1.07               \\
AIC                      & 73.3                & 74.6                            & 56.9                            & 32.5               \\
BIC                      & 75.8                & 78.3                            & 61.7                            & 35.0               \\
rms of the residuals (s) & 91.0                & 91.4                            & 83.1                            & 56.7               \\
\hline
\end{tabular}
\end{table*}
\\ [6pt]
{\bf HAT-P-29.} 
Photometric follow-up of HAT-P-29\,b transit events were performed by \citet{wang:2018} and \citet{mallonn:2019}. In particular, \citet{mallonn:2019} recorded two incomplete transit light curves with the Stella 1.2\,m telescope at the Izana Observatory, while \citet{wang:2018} reported the observations of one complete and six incomplete transit light curves with the Schmidt telescope at the Xinglong Station; these authors also observed a complete light curve with the 1\,m telescope operated at the Weihai Observatory. All these light curves have point-to-point scatters larger than 2\,mmag. 

The two timings reported by \citet{mallonn:2019} are early by about 900\,s. Since they were both based on transit events in which the egress was not observed, their reliability is reduced (e.g., \citealt{gibson:2009}) and we decided to exclude them from the analysis.

We joined our new timings (see Fig\,\ref{fig:hatp29_lc}) with that from the discovery paper \citep{buchhave:2011} and the eight ones from \citet{wang:2018}, obtaining a total of 18 mid-transit times. We tried to model the data by using both a linear and a quadratic ephemeris, the latter in the form 
\begin{equation}
T_{\rm mid}=T_0+P_{\rm orb}E+\frac{1}{2}\frac{dP_{\rm orb}}{dE}E^2\,,
\end{equation}
where $\frac{dP_{\rm orb}}{dE}$ is the change in the orbital period between succeeding transits. The fit of the mid-transit times with a straight line gave 
\begin{equation}
\label{eq:fit_linear_1}
T_{\rm mid} = {\rm BJD_{TDB}}\, 2\,455\,838.59462\,(61) + 5.7233746\,(32) \, E ,
\end{equation}
with a $\chi_{\nu}^2=6.4$ and a root-mean-square deviation (rmsd) scatter of 244\,s. The residuals are plotted in Fig.\,\ref{fig:OC_plots}.
Instead, the best-fitting quadratic ephemeris returned 
\begin{equation}
\label{eq:fit_quadratic_1}
\begin{split}
T_{\rm mid}={\rm BJD_{TDB}}\,2\,455\,838.59442\,(59) + 5.7233823\,(59) \,E + \\ 
-(2.4 \pm 1.6)\times 10^{-8} E^2 \,, ~~~~~~~~~~~~~~~ \\
\end{split}
\end{equation}
with a $\chi_{\nu}^2=5.6$ and ${\rm rmsd}=267$\,s. We also estimated the Akaike Information Criterion (AIC) and the Bayesian Information Criterion (BIC). Both these criteria slightly prefer the quadratic model over the linear one. The rms is larger for the quadratic ephemeris versus the linear one only because it does not depend on the errorbars of the measured transit time.
As in previous works of our series (e.g., \citealt{southworth:2012,mancini:2013}) such large values of the $\chi_{\nu}$ should not be interpreted as a suggestion of TTVs, but as an indication that the uncertainties in the various $T_0$ measurements are too small. As a matter of fact, if we exclude from the analysis the five timings measured by \citet{wang:2018} from noisy data covering only part of transits, both the fits have a lower and similar reduced chi-square. We found $\chi_{\nu}^2=3.9$ and ${\rm rmsd}=249.7$\,s for the linear model and $\chi_{\nu}^2=3.9$ and ${\rm rmsd}=222.0$\,s for the quadratic model. This time, both the AIC and BIC criteria prefer the linear model over the quadratic one.

In conclusion, considering the amount of available data and their quality, we did not find a clear indication of the existence of TTVs. Further investigation needs more photometric follow-up observations of transits by HAT-P-29\,b. We stress that the new linear ephemeris that we determined (Eq.\,\ref{eq:fit_linear_1}; Fig.\,\ref{fig:OC_plots}) is such that the orbital period is $16.3\pm4.5$ seconds longer than that of the discovery paper, in good agreement with what was found by \citet{wang:2018}.

\section{Frequency analysis of the time-series light curves}
\label{Frequency_analysis_time-series}
Knowledge of the stellar rotational period $P_{\rm rot}$ is important for better characterization of a star-planet system. In particular, by combining the value of $P_{\rm rot}$ with those of the stellar radius $R_\star$ and projected rotational velocity $v\sin{i_\star}$, we can determine the inclination of the stellar spin axis $i_\star$. In turn, $i_\star$, together with the projected spin-orbit misalignment angle $\lambda$, yields an estimation of the true obliquity angle $\psi$ \citep{winn:2007}.

\begin{figure}
\resizebox{\hsize}{!}{\includegraphics{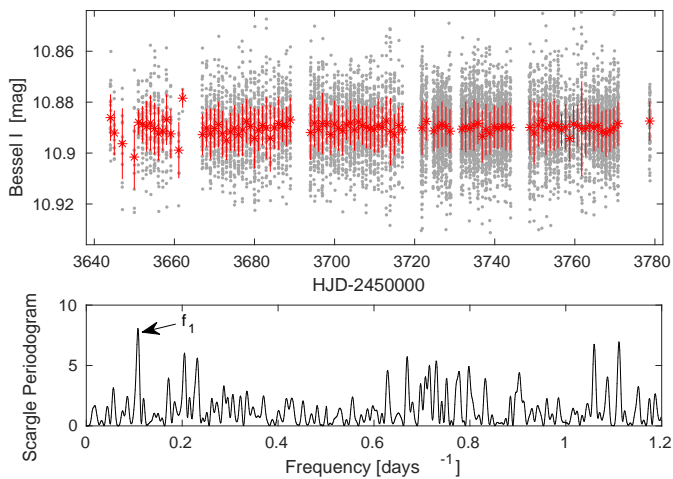}}
\caption{Top panel: HATNet photometric time series of HAT-P-15. Grey points are the original TFA3 data set, red asterisks are the data binned on a nightly base. Bottom panel: Scargle periodogram of the original data set.}
\label{Fig:HATP15_Phot}
\end{figure}

\begin{figure}
\resizebox{\hsize}{!}{\includegraphics{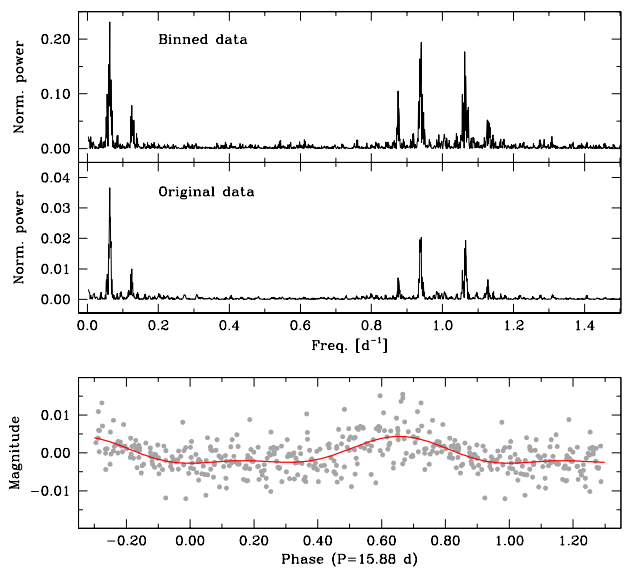}}
\caption{Detection of the rotational period of HAT-P-21. {\it Top panel}: power spectrum of the photometric measurements. The first two peaks corresponding to $f=0.063$\,d$^{-1}$ and $2f$, while the other peaks are their aliases. {\it Bottom panel}: photometric measurements (in grey) folded with $P_{\rm rot}=15.88$\,days. The error bars have been suppressed for clarity.}
\label{Fig:HATP21_Phot}
\end{figure}

We have analysed the HATNet photometric data\footnote{The original  HATNet lightcurves are publicly available at https://hatnet.org/planets/discovery-hatlcs.html} to look for possible periodic variations induced by activity and modulated by stellar rotation.  
We applied the Trend Filtering Algorithm (TFA; \citealt{kovacs:2005}) to the photometric time-series produced by the HATNet pipeline and examined all three data sets corresponding to different apertures.
\\ [6pt]
{\bf HAT-P-15} was observed on 106 different nights between September 2005 and February 2006. The typical standard deviation of photometric measurements within the same night is $\sigma \sim 9$\,mmag.
Nightly averaged photometric values have very small dispersion ($\sigma \sim 2$\,mmag), indicating
that the star had a low activity level (see the upper panel in Fig.~\ref{Fig:HATP15_Phot}).
After removing in-transit data points (190 out of 8945), we calculated the Scargle periodogram (see the bottom panel in Fig. \ref{Fig:HATP15_Phot}) and found that the highest peak occurs at $P_{1}\equiv1/f_{1}=9.3$\,days. By using the bootstrap method, we estimated for the peak at $P_{1}$ a false alarm probability (FAP) of $7.6\%$ and concluded that it is not statistically significant.
\\ [6pt]
{\bf HAT-P-17} was monitored by four different telescopes of the HAT network from June 2004 to October 2005 for a total of $\sim 250$ nights. Within the same night, photometric measurements have an average standard deviation of $\sigma \sim 6$\,mmag. We have considered first the full dataset and then, independently, the first and last longer observing seasons 
with red asterisks and blue circles respectively); in no case does the periodogram analysis show significant peaks. The low dispersion ($\sigma=1.8$\,mmag) of the nightly binned photometry indicates a very low activity level.
\\ [6pt]
{\bf HAT-P-21} was monitored from November 2006 to June 2008, for a total of 24633 single measurements. The power spectrum of the 288 daily means  clearly shows two peaks at $f=0.063$\,d$^{-1}$ and $2f=0.126$\,d$^{-1}$ (Fig.~\ref{Fig:HATP21_Phot}, top panel). 
Interpreting the signal as due to the stellar rotation, we searched for the best fit by fixing simultaneously the frequency $f$ and its harmonic $2f$, thus obtaining $P_{\rm rot}=15.88 \pm 0.02$\, days.
The folded light curve shows a flat part and a full-amplitude of $7.1$\,mmag (Fig.~\ref{Fig:HATP21_Phot},  bottom panel). Since the noise level in the power spectrum corresponds to 0.28\,mmag, the signal has  to be considered highly significant (${\rm S/N}=12.7$). 
\\ [6pt]
{\bf HAT-P-26} was observed from January to August 2009, for a total of 12223 measurements. The frequency analysis of the 150 daily means does not show any significant peak above the noise level of  0.21\,mmag. 
\\ [6pt]
{\bf HAT-P-29} was observed from October 2008 to March 2009, for a total of 3128 measurements. The frequency analysis of the 76 daily means does not show any significant peak above the  noise level of 0.25\,mmag. 
%

\section{Analysis of stellar parameters}
\label{sec:EXOFAST}
We fit the Spectral Energy Distribution (SED) via the MESA Isochrones and Stellar Tracks (MIST) \citep{dotter:2016,choi:2016} through the \texttt{EXOFASTv2} suite \citep{eastman:2019}. We fit the available archival magnitudes imposing gaussian priors on $T_{\rm eff }$ and [Fe/H] based on spectroscopic measurements and on parallax $\pi$ based on the Gaia~EDR3 astrometric measurement \citep{gaia:2016,gaia:2021}; this astrometric prior helps constraining the stellar radius and improves the precision of the stellar parameters resulting from the SED fitting procedure. The stellar parameters were simultaneously constrained by the SED and the MIST isochrones, as the SED primarily constrains $R_\star$ and $T_{\rm eff}$, and a penalty for straying from the MIST evolutionary tracks ensures that the resulting star is physical in nature. The results are shown in the Tables reported in Appendix\,\ref{appendix_Revised_physical_parameters} and they were used for the best-modelling fit of the RM effects (Sect.~\ref{subsec:spin-orbit_alignment}) as well as for reviewing the physical parameters of the systems (Sect.~\ref{sec:Final_Physical_parameters}).

\section{HARPS-N spectra analysis}
\label{sec:HARPS-N_spectra_analysis}
\subsection{Stellar atmospheric parameters}
Stellar atmospheric parameters were derived using the weighted means of all HARPS-N spectra available for the five targets. We therefore measured the equivalent widths (EWs) of iron lines taken from the list by \cite{biazzoetal2015}, and, together with the {\it abfind} driver of the MOOG code (\citealt{sneden1973}, version 2013) and the \cite{castellikurucz2004} grid of model atmospheres, we obtained effective temperature ($T_{\rm eff}$), surface gravity ($\log g$), microturbulence velocity ($\xi$), and iron abundance ([Fe/H]). In particular, we imposed the independence of the iron abundance on the line excitation potentials (for $T_{\rm eff}$) and EWs (for $\xi$), and the ionization equilibrium between \ion{Fe}{i} and \ion{Fe}{ii} (for $\log g$). All the analysis was performed differentially with respect to the Sun, thanks to a mean Vesta spectrum acquired with HARPS-N.

After fixing the stellar parameters ($T_{\rm eff}$, $\log g$, $\xi$, [Fe/H]) at the values derived through the iron line EWs, we applied the spectral synthesis method to derive the projected rotational velocity ($v \sin i$), as done in \cite{barbatoetal2019}. We therefore considered two spectral regions around 6200 and 6700\,\AA\ and used both the {\it synth} driver of the same MOOG code and the model atmospheres. 

We refer to the mentioned papers (and references therein) for further details on the procedures. The final results of the spectroscopic analysis applied here to determine the stellar atmospheric parameters are listed in Table~\ref{tab:star-par}.

\setlength{\tabcolsep}{4pt}
\begin{table}
\tiny{
\caption{Stellar atmospheric parameters determined from HARPS-N spectra.}
\label{tab:star-par}
\centering
\begin{tabular}{c c c c c c}     
\hline\hline \\ [-6pt]
Object & $T_{\rm eff}$ & $\log{g}$ & $\xi$ & [Fe/H] & $v \sin{i_{\star}}$\\
& (K) & (dex) & (km\,s$^{-1}$) & (dex)   & (km\,s$^{-1}$) \\
\hline \\ [-6pt]
HAT-P-15 & $5620 \pm 20$ & $4.45 \pm 0.15$ & $0.77\pm0.18$ & $+0.24 \pm 0.10$ & $2.3 \pm 0.5$ \\
HAT-P-17 & $5350 \pm 20$ & $4.55 \pm 0.20$ & $0.80\pm0.30$ & $+0.02 \pm 0.09$ & $0.5 \pm 0.5$ \\
HAT-P-21 & $5695 \pm 45$ & $4.28 \pm 0.15$ & $0.98\pm0.05$ & $+0.04 \pm 0.09$ & $3.5 \pm 0.5$ \\
HAT-P-26 & $5100 \pm 20$ & $4.51 \pm 0.13$ & $0.30\pm0.30$ & $+0.05 \pm 0.10$ & $1.9 \pm 0.4$ \\
HAT-P-29 & $6140 \pm 30$ & $4.39 \pm 0.14$ & $1.27\pm0.02$ & $+0.25 \pm 0.08$ & $4.5 \pm 0.8$ \\
\hline
\end{tabular}
}
\end{table}
\subsection{Determination of the spin-orbit alignment}
\label{subsec:spin-orbit_alignment}
The modelling and fitting of the RV measurements were performed by using a code that we developed within the MATLAB software ambient\footnote{MATLAB R2015b, Optimization Toolbox 7.3 and Curve Fitting Toolbox 3.5.2, The MathWorks, Inc., Natick, Massachusetts, United States.}.
A thorough description of the code was already given in \citet{esposito:2017}. Practically, we derive the best-fitting values for three parameters: the stellar projected rotational velocity $v \sin{i_{\star}}$, the systemic RV $\gamma$ and the sky-projected orbital obliquity angle $\lambda$. The other pertinent parameters (see \citealt{esposito:2017}) are kept fixed to the values found in the photometric and spectroscopic analysis, while their uncertainties are propagated for determining the error bars of $v \sin{i_{\star}}$, $\gamma$ and $\lambda$. We have upgraded the code to include the possibility to model and fit the effect of the stellar convective blueshift (CB) on the in-transit RV curve; we used a simple one-parameter model introduced by \citet{shporer:2011}. The results of the fits are summarised in Table~\ref{tab:RM}, while the best-fitting RV models are shown in Figs.~\ref{fig:RM_hatp15}, \ref{fig:RM_hatp17}, \ref{fig:RM_hatp21}, \ref{fig:RM_hatp26} and \ref{fig:RM_hatp29}, superimposed on the data points.
%
\begin{table}
\centering
\caption{Parameters from the best-fitting models of the RM effect for the five planetary systems.}
\label{tab:RM}
\resizebox{\hsize}{!}{
\begin{tabular}{c c c c c}
\hline\hline \\ [-6pt]
Object & $\lambda$ & $v \sin{i_{\star}}$ & $\gamma$       & CBV    \\
       & (degree)  & (km\,s$^{-1}$)      & (km\,s$^{-1}$) & (km\,s$^{-1}$)  \\
\hline \\ [-6pt]
HAT-P-15  & 
$\begin{tabular}{c}
$25 \pm 23$ \\
${\bf 13 \pm 6}$ \\
\end{tabular}$ &
$\begin{tabular}{c}
$1.58 \pm 0.29$ \\
${\bf 1.53 \pm 0.25}$ \\
\end{tabular}$ &
$\begin{tabular}{c}
$31.7627 \pm 0.0014$ \\
${\bf 31.7622 \pm 0.0009}$ \\
\end{tabular}$ &
$\begin{tabular}{c}
$-1.40 \pm 0.85$ \\
-- \\
\end{tabular}$ \\  [2pt]
\hline \\ [-6pt]
HAT-P-17  & 
$\begin{tabular}{c}
${\bf -27.5 \pm  6.7}$ \\
$-41.1 \pm 3.6$ \\
\end{tabular}$ &
$\begin{tabular}{c}
 ${\bf 0.84 \pm 0.07}$ \\
$1.00 \pm 0.09$ \\
\end{tabular}$ & 
$\begin{tabular}{c}
${\bf 20.3141 \pm 0.0021}$ \\
$20.3125 \pm 0.0021$ \\
\end{tabular}$ &
$\begin{tabular}{c}
${\bf -0.46 \pm 0.11}$ \\
-- \\
\end{tabular}$ \\  [2pt]
\hline \\ [-6pt]
HAT-P-21  & 
$\begin{tabular}{c}
$-0.5 \pm 12.4$ \\
${\bf -0.7 \pm 12.5}$ \\
\end{tabular}$ &
$\begin{tabular}{c}
$3.9 \pm 0.9$ \\
${\bf 3.9 \pm 0.9}$ \\
\end{tabular}$ &
$\begin{tabular}{c}
$-53.018 \pm 0.005$ \\
${\bf -53.018 \pm 0.005}$ \\
\end{tabular}$ &
$\begin{tabular}{c}
 undetected \\
-- \\
\end{tabular}$ \\  [2pt]
\hline \\ [-6pt]
HAT-P-26  & $18 \pm 49$ & -- & -- & -- \\  [2pt]%
\hline \\ [-6pt]
HAT-P-29  & 
$\begin{tabular}{c}
$-21 \pm 28$ \\
 ${\bf -26 \pm  16}$  \\
\end{tabular}$ &
$\begin{tabular}{c}
$5.1 \pm 0.9$ \\
${\bf 5.2 \pm 0.7}$  \\
\end{tabular}$ &
$\begin{tabular}{c}
$-21.6511\pm 0.0020$ \\
${\bf -21.6513 \pm  0.0019}$ \\
\end{tabular}$ &
$\begin{tabular}{c}
$ > -0.58$ \\
-- \\
\end{tabular}$ \\
\hline
\end{tabular}
}
\tablefoot{{Except for the first, the columns contain two values. Those on the top are from the fit in which we considered the stellar convective blueshift (CB), whereas those on the bottom did not. The preferred values are given in bold font for each target; see the text for details. Due to the low quality of the HAT-P-26 data, we were not able to well constrain $\lambda$ for this system, which was merely estimated by fixing the value of $v \sin{i_{\star}}$.}}
\end{table}
\\ [6pt]
{\bf HAT-P-15.} For this target, we only have the RV time-series spanning one transit and no photometric follow-up observations. Therefore, for many relevant parameters, we had to adopt values from the literature as well as from our analysis of the stellar parameters (see Sect.~\ref{sec:EXOFAST} and Table~\ref{tab:hatp15_final_parameters}).
The free parameters in our fit are $\lambda$, $v \sin{i_{\star}}$ and $\gamma$. Since the uncertainties on the ephemerides reported in \citet{kovacs:2010} propagated to an uncertainty of 10 minutes on the mid-transit time at the epoch of our observations, we also included the time of periastron as a free parameter. 

We considered both models with and without the CB effect and show the results of the fits in Fig. \ref{fig:RM_hatp15}. We used the Bayesian Information Criterion (BIC) to compare the two models. With $\Delta{\rm BIC}=-0.95$, the model with CB is only marginally better. The best fit value of CBV $= -1.40 \pm 0.85$ km\,s$^{-1}$ is suspiciously high for a G5 star \citep{dravins:1990}, and we think it is driven by the first three in-transit data points. Therefore we prefer
to adopt the best fit values of the model without the CB effect, for which $\lambda=13^{\circ} \pm 6^{\circ}$.
\\ [6pt]
{\bf HAT-P-17} was already observed with the Keck/HIRES for detecting the RM effect, as reported by \citet{fulton:2013}. Due to the slow stellar rotational velocity ($v\sin{i_\star} = 0.56_{-0.14}^{+0.12}$\,km\,s$^{-1}$), they estimated the amplitude of the RM effect to be only $\sim$7\,m\,s$^{-1}$. As a consequence, they remarked on the need to model also the effect of the CB in order to derive a correct estimation of $\lambda$. Indeed, without the CB effect modelling, they  obtained $\lambda= 37 \pm 12$\,deg, whereas with the CB effect they found $\lambda= 19_{-16}^{+14}$\,deg, and a CB velocity parameter ${\rm CBV} = -0.65 \pm 0.23$\,km\,s$^{-1}$.

First, we used our code to make an independent fit of the HIRES RVs. 
Without modelling the CB effect, we obtained $\lambda = -28$\,deg and $v\sin{i_\star} = 0.8$\,km\,s$^{-1}$; accounting for the CB effect we derived $\lambda=-17$\,deg, $v\sin{i_\star}=0.7$\,km\,s$^{-1}$, and ${\rm CBV} = -0.33$\,km\,s$^{-1}$. Provided that \citet{fulton:2013} are most likely using a different convention for the sign of $\lambda$ (compare their Fig.~3 with our Fig.~\ref{fig:RM_hatp17}), our results are compatible with theirs: considering the CB effect results in a value of $\lambda$ closer to zero.
However we obtain a significantly smaller value for CBV.
Unlike \citet{fulton:2013}, we have rejected two of their data points as outliers (see Fig.~\ref{fig:RM_hatp17}) because they showed residuals larger than $4\,\sigma$. Correspondingly, we find a value for the CBV closer to zero, that is ${\rm CBV} = -0.33$\,km\,s$^{-1}$. Instead, by including all the data points, our best-fitting value for the CBV is $-0.58$\,km\,s$^{-1}$, which is similar to the result found by \citet{fulton:2013}.

Next, we analysed our HARPS-N RV data set, by using the same approach as for the HIRES data. The best fit values, without the CB effect, are $\lambda=-55$\,deg and $v\sin{i_\star}=1.6$\,km\,s$^{-1}$. By modelling also the CB effect, we obtained $\lambda=-37$\,deg, $v\sin{i_\star}=1.1$\,km\,s$^{-1}$, and ${\rm CBV} = -0.57$\,km\,s$^{-1}$. Although with a marginal statistical significance, both for HIRES and HARPS-N data sets the model including the CB effect is to be preferred, as we obtain smaller BIC: $\Delta$BIC $= -0.9$ and $-4.1$ for HIRES and HARPS-N, respectively.

Finally, we made a combined fit of the HIRES and HARPS-N RVs, and estimated the uncertainties on the best-fit parameters using the bootstrap method. The best-fit RV curve models are displayed, together with the RV data sets, in Fig.~\ref{fig:RM_hatp17}. The left panels shows the model considering the RM effect only, whereas the model in the right panels accounts also for the CB effect. In the first case we derive $v\sin{i_\star} = 1.00\pm0.09$\,km\,s$^{-1}$, $\lambda = -41.1\pm3.6$\,deg, while in the second case we obtain ${\rm CBV}=-0.46 \pm 0.11$\,km\,s$^{-1}$, $v\sin{i_\star}=0.84\pm 0.07$\,km\,s$^{-1}$, $\lambda = -27.5 \pm 6.7$\,deg. Also for the combined fit, the model that includes the CB effect is to be preferred as we obtain $\Delta{\rm BIC}=-6.7$. We adopt this latter value as our final estimation of $\lambda$ for HAT-P-17\,b.
\\ [6pt]
{\bf HAT-P-21} For this target, we adopted all
the planetary and stellar relevant parameters
as obtained from the analysis of the light curves
and the stellar spectra (see Table \ref{tab:hatp21_final_parameters}).
The best-fitting values of the parameters determined 
by the analysis of the in-transit RV curve are reported
in Table \ref{tab:RM}. In particular, we obtained $\lambda=-0.7^{\circ}\pm12.5^{\circ}$.

The fit using the model which include the CB effect gives a value of CBV very close to zero, and 
correspondingly the other fitted parameters have the same values as in the fit  without CB. We notice that HAT-P-21\,b has an impact parameter of $b=0.62$, i.e.\ the planet never occults the central part of the stellar disk and therefore we expect the RV variations due to the CB effect to be small. We conclude that the data do not have the quality needed for a solid detection of the CB effect.

As we know the rotational period of its parent star (see Sect.~\ref{Frequency_analysis_time-series}), we are able to estimate the quantity $i_{\star}$ using the following equation 
\begin{equation}
P_{\mathrm{rot}}\approx \frac{2 \pi R_{\star}}{v\sin{i_{\star}}}\sin{i_{\star}}, %
\label{Eq:1}
\end{equation}
which resulted to be $62^{\circ}\pm16^{\circ}$. Knowing $i_\star$, $i$ and $\lambda$, we can calculate the true misalignment angle via \citep{winn:2007}
\begin{equation}
\cos{\psi}=\cos{i_{\star}}\cos{i}+\sin{i_{\star}}\sin{i}\cos{\lambda}. %
\label{Eq:2}
\end{equation}
The value that we obtained is $\psi=25^{\circ} \pm 16^{\circ}$.
\\ [6pt]
{\bf HAT-P-26} Based on published ephemerides, the first point of the HARPS-N time-series observations was taken $\sim 15$ minutes after the transit had already started. Instead, the last points present a large scatter due to worsening weather conditions (see Fig.~\ref{fig:RM_hatp26}). Therefore, the fit of the data does not allow to constrain $\lambda$ at the required precision as in the other four cases presented in this work. We made a putative estimate of $\lambda$ by fixing the value of $v \sin{i_{\star}}$ to that estimated from the spectroscopy (see Table~\ref{tab:star-par}) and finding $\lambda=18^{\circ} \pm 49^{\circ}$, which suggests a a prograde orbit for HAT-P-26\,b.  
\\ [6pt]
{\bf HAT-P-29}  We find that the fit of the data using a model which includes the CB effect is slightly disfavoured with respect to the fit without CB  ($\Delta{\rm BIC}=3.0$); given the relatively large value of $v \sin{i_{\star}}=5.2$\,km\,s$^{-1}$, we expected that
the CB effect can only have a minor impact on the shape of the in-transit RV curve.
Therefore, for HAT-P-29 we adopt the results of the fit without CB, that is $\lambda=-26^{\circ} \pm 16^{\circ}$. However, the analysis of the fit with the CB provides us with a useful $1\,\sigma$ lower limit ($>-0.58$\,km\,s$^{-1}$) to the value of CBV.
%
\section{Physical parameters}
\label{sec:Final_Physical_parameters}
Considering the new photometric data available for HAT-P-17, HAT-P-21, HAT-P-26 and HAT-P-29, we reviewed the physical properties of these planetary systems. We followed the {\it Homogeneous Studies} approach (\citealp{southworth:2012} and references therein) and combined the parameters obtained from the light curves and spectroscopic observations, placing constraints on the properties of the host stars, which we can deduce from theoretical evolutionary models of stars. In particular, we used the following spectroscopic properties of the host stars, which we obtained from the analysis of the stellar spectra (see Sect.~\ref{sec:HARPS-N_spectra_analysis}): the projected rotational velocity $v \sin{i_{\star}}$, the effective temperature $T_{\rm eff}$, the logarithmic surface gravity $\log{g}$ and the iron abundance. 

Since the HARPS-N data were collected during transit events for measuring the RM effect, we do not have new out-of-transit RV points for redetermining the velocity amplitude, $K_{\star}$, of the RV curves. Therefore, we adopted the values from the literature; they are reported in Tables~\ref{tab:hatp15_final_parameters}, \ref{tab:hatp17_final_parameters}, \ref{tab:hatp21_final_parameters}, \ref{tab:hatp26_final_parameters} and \ref{tab:hatp29_final_parameters}, together with the other relevant parameters. 

With these input parameters, as in \citet{mancini:2018}, we made use of the {\sc jktabsdim} code \citep{southworth:2009} to make new estimates of the main physical properties of the planetary systems HAT-P-17, HAT-P-21, HAT-P-26 and HAT-P-29. By iteratively modifying the velocity amplitude of the planet, {\sc jktabsdim} maximizes the agreement between the measured $T_{\rm eff}$ and $R_{\star}/a$ and  with those predicted by a set of five theoretical models. A wide range of possible ages for each of the host stars was considered. The code returned five different estimates for each of the output parameters, one for each set of theoretical models, and we took the unweighted means as the final values of the parameters. The corresponding systematic uncertainties, caused by the use of theoretical models, were calculated considering the maximum deviation between the values of the final parameters and the single ones coming from the five theoretical models. Instead, statistical uncertainties were propagated from the error bars in the values of the input parameters. The final values are reported in Tables~\ref{tab:hatp17_final_parameters}, \ref{tab:hatp21_final_parameters}, \ref{tab:hatp26_final_parameters} and \ref{tab:hatp29_final_parameters}, together with values taken from the literature, which are shown there for comparison purposes. Our estimations of the radii and masses for the stars and planets are 
all within the error bars of literature determinations, but slightly more precise.

\section{Discussion}
\label{discussion}

\subsection{On the spin-orbit alignment of hot Jupiters}
\begin{figure*}
\resizebox{\hsize}{!}{\includegraphics{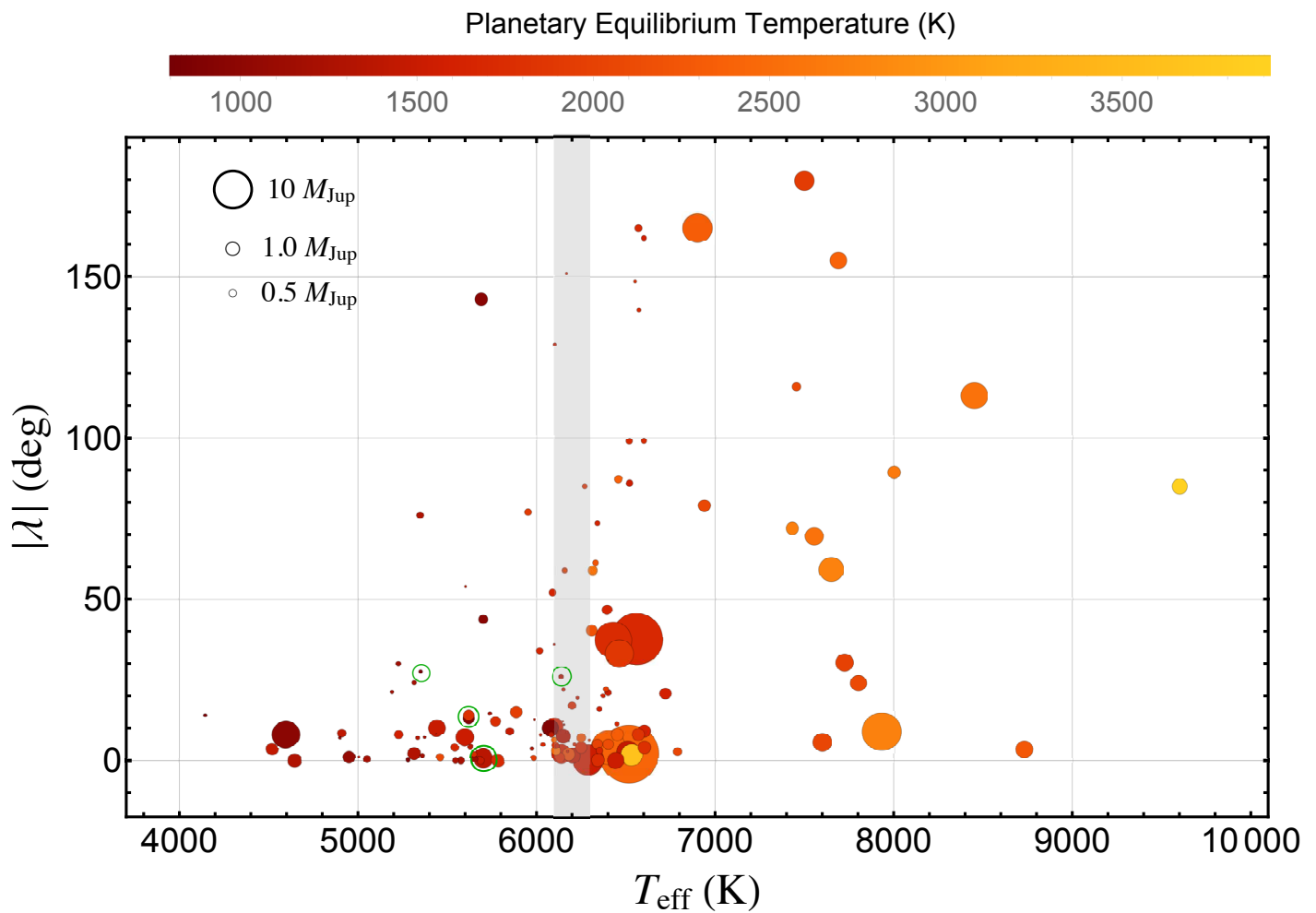}}
\caption{Absolute values of the sky-projected orbital obliquity angles of close-in exoplanets (with mass $0.3\,M_{\rm Jup} < M_{\rm p} < 13\,M_{\rm Jup}$ and $a/R_{\star}<25$), as a function of the host star's effective temperature. The planets are represented by circles, whose sizes is proportional to their mass. The error bars have been suppressed for clarity. Colour indicates their equilibrium temperature. The grey zone should discriminate two different populations of hot Jupiters, according to several authors (e.g., \citealt{winn:2010,albrecht:2012}). The planets surrounded by the green circles are those examined in this work, except for HAT-P-26\,b. The other data were taken from TEPCat in February 2022.}
\label{Fig:diagram_obliquity_1}
\end{figure*}

At present (February 2022), TEPCat lists determinations of $\lambda$ for roughly 170 transiting exoplanets, while $\psi$ has been constrained for only 39. Most of them are hot Jupiters, which represents the only class for which we have a good statistical sample. What can be deduced from this collection of $\lambda$ measurements for hot Jupiters is a matter of debate. So far, no convincing correlations have emerged from plotting the projected obliquity versus other parameters, such as planetary radius and mass, orbital separation, stellar age, etc. Early studies (e.g., \citealt{winn:2010,albrecht:2012,dawson:2014,{tregloan:2015}}), which were based on a smaller sample of $\lambda$ measurements, have tentatively identified two populations of more or less aligned hot Jupiters based on the effective temperature of their parent stars. Considering our new data and those listed in TEPCat, we plotted the absolute values of $\lambda$ of hot Jupiters  ($0.3\,M_{\rm Jup} < M_{\rm p} < 13\,M_{\rm Jup}$ and $a/R_{\star}<25$) versus $T_{\rm eff}$ in Fig.~\ref{Fig:diagram_obliquity_1}. Of the more than 130 planets appearing in the diagram, only 14 have an eccentricity $e>0.1$; four of them are the planets presented in this study. 

The separation line between the two groups is related to the Kraft break (the remarkable decrease in the rotation velocities observed in main-sequence stars later than F5, \citealt{kraft:1967}), and falls somewhere between 6090 and 6300\,K (see the gray zone in Fig.~\ref{Fig:diagram_obliquity_1}); planets orbiting stars with mostly convective (radiative) outer envelopes are on the left (right) side of this plot.

Cooler than radiative stars, the convective stars are supposed to have a rapid tidal dissipation as the convective cells produce the turbulent cascades that lead to energy loss. Instead, the radiative stars are thought to have much weaker tidal dissipation. Consequently, the orbit of hot Jupiters hosted by relatively cool stars ($T_{\rm eff} < 6100$\,K) should be much more aligned with the spin of their hosts because tides limit possible obliquity on timescales much shorter than those related to the orbital decay of hot Jupiters. Specifically, the tidal modes responsible for the damping of the obliquity may be different from those producing the orbital decay and can act on a remarkably shorter timescale producing a spin-orbit alignment without a significant decrease of the orbital semi-major axis \citep{lai:2012,valsecchi:2014,Lanza22}. As we can see from Fig.~\ref{Fig:diagram_obliquity_1}, although high-obliquity hot Jupiters were found regularly above the Kraft break, there are several exceptions that challenge this theory.

\begin{figure*}
\centering
\resizebox{.8\hsize}{!}{\includegraphics{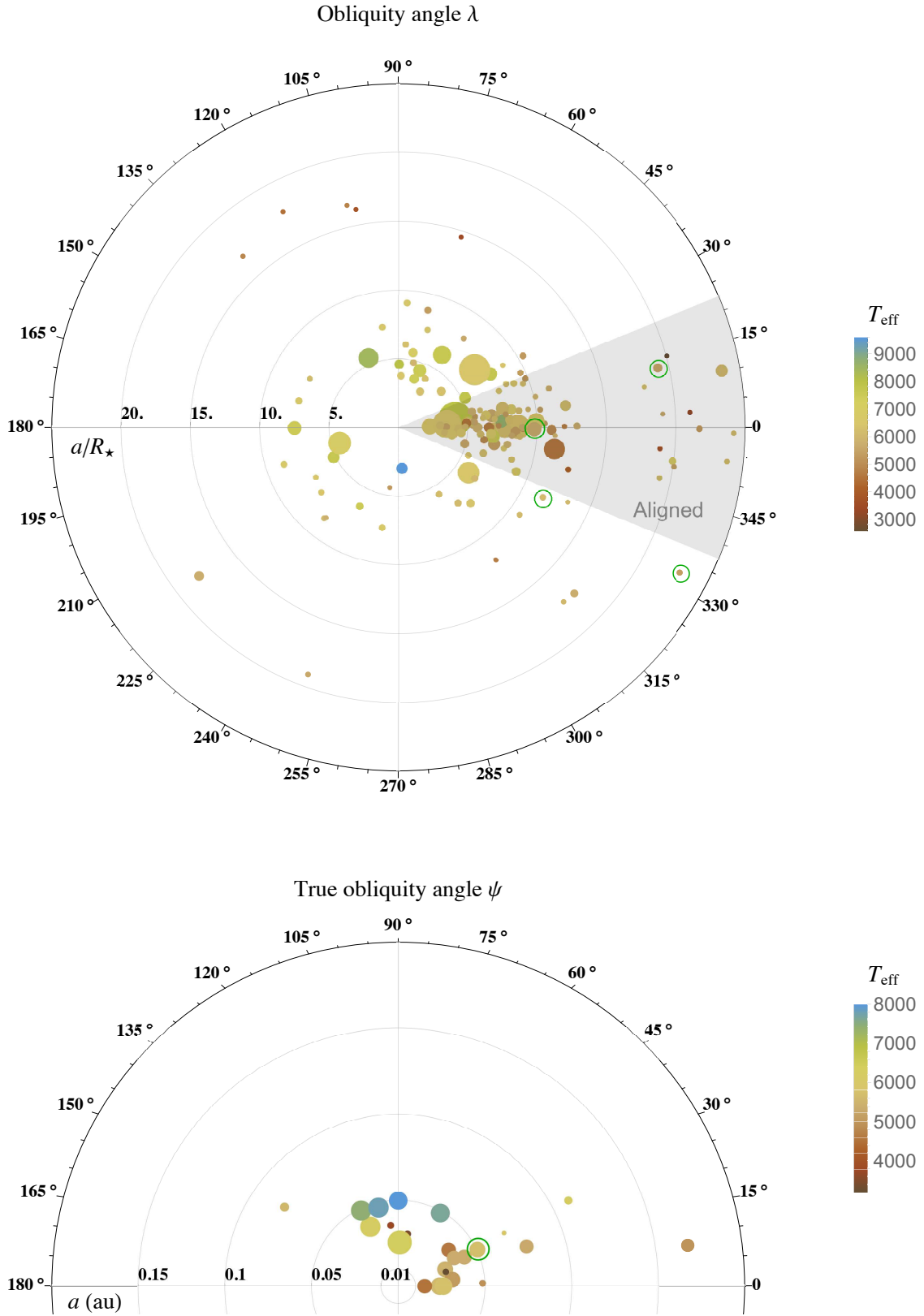}}
\caption{{\bf Top panel:} Sky-projected orbital obliquity of known exoplanets as a function of their scaled orbital distance $a/R_{*}$ from the host star. The plot includes all planets with $M_{\rm p} < 13\, M_{\rm Jup}$ and $a/R_{\star}<25$. They are represented by circles, whose sizes is proportional to their mass. The planets surrounded by the green circles are those examined in this work, except for HAT-P-26\,b.
{\bf Bottom panel:} True orbital obliquity of known exoplanets. They are represented by circles, whose size is proportional to their radius. The planet surrounded by the green circle is HAT-P-21\,b, which was examined in this work.
{\bf Both panels:} The error bars have been suppressed for clarity. Colour indicates the effective temperature of their parent stars. The data were taken from TEPCat in February 2022. Figure inspired by similar plots from J. Winn; see also \citet{zhou:2019}.}
\label{Fig:diagram_obliquity_2}
\end{figure*}

In order to have a more general picture of the current situation, we refer the reader to the top panel of Fig.~\ref{Fig:diagram_obliquity_2}, where we made a polar plot of the measured sky-projected obliquities of all known systems hosting a planet with $M_{\rm p}<13\,M_{\rm Jup}$ and a scaled orbital distance $a/R_{*}$ from the host star up to 25.
Even for systems hosting smaller planets, it is difficult to see a correlation between $\lambda$ and $T_{\rm eff}$.
\begin{figure}
\resizebox{\hsize}{!}{\includegraphics{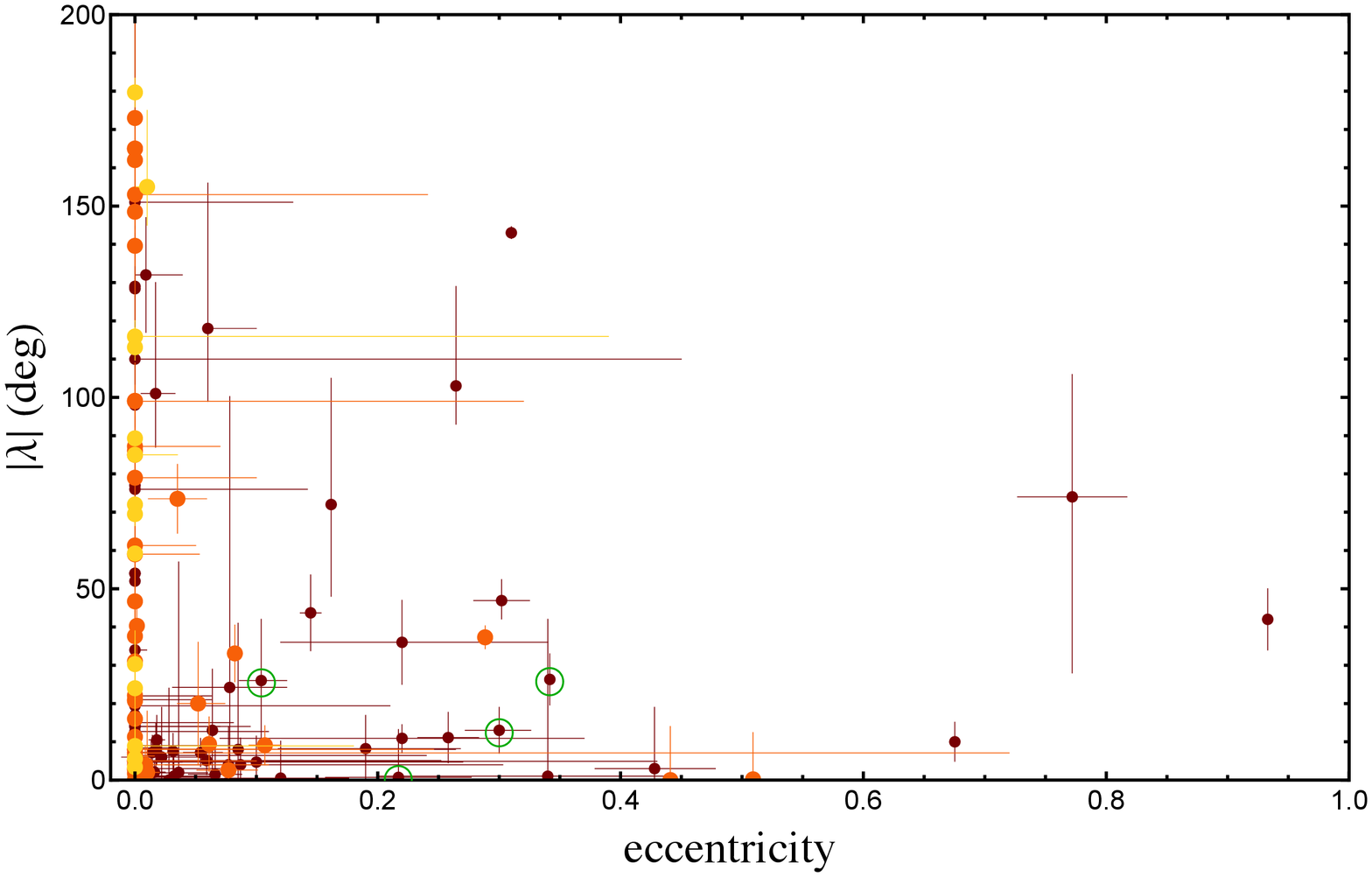}}
\caption{Sky-projected orbital obliquity of known exoplanets as a function of their orbital eccentricity. Different colours indicate a different value of $T_{\rm eff}$: dark red are for cool hosts ($T_{\rm eff}<6250$\,K); orange are for hot hosts $6250<T_{\rm eff}<7000$\,K; yellow are for very hot hosts $T_{\rm eff}<7000$\,K. Many of the planet with eccentricity equal to zero does not have horizontal error bars. The planets surrounded by the green circles are those examined in this work, except for HAT-P-26\,b. The other data were taken from TEPCat in February 2022}
\label{Fig:diagram_obliquity_e}
\end{figure}

Based on the expectations for the rotation velocities of stars with effective temperatures between 5900 and 6600\,K from \citet{louden:2021}, \citet{albrecht:2021} calculated the expected $\psi$ for a sample of 57 planetary systems. They found that perpendicular orbits ($\psi= 80^{\circ}-125^{\circ}$) are statistically favoured.

According to TEPCat, there are now 25 exoplanets for which we know their radius and have the measurement of $\psi$. They are shown in another polar plot (bottom panel of Fig.~\ref{Fig:diagram_obliquity_2}), in which most of the exoplanets orbiting cool stars have $\psi<30^{\circ}$, whereas the four exoplanets orbiting hot stars ($T_{\rm eff}\geq 7650$\,K) have $60^{\circ}<\psi<135^{\circ}$. However, since the statistical sample is not yet significant, it is hard to make strong assertions.

\subsection{Tidal-alignment timescales for the five systems of our study}
An estimate of the tidal-alignment timescales in the specific cases of our four systems (HAT-P-15, HAT-P-17, HAT-P-21, and HAT-P-29), based on an adapted version of the tidal model of \citet{Leconteetal10} already used in our previous investigations \citep[e.g.,][]{esposito:2017}, shows that only HAT-P-21 has an e-folding timescale for the damping of its obliquity of $\sim 0.3$~Gyr, shorter than the main-sequence lifetime of the star, when we adopt a stellar modified tidal quality factor $Q^{\prime}_{\rm s}=10^{6}$, as expected for an efficient dissipation of the obliquity tides as in the model by \citet{lai:2012}. The age estimated from the observed rotation period of HAT-P-21 using gyrochronology, that is $ \sim 1.5-2.0$~Gyr, is in tension with the age estimated from the isochrone fitting ($\sim 7.8 \pm 2.6$~Gyr). Its relatively fast rotation could be due to the tides produced by the close-by massive planet that tend to spin-up HAT-P-21 with a characteristic e-folding timescale of $\sim 4$~Gyr, if we assume a stellar modified tidal quality factor $Q_{\rm s}^{\prime} = 10^{7}$ as suggested, for example, by \citet{Jacksonetal09}.  The e-folding timescale for the damping of the orbital eccentricity is $\sim 3.3$~Gyr, when we adopt modified tidal quality factors $Q_{\rm s}^{\prime} = 10^{7}$ for the star and $Q_{\rm p}^{\prime} =10^{7}$ for the planet, respectively. The modified tidal quality factor of the planet is obtained by scaling the value of Jupiter to the slow rotation of HAT-P-21\,b assumed to be synchronized with its orbital motion \citep[cf.][]{Ogilvie14}. These considerations suggest that HAT-P-21 is indeed an old star and that the planet could have migrated close to its host through an orbit of initially high eccentricity that was significantly reduced by tides during the main-sequence lifetime of the system together with any initially large obliquity. 

Considering the other systems, their orbits also show a significant eccentricity. Assuming $Q_{\rm s}^{\prime}=10^{7}$ and $Q_{\rm p}^{\prime} = 10^{7}$ as in the case of HAT-P-21 and HAT-P-21b, respectively, the e-folding timescale for the damping of the eccentricity, $\tau_{\rm e}$, is  comparable with the estimated age of the system only in the case of HAT-P-26 ($\tau_{\rm e} \sim 6$~Gyr), while it is longer for the other systems. The rotation of the hosts and the obliquity of the planetary orbits were  not significantly affected by the tides during their main-sequence evolution, even assuming a strong interaction with $Q_{\rm s}^{\prime} = 10^{6}$. In conclusion, these considerations suggest that, with the possible exception of HAT-P-21, the  rather small spin-orbit misalignments observed in the other systems are likely to be primordial as well as their significant eccentricities. The only possible exception could be the eccentricity of HAT-P-26, which might require an excitation  by a third body to account for the observed values.

\section{Summary}
\label{summary}
Within the GAPS programme, we are observing a sample of transiting-exoplanet systems, mostly hosting hot Jupiters, with the HARPS-N spectrograph, supported by an array of medium-class telescopes. The aim is to better characterize these planetary systems and get information about the degree of orbital alignment of this class of planets, according to the characteristics of their parent stars.

We divided in two groups the five targets under study in this work. In the first group there are four hot Jupiters (HAT-P-15, HAT-P-17, HAT-P-21, and HAT-P-29), while in the second group there is one Neptune-mass planet (HAT-P-26). Details of our new observations are reported in Tables~\ref{tab:logspec} and \ref{tab:logs}. Thanks to new spectroscopic and photometric observations that we collected, and the public data from TESS and \textit{Gaia}, we were able to ($i$) review their physical and orbital parameters and ($ii$) reveal the RM effect, during transit events, and measure the spin-orbit alignment of these systems. 

Our main results are as follows.
\begin{itemize}
\item[$\bullet$] We revised most of the physical parameters of the five planetary systems. Our results are reported in Tables~\ref{tab:hatp15_final_parameters}, \ref{tab:hatp17_final_parameters}, \ref{tab:hatp21_final_parameters}, \ref{tab:hatp26_final_parameters} and \ref{tab:hatp29_final_parameters}, and are in good agreement (and, in general, slightly more accurate) with those obtained previously by other authors. \\ [-6pt]
\item[$\bullet$] 
We estimated new mid-transit times for four of the systems (HAT-P-17, HAT-P-21, HAT-P-26, and HAT-P-29) and augmented them with published values to obtain lists of transit times. They are reported in Tables~\ref{tab:transit_times_hatp17}, \ref{tab:transit_times_hatp21}, \ref{tab:transit_times_hatp26} and \ref{tab:transit_times_hatp29}. They were used for updating the orbital periods and expected mid-transit times of the systems. We also searched for evidence of TTVs. Our analysis shows an indication of possible TTVs in the HAT-P-29 planetary system, which must be verified with more data. We determined a new linear ephemeris with an orbital period $17\pm4.2$\,s longer than that found by \citet{buchhave:2011}. We also confirmed the much stronger indication of TTVs that was found by \citet{vonEssen:2019} for the HAT-P-26 system. This sinusoidal variation may be caused by a third body in the system. More follow-up observations are required to confirm its existence.
\\ [-6pt]
\item[$\bullet$] 
The frequency analysis of the HATNet photometric time-series for HAT-P-21 highlighted a modulation caused by stellar activity, allowing us to get a measurement of the rotational period of the star of $P_{\rm rot}=15.88 \pm 0.02$\,d. A similar analysis, performed for the HATNet light curves of the other four stars, did not unearth out any clear photometric modulation.
\\ [-6pt]
\item[$\bullet$] 
We used the HARPS-N spectrograph to monitor one transit for each of the five planets. The RM effect was completely covered for HAT-P-15, HAT-P-17, HAT-P-21, HAT-P-29 and only partially for HAT-P-26, which also suffered from adverse weather conditions at the end of the observations.
We successfully measured the sky-projected orbital obliquity for four of the systems, obtaining $\lambda=13^{\circ}\pm6^{\circ}$, $\lambda=-26.3^{\circ}\pm6.7^{\circ}$, $\lambda=-0.7^{\circ}\pm12.5^{\circ}$, $\lambda=-26^{\circ}\pm16^{\circ}$, for HAT-P-15\,b, HAT-P-17\,b, HAT-P-21\,b and  HAT-P-29\,b, respectively, all indicating good spin-orbit alignments within the uncertainties. Even though we were not able to constrain $\lambda$ for HAT-P-26\,b, the modelling of the data returns a value, $\lambda=18^{\circ}\pm49^{\circ}$, that also suggests a prograde orbit for this planet; this is also supported by the shape of the RM effect, see Fig.~\ref{fig:RM_hatp26}. Finally, for the HAT-P-21 system, we were able to determine its true obliquity, obtaining $\psi=25^{\circ} \pm 16^{\circ}$.
\end{itemize}

We also discuss the case that the sky projected spin-orbit misalignment of exoplanets, especially hot Jupiters, can be correlated with the temperature of their parent stars.
We confirm that hot Jupiters with low-obliquity are regularly found orbiting convective stars, with effective temperature below the Kraft break, which have a rapid tidal dissipation when compared with radiative stars. However, the existence of several exceptions makes things not so clear. A further effort to enlarge the sample is required in order to shed new light on the matter.

Finally, we roughy estimated the tidal-alignment timescales of the systems under study and made some deductions about the origin of their rather small spin-orbit misalignments that we measured.

\begin{acknowledgements}
This paper is based on observations collected with the following telescopes: the 3.58\,m Telescopio Nazionale Galileo (TNG), operated on the island of La Palma (Spain) by the Fundaci\'{o}n Galileo Galilei of the INAF (Istituto Nazionale di Astrofisica) at the Spanish Observatorio del Roque de los Muchachos, in the frame of the programme Global Architecture of Planetary Systems (GAPS); the Zeiss 1.23\,m telescope at the Centro Astron\'{o}mico Hispano Alem\'{a}n (CAHA) in Calar Alto (Spain); the Cassini 1.52\,m telescope at the Astrophysics and Space Science Observatory of Bologna in Loiano (Italy); the Copernico telescope (Asiago, Italy) of the INAF - Osservatorio Astronomico di Padova; the 0.82\,m IAC\,80 Telescope, operated on the island of Tenerife by the Instituto de Astrof\'{i}sica de Canarias in the Spanish Observatorio del Teide. The HARPS-N instrument has been built by the HARPS-N Consortium, a collaboration between the Geneva Observatory (PI Institute), the Harvard-Smithonian Center for Astrophysics, the University of St. Andrews, the University of Edinburgh, the Queen's University of Belfast, and INAF. This research made use of Lightkurve, a Python package for Kepler and TESS data analysis (Lightkurve Collaboration, 2018). The other reduced light curves presented in this work will be made available at the CDS (http://cdsweb.u-strasbg.fr/).
This work has made use of data from the European Space Agency (ESA) mission {\it Gaia} (\url{https://www.cosmos.esa.int/gaia}), processed by the {\it Gaia} Data Processing and Analysis Consortium (DPAC, \url{https://www.cosmos.esa.int/web/gaia/dpac/consortium}). Funding for the DPAC has been provided by national institutions, in particular the institutions participating in the {\it Gaia} Multilateral Agreement.
We thank Roberto Gualandi for his technical assistance at the Cassini telescope.
We thank the support astronomers of CAHA for their technical assistance at the Zeiss telescope.
L.\,M. acknowledges support from the ``Fondi di Ricerca Scientifica d'Ateneo 2021'' of the University of Rome ``Tor Vergata''.
We acknowledge the use of the following internet-based resources: the ESO Digitized Sky Survey;
the TEPCat catalogue; the SIMBAD database operated at CDS, Strasbourg, France; and the arXiv scientific paper preprint service operated by Cornell University.
ME acknowledges the support of the DFG priority program SPP  992  ``Exploring the Diversity of Extrasolar Planets'' (HA 3279/12-1).
\end{acknowledgements}

%
%

\begin{appendix}

\section{HARPS-N RV measurements} 
\label{appendix_RV}
The RV measurements, which were obtained with HARPS-N (this work), are reported in this appendix.
%
\begin{table*}[ht]
\caption{HARPS-N RV data for HAT-P-15.}
\label{tab:RV_HAT-P-15}
\centering
\begin{tabular}{rrrrrrrc}
\hline
\hline
 BJD (TDB)      & T$_{\rm exp}$   &  RV      & error  &    FWHM   & Bis. Span &  Airmass & Flag \\ 
                & [sec] &[km\,s$^{-1}$] & [km\,s$^{-1}$] & [km\,s$^{-1}$] & [km\,s$^{-1}$] & &  \\
\hline
2\,457\,343.405902   & 900 &  31.7623 & 0.0040 & 7.20 &  $-$0.018 & 1.62  &  o \\ 
2\,457\,343.416435   & 900 &  31.7730 & 0.0041 & 7.19 &  $-$0.025 & 1.52  &  o \\ 
2\,457\,343.427373   & 900 &  31.7659 & 0.0040 & 7.20 &  $-$0.012 & 1.44  &  o \\ 
2\,457\,343.438021   & 900 &  31.7601 & 0.0034 & 7.18 &  $-$0.031 & 1.37  &  o \\ 
2\,457\,343.448820   & 900 &  31.7675 & 0.0034 & 7.19 &  $-$0.015 & 1.30  &  i \\ 
2\,457\,343.459318   & 900 &  31.7639 & 0.0034 & 7.19 &  $-$0.036 & 1.25  &  i \\ 
2\,457\,343.470117   & 900 &  31.7646 & 0.0042 & 7.19 &  $-$0.030 & 1.21  &  i \\ 
2\,457\,343.481032   & 900 &  31.7661 & 0.0042 & 7.18 &  $-$0.032 & 1.17  &  i \\ 
2\,457\,343.491564   & 900 &  31.7729 & 0.0032 & 7.21 &  $-$0.017 & 1.13  &  i \\ 
2\,457\,343.502213   & 900 &  31.7681 & 0.0033 & 7.20 &  $-$0.027 & 1.11  &  i \\ 
2\,457\,343.512827   & 900 &  31.7708 & 0.0034 & 7.21 &  $-$0.021 & 1.08  &  i \\ 
2\,457\,343.524991   & 900 &  31.7655 & 0.0038 & 7.19 &  $-$0.021 & 1.06  &  i \\ 
2\,457\,343.535397   & 900 &  31.7650 & 0.0037 & 7.19 &  $-$0.018 & 1.05  &  i \\ 
2\,457\,343.546381   & 900 &  31.7629 & 0.0034 & 7.21 &  $-$0.010 & 1.03  &  i \\ 
2\,457\,343.556798   & 900 &  31.7546 & 0.0033 & 7.19 &  $-$0.023 & 1.02  &  i \\ 
2\,457\,343.567423   & 900 &  31.7481 & 0.0035 & 7.21 &  $-$0.017 & 1.02  &  i \\ 
2\,457\,343.578222   & 900 &  31.7522 & 0.0037 & 7.20 &  $-$0.010 & 1.02  &  i \\ 
2\,457\,343.589160   & 900 &  31.7487 & 0.0036 & 7.19 &  $-$0.014 & 1.02  &  i \\ 
2\,457\,343.599773   & 900 &  31.7468 & 0.0032 & 7.20 &  $-$0.033 & 1.02  &  i \\ 
2\,457\,343.610468   & 900 &  31.7397 & 0.0030 & 7.18 &  $-$0.025 & 1.03  &  i \\ 
2\,457\,343.620955   & 900 &  31.7385 & 0.0030 & 7.21 &  $-$0.027 & 1.04  &  i \\ 
2\,457\,343.631846   & 900 &  31.7378 & 0.0032 & 7.21 &  $-$0.008 & 1.06  &  i \\  
2\,457\,343.642668   & 900 &  31.7397 & 0.0031 & 7.19 &  $-$0.028 & 1.08  &  i \\ 
2\,457\,343.653513   & 900 &  31.7355 & 0.0030 & 7.20 &  $-$0.028 & 1.10  &  i \\ 
2\,457\,343.664069   & 900 &  31.7428 & 0.0027 & 7.19 &  $-$0.020 & 1.13  &  i \\ 
2\,457\,343.674741   & 900 &  31.7517 & 0.0027 & 7.18 &  $-$0.024 & 1.16  &  i \\ 
2\,457\,343.685401   & 900 &  31.7448 & 0.0027 & 7.18 &  $-$0.025 & 1.19  &  o \\ 
2\,457\,343.696188   & 900 &  31.7463 & 0.0028 & 7.20 &  $-$0.016 & 1.24  &  o \\ 
2\,457\,343.706906   & 900 &  31.7447 & 0.0027 & 7.20 &  $-$0.025 & 1.29  &  o \\ 
2\,457\,343.717450   & 900 &  31.7408 & 0.0030 & 7.19 &  $-$0.008 & 1.35  &  o \\ 
2\,457\,343.728029   & 900 &  31.7403 & 0.0035 & 7.19 &  $-$0.023 & 1.42  &  o \\ 
2\,457\,343.739465   & 900 &  31.7464 & 0.0034 & 7.19 &  $-$0.026 & 1.50  &  o \\ 
2\,457\,343.749986   & 900 &  31.7417 & 0.0036 & 7.20 &  $-$0.024 & 1.59  &  o \\ 
\hline 
\end{tabular}
\tablefoot{{The columns report: BJD (TDB), the mid-exposure Barycentric Julian Dates in Barycentric Dynamical Time; T$_{\rm exp}$, the exposure time; RV and error are the radial velocity measurement and its estimated uncertainty; FWHM, the Full Width at Half Maximum of the Cross-Correlation Function; Bis. Span, the radial velocity bisector span of the CCF; Airmass, the airmass of the star at the beginning of the exposure;
Flag, indicating wether the spectrum was taken in-transit (i) or off-transit (o).}}
\end{table*}

\begin{table*}[ht]
\caption{HARPS-N RV data for HAT-P-17. Same columns as in Table \ref{tab:RV_HAT-P-15}.}
\label{tab:RV_HAT-P-17}
\centering
\begin{tabular}{rrrrrrrc}
\hline
\hline
 BJD (TDB) & T$_{\rm exp}$ & RV & error & FWHM & Bis. Span & Airmass & Flag \\ 
           & [sec] &[km\,s$^{-1}$] & [km\,s$^{-1}$] & [km\,s$^{-1}$] & [km\,s$^{-1}$] & & \\
\hline
2\,456\,579.321287  &  900 &   20.3041 &  0.0013 &  6.74  &    $-$0.043  &    1.07  &  i  \\
2\,456\,579.333644  &  900 &   20.3097 &  0.0015 &  6.74  &    $-$0.041  &    1.04  &  i  \\
2\,456\,579.344361  &  900 &   20.3111 &  0.0015 &  6.74  &    $-$0.034  &    1.03  &  i  \\
2\,456\,579.355070  &  900 &   20.3122 &  0.0012 &  6.74  &    $-$0.040  &    1.02  &  i  \\
2\,456\,579.365796  &  900 &   20.3140 &  0.0013 &  6.73  &    $-$0.037  &    1.01  &  i  \\
2\,456\,579.375810  &  900 &   20.3115 &  0.0014 &  6.73  &    $-$0.036  &    1.00  &  i  \\
2\,456\,579.387239  &  900 &   20.3065 &  0.0015 &  6.74  &    $-$0.041  &    1.00  &  i  \\
2\,456\,579.397948  &  900 &   20.3047 &  0.0014 &  6.73  &    $-$0.040  &    1.00  &  i  \\
2\,456\,579.408740  &  900 &   20.3018 &  0.0012 &  6.75  &    $-$0.035  &    1.01  &  i  \\
2\,456\,579.419453  &  900 &   20.3010 &  0.0013 &  6.73  &    $-$0.044  &    1.02  &  i  \\
2\,456\,579.430166  &  900 &   20.2960 &  0.0013 &  6.74  &    $-$0.045  &    1.03  &  i  \\
2\,456\,579.440874  &  900 &   20.2920 &  0.0012 &  6.74  &    $-$0.036  &    1.04  &  i  \\
2\,456\,579.451587  &  900 &   20.2912 &  0.0013 &  6.74  &    $-$0.036  &    1.06  &  i  \\
2\,456\,579.462300  &  900 &   20.2893 &  0.0013 &  6.74  &    $-$0.048  &    1.09  &  i  \\
2\,456\,579.473017  &  900 &   20.2968 &  0.0013 &  6.73  &    $-$0.039  &    1.12  &  i  \\
2\,456\,579.483726  &  900 &   20.2965 &  0.0013 &  6.74  &    $-$0.038  &    1.15  &  i  \\
2\,456\,579.494438  &  900 &   20.2970 &  0.0013 &  6.74  &    $-$0.039  &    1.20  &  o  \\
2\,456\,579.505160  &  900 &   20.2976 &  0.0013 &  6.74  &    $-$0.038  &    1.25  &  o  \\
2\,456\,579.515868  &  900 &   20.2955 &  0.0013 &  6.74  &    $-$0.039  &    1.30  &  o  \\
2\,456\,579.526586  &  900 &   20.2946 &  0.0013 &  6.74  &    $-$0.039  &    1.37  &  o  \\
2\,456\,579.537298  &  900 &   20.2941 &  0.0013 &  6.74  &    $-$0.042  &    1.45  &  o  \\
2\,456\,579.548011  &  900 &   20.2938 &  0.0013 &  6.74  &    $-$0.041  &    1.54  &  o  \\
2\,456\,579.560168  &  900 &   20.2962 &  0.0016 &  6.74  &    $-$0.045  &    1.67  &  o  \\
\hline 
\end{tabular}
\end{table*}

\begin{table*}[ht]
\caption{HARPS-N RV data for HAT-P-21. Same columns as in Table \ref{tab:RV_HAT-P-15}.}
\label{tab:RV_HAT-P-21}
\centering
\begin{tabular}{rrrrrrrc}
\hline
\hline
BJD (TDB) & T$_{\rm exp}$ & RV & error & FWHM & Bis. Span & Airmass & Flag \\ 
          & [sec] &[km\,s$^{-1}$] & [km\,s$^{-1}$] & [km\,s$^{-1}$] & [km\,s$^{-1}$] & &  \\
\hline
2\,456\,724.442357  &    600  &   $-$52.8637  &  0.0064  &  8.21  &  0.021  &  1.28  & o 	 \\
2\,456\,724.449582  &    600  &   $-$52.8660  &  0.0063  &  8.26  &  0.002  &  1.24  & o 	 \\
2\,456\,724.456804  &    600  &   $-$52.8727  &  0.0060  &  8.20  &  0.043  &  1.21  & o 	 \\
2\,456\,724.464026  &    600  &   $-$52.8741  &  0.0057  &  8.21  &  0.025  &  1.19  & o 	 \\
2\,456\,724.471252  &    600  &   $-$52.8720  &  0.0060  &  8.22  &  0.017  &  1.16  & o 	 \\
2\,456\,724.478478  &    600  &   $-$52.8810  &  0.0069  &  8.21  &  0.034  &  1.14  & o 	 \\
2\,456\,724.485705  &    600  &   $-$52.8845  &  0.0075  &  8.23  &  0.042  &  1.12  & o 	 \\
2\,456\,724.492931  &    600  &   $-$52.9000  &  0.0062  &  8.21  &  0.030  &  1.10  & i 	 \\
2\,456\,724.500144  &    600  &   $-$52.8909  &  0.0063  &  8.20  &  0.005  &  1.09  & i 	 \\
2\,456\,724.507366  &    600  &   $-$52.8949  &  0.0060  &  8.22  &  0.030  &  1.07  & i 	 \\
2\,456\,724.514595  &    600  &   $-$52.8935  &  0.0059  &  8.21  &  0.014  &  1.06  & i 	 \\
2\,456\,724.521825  &    600  &   $-$52.8871  &  0.0062  &  8.20  &  0.021  &  1.05  & i 	 \\
2\,456\,724.529047  &    600  &   $-$52.8948  &  0.0050  &  8.18  &  0.020  &  1.04  & i 	 \\
2\,456\,724.536282  &    600  &   $-$52.8960  &  0.0055  &  8.20  &  0.012  &  1.04  & i 	 \\
2\,456\,724.543512  &    600  &   $-$52.9001  &  0.0069  &  8.20  &  0.017  &  1.03  & i 	 \\
2\,456\,724.550742  &    600  &   $-$52.9140  &  0.0070  &  8.23  &  0.010  &  1.03  & i 	 \\
2\,456\,724.557982  &    600  &   $-$52.9195  &  0.0067  &  8.23  &  0.037  &  1.02  & i 	 \\
2\,456\,724.565212  &    600  &   $-$52.9277  &  0.0053  &  8.23  &  0.021  &  1.02  & i 	 \\
2\,456\,724.572433  &    600  &   $-$52.9349  &  0.0046  &  8.24  &  0.016  &  1.02  & i 	 \\
2\,456\,724.579654  &    600  &   $-$52.9463  &  0.0058  &  8.22  &  0.031  &  1.02  & i 	 \\
2\,456\,724.586872  &    600  &   $-$52.9542  &  0.0095  &  8.26  &  0.047  &  1.03  & i 	 \\
2\,456\,724.594098  &    600  &   $-$52.9690  &  0.0064  &  8.24  &  0.016  &  1.03  & i 	 \\
2\,456\,724.601319  &    600  &   $-$52.9625  &  0.0066  &  8.23  &  0.034  &  1.04  & i 	 \\
2\,456\,724.608540  &    600  &   $-$52.9645  &  0.0061  &  8.21  &  0.030  &  1.05  & i     \\   
2\,456\,724.615766  &    600  &   $-$52.9720  &  0.0079  &  8.22  &  0.037  &  1.05  & i     \\   
2\,456\,724.622988  &    600  &   $-$52.9893  &  0.0089  &  8.17  &  0.032  &  1.07  & i     \\   
2\,456\,724.630200  &    600  &   $-$52.9778  &  0.0433  &  8.29  &  0.169  &  1.08  & i     \\   
2\,456\,724.637426  &    600  &   $-$52.9401  &  0.0679  &  8.32  &  0.186  &  1.09  & i     \\   
2\,456\,724.644648  &    600  &   $-$52.9836  &  0.0136  &  8.23  &  0.000  &  1.11  & i     \\   
2\,456\,724.651878  &    600  &   $-$52.9986  &  0.0158  &  8.21  &  0.033  &  1.13  & i     \\   
2\,456\,724.659090  &    600  &   $-$52.9779  &  0.0094  &  8.18  &  0.024  &  1.15  & i     \\   
2\,456\,724.666316  &    600  &   $-$52.9999  &  0.0086  &  8.19  &  0.019  &  1.17  & i     \\   
2\,456\,724.673543  &    600  &   $-$52.9878  &  0.0068  &  8.23  &  0.080  &  1.20  & o     \\  
2\,456\,724.680770  &    600  &   $-$53.0202  &  0.0093  &  8.20  &  0.058  &  1.22  & o     \\   
2\,456\,724.687986  &    600  &   $-$53.0051  &  0.0097  &  8.20  &  0.050  &  1.25  & o     \\   
2\,456\,724.695247  &    430  &   $-$52.9971  &  0.0411  &  8.20  &  0.000  &  1.28  & o     \\   
\hline 
\end{tabular}
\end{table*}

\begin{table*}[ht]
\caption{
HARPS-N RV data for HAT-P-26. Same columns as in Table \ref{tab:RV_HAT-P-15}.
}
\label{tab:RV_HAT-P-26}
\centering
\begin{tabular}{rrrrrrrc}
\hline
\hline
BJD (TDB) & T$_{\rm exp}$ & RV & error & FWHM & Bis. Span & Airmass & Flag \\ 
          & [sec] &[km\,s$^{-1}$] & [km\,s$^{-1}$] & [km\,s$^{-1}$] & [km\,s$^{-1}$] & &  \\
\hline
2\,457\,108.501404  &   600  & 13.8467  &  0.0049	 &  6.01    &      0.007   &  1.60   &  i  \\
2\,457\,108.508661  &   600  & 13.8524  &  0.0059	 &  6.02    &      0.015   &  1.53   &  i  \\
2\,457\,108.515896  &   600  & 13.8539  &  0.0075	 &  5.99    &      0.020   &  1.47   &  i  \\
2\,457\,108.523127  &   600  & 13.8450  &  0.0057	 &  6.01    &   $-$0.003   &  1.41   &  i  \\
2\,457\,108.530357  &   600  & 13.8499  &  0.0052	 &  6.01    &      0.000   &  1.37   &  i  \\
2\,457\,108.537592  &   600  & 13.8450  &  0.0045	 &  6.03    &   $-$0.004   &  1.32   &  i  \\
2\,457\,108.544827  &   600  & 13.8477  &  0.0045	 &  6.00    &   $-$0.015   &  1.29   &  i  \\
2\,457\,108.552079  &   600  & 13.8400  &  0.0063	 &  6.00    &      0.016   &  1.25   &  i  \\
2\,457\,108.559314  &   600  & 13.8453  &  0.0072	 &  5.98    &      0.011   &  1.23   &  i  \\
2\,457\,108.566541  &   600  & 13.8462  &  0.0070	 &  6.02    &      0.016   &  1.20   &  i  \\
2\,457\,108.573780  &   600  & 13.8351  &  0.0070	 &  5.98    &      0.001   &  1.18   &  i  \\
2\,457\,108.581010  &   600  & 13.8419  &  0.0069	 &  6.00    &      0.019   &  1.16   &  i  \\
2\,457\,108.588255  &   600  & 13.8465  &  0.0056	 &  6.00    &   $-$0.018   &  1.14   &  i  \\
2\,457\,108.595494  &   600  & 13.8408  &  0.0049	 &  6.02    &      0.001   &  1.13   &  i  \\
2\,457\,108.602729  &   600  & 13.8487  &  0.0048	 &  6.00    &   $-$0.003   &  1.12   &  i  \\
2\,457\,108.609973  &   600  & 13.8466  &  0.0050	 &  6.00    &      0.008   &  1.11   &  i  \\
2\,457\,108.617212  &   600  & 13.8386  &  0.0048	 &  6.00    &   $-$0.010   &  1.11   &  o  \\
2\,457\,108.624457  &   600  & 13.8432  &  0.0050	 &  5.99    &   $-$0.001   &  1.10   &  o  \\
2\,457\,108.631683  &   600  & 13.8432  &  0.0048	 &  6.01    &      0.023   &  1.10   &  o  \\
2\,457\,108.638927  &   600  & 13.8476  &  0.0062	 &  6.01    &   $-$0.010   &  1.10   &  o  \\
2\,457\,108.646180  &   600  & 13.8526  &  0.0060	 &  6.02    &      0.017   &  1.10   &  o  \\
2\,457\,108.653455  &   600  & 13.8406  &  0.0090	 &  5.98    &   $-$0.010   &  1.11   &  o  \\
2\,457\,108.660704  &   600  & 13.8463  &  0.0139	 &  5.97    &      0.024   &  1.12   &  o  \\
2\,457\,108.667931  &   600  & 13.8507  &  0.0129	 &  6.00    &      0.009   &  1.13   &  o  \\
2\,457\,108.675161  &   600  & 13.8481  &  0.0281	 &  6.01    &   $-$0.028   &  1.14   &  o  \\
2\,457\,108.682391  &   600  & 13.8284  &  0.0170	 &  5.98    &   $-$0.004   &  1.16   &  o  \\
2\,457\,108.689625  &   600  & 13.8211  &  0.0202	 &  6.04    &   $-$0.073   &  1.17   &  o  \\   
\hline 
\end{tabular}
\end{table*}

\begin{table*}[ht]
\caption{HARPS-N RV data for HAT-P-29. Same columns as in Table \ref{tab:RV_HAT-P-15}.}
\label{tab:RV_HAT-P-29}
\centering
\begin{tabular}{rrrrrrrc}
\hline
\hline
BJD (TDB) & T$_{\rm exp}$ & RV & error & FWHM & Bis. Span & Airmass & Flag \\ 
          & [sec] &[km\,s$^{-1}$] & [km\,s$^{-1}$] & [km\,s$^{-1}$] & [km\,s$^{-1}$] & &  \\
\hline
2\,456\,582.520463  &  900   &  $-$21.6462  &  0.0054  &  9.14  &  0.032  &  1.12  &  o \\
2\,456\,582.532879  &  900   &  $-$21.6450  &  0.0053  &  9.17  &  0.033  &  1.10  &  o \\
2\,456\,582.543602  &  900   &  $-$21.6329  &  0.0056  &  9.17  &  0.034  &  1.10  &  o \\
2\,456\,582.554320  &  900   &  $-$21.6450  &  0.0056  &  9.15  &  0.030  &  1.09  &  i \\
2\,456\,582.565039  &  900   &  $-$21.6208  &  0.0055  &  9.10  &  0.040  &  1.09  &  i \\
2\,456\,582.575761  &  900   &  $-$21.6168  &  0.0051  &  9.15  &  0.025  &  1.09  &  i \\
2\,456\,582.586475  &  900   &  $-$21.6169  &  0.0045  &  9.11  &  0.027  &  1.09  &  i \\
2\,456\,582.597193  &  900   &  $-$21.6130  &  0.0047  &  9.14  &  0.015  &  1.09  &  i \\
2\,456\,582.607902  &  900   &  $-$21.6223  &  0.0048  &  9.16  &  0.037  &  1.10  &  i \\
2\,456\,582.618616  &  900   &  $-$21.6420  &  0.0048  &  9.18  &  0.025  &  1.12  &  i \\
2\,456\,582.629330  &  900   &  $-$21.6328  &  0.0050  &  9.17  &  0.022  &  1.13  &  i \\
2\,456\,582.640048  &  900   &  $-$21.6488  &  0.0052  &  9.17  &  0.047  &  1.15  &  i \\
2\,456\,582.650837  &  900   &  $-$21.6642  &  0.0071  &  9.17  &  0.032  &  1.17  &  i \\
2\,456\,582.661551  &  900   &  $-$21.6795  &  0.0078  &  9.16  &  0.017  &  1.20  &  i \\
2\,456\,582.672260  &  900   &  $-$21.6765  &  0.0123  &  9.12  &  0.033  &  1.23  &  i \\
2\,456\,582.682983  &  900   &  $-$21.6745  &  0.0082  &  9.11  &  0.005  &  1.26  &  i \\
2\,456\,582.693692  &  900   &  $-$21.6841  &  0.0060  &  9.11  &  0.024  &  1.30  &  i \\
2\,456\,582.704406  &  900   &  $-$21.6777  &  0.0054  &  9.13  &  0.026  &  1.35  &  i \\
2\,456\,582.715119  &  900   &  $-$21.6559  &  0.0052  &  9.13  &  0.023  &  1.40  &  i \\
2\,456\,582.725833  &  900   &  $-$21.6533  &  0.0057  &  9.14  &  0.046  &  1.46  &  o \\
2\,456\,582.736547  &  900   &  $-$21.6594  &  0.0062  &  9.17  &  0.042  &  1.53  &  o \\
2\,456\,582.747265  &  900   &  $-$21.6641  &  0.0067  &  9.15  &  0.002  &  1.61  &  o \\
2\,456\,582.759854  &  900   &  $-$21.6700  &  0.0072  &  9.17  &  0.053  &  1.72  &  o \\
\hline 
\end{tabular}
\end{table*}

\section{Photometric light curves} 
\label{appendix_Photometric_light_curves}
The light curves, which were analysed in this work, are plotted in this appendix. A table with the details of the ground-based photometric follow-up observations is also reported.

\begin{table*}
\caption{Details of the photometric follow-up observations presented in this work.} %
\label{tab:logs} %
\centering     %
\tiny          %
\setlength{\tabcolsep}{5pt}
\begin{tabular}{lccccccc}
\hline\hline\\[-6pt]
Telescope & Date of   & Start time & End time  & $N_{\rm obs}$ & $T_{\rm exp}$ & Filter & Scatter \\
          & first obs &    (UT)    &   (UT)    &               & (s)           &        & (mmag)  \\
\hline \\[-6pt]
\multicolumn{7}{l}{\textbf{HAT-P-17:}} \\
Cassini\,1.52\,m & 2012.07.15 & 21:11 & 01:08 & 146 & $60-120$ & Gunn $i$ & 0.93 \\
CA\,1.23\,m      & 2014.07.20 & 00:55 & 04:05 & 117 & $60-100$ & Cousins $I$ & 0.79 \\[2pt]
\multicolumn{7}{l}{\textbf{HAT-P-21:}} \\
Cassini\,1.52\,m & 2012.03.03 & 21:54 & 00:47 & 211 & $60$ & Gunn $r$ & 4.23 \\
CA\,1.23\,m      & 2012.03.07 & 00:55 & 05:10 & 145 & $110-120$ & Cousins $I$ & 0.99 \\ [2pt]
\multicolumn{7}{l}{\textbf{HAT-P-26:}}  \\
CA\,1.23\,m & 2012.03.04 & 01:21 & 05:50 &  67 & $120$    & Cousins $I$ & 0.69 \\
CA\,1.23\,m & 2014.04.17 & 21:54 & 04:40 & 209 & $90-120$ & Cousins $I$ & 0.75 \\
CA\,1.23\,m & 2016.03.14 & 22:14 & 03:07 & 190 & $110$    & Cousins $I$ & 1.01 \\
CA\,1.23\,m & 2017.04.22 & 20:52 & 04:08 & 171 & $120$    & Cousins $I$ & 1.70 \\
CA\,1.23\,m & 2018.02.26 & 00:23 & 06:06 & 282 & $45-100$ & Cousins $I$ & 0.95 \\ [2pt]
\multicolumn{7}{l}{\textbf{HAT-P-29:}}  \\
CA\,1.23\,m  & 2011.10.03 & 22:13 & 05:07 & 121 & $120$      & Cousins $R$ & 0.58 \\
TNG\,3.58\,m & 2012.10.10 & 01:05 & 05:25 & 783 & $ 15$      & Johnson $R$ & 0.33 \\
CA\,1.23\,m  & 2012.11.01 & 20:37 & 04:12 & 180 & $105-180$  & Cousins $R$ & 1.18 \\
IAC\,80\,cm  & 2013.10.16 & 23:21 & 04:53 & 234 & $60$       & Cousins $R$ & 1.01 \\
Copernico\,1.80\,m & 2014.10.29 & 17:54 & 00:02 & 2303 & $7$ & Sloan   $r$ & 0.67 \\
Copernico\,1.80\,m & 2016.01.07 & 17:20 & 23:26 & 2062 & $7$ & Sloan   $r$ & 0.54 \\
CA\,1.23\,m  & 2014.10.23 & 21:52 & 05:19 & 254 & $95-120$   & Cousins $I$ & 0.58 \\
CA\,1.23\,m  & 2020.10.23 & 21:04 & 05:40 & 564 & $30-70$    & Cousins $I$ & 0.95 \\
\hline
\end{tabular}
\tablefoot{$N_{\rm obs}$ is the number of observations, $T_{\rm
exp}$ is the exposure time.
Scatter is the \emph{rms} scatter of the data versus a fitted model.}
\end{table*}
\begin{figure}
\centering
\includegraphics[width=\hsize]{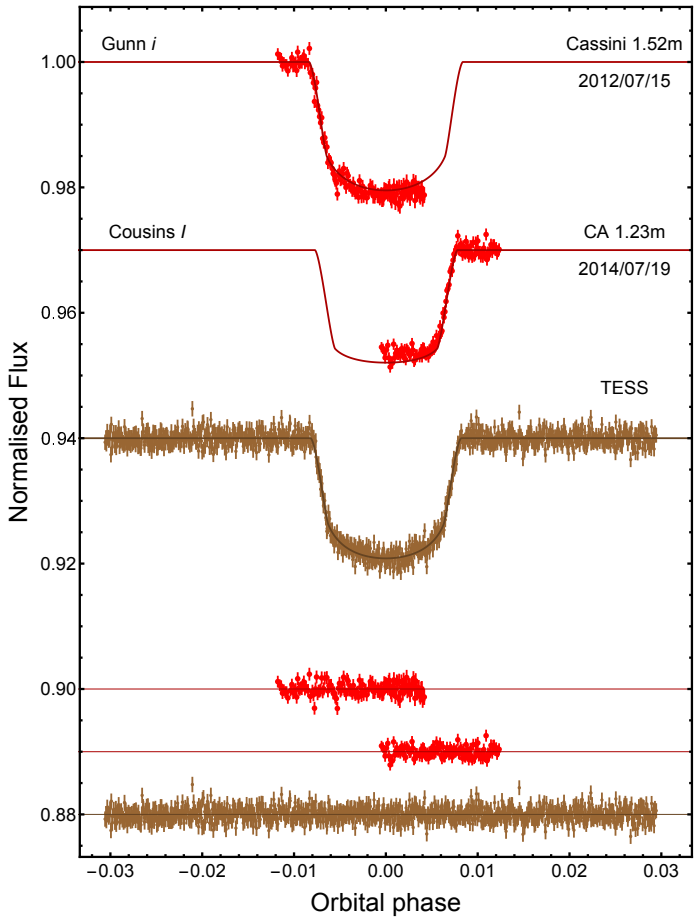}
\caption{Phased light curves of HAT-P-17\,b transits presented in this work. Two incomplete transits were observed with ground-based telescopes. Two complete transits were observed with TESS; they are plotted phased. These light curves are compared with the best {\sc jktebop} fits. The dates, telescopes, and filters related to the observation of each transit event are indicated. Residuals from the fits are plotted at the bottom of the figure.}
\label{fig:hatp17_lc}
\end{figure}
\begin{figure}
\centering
\includegraphics[width=\hsize]{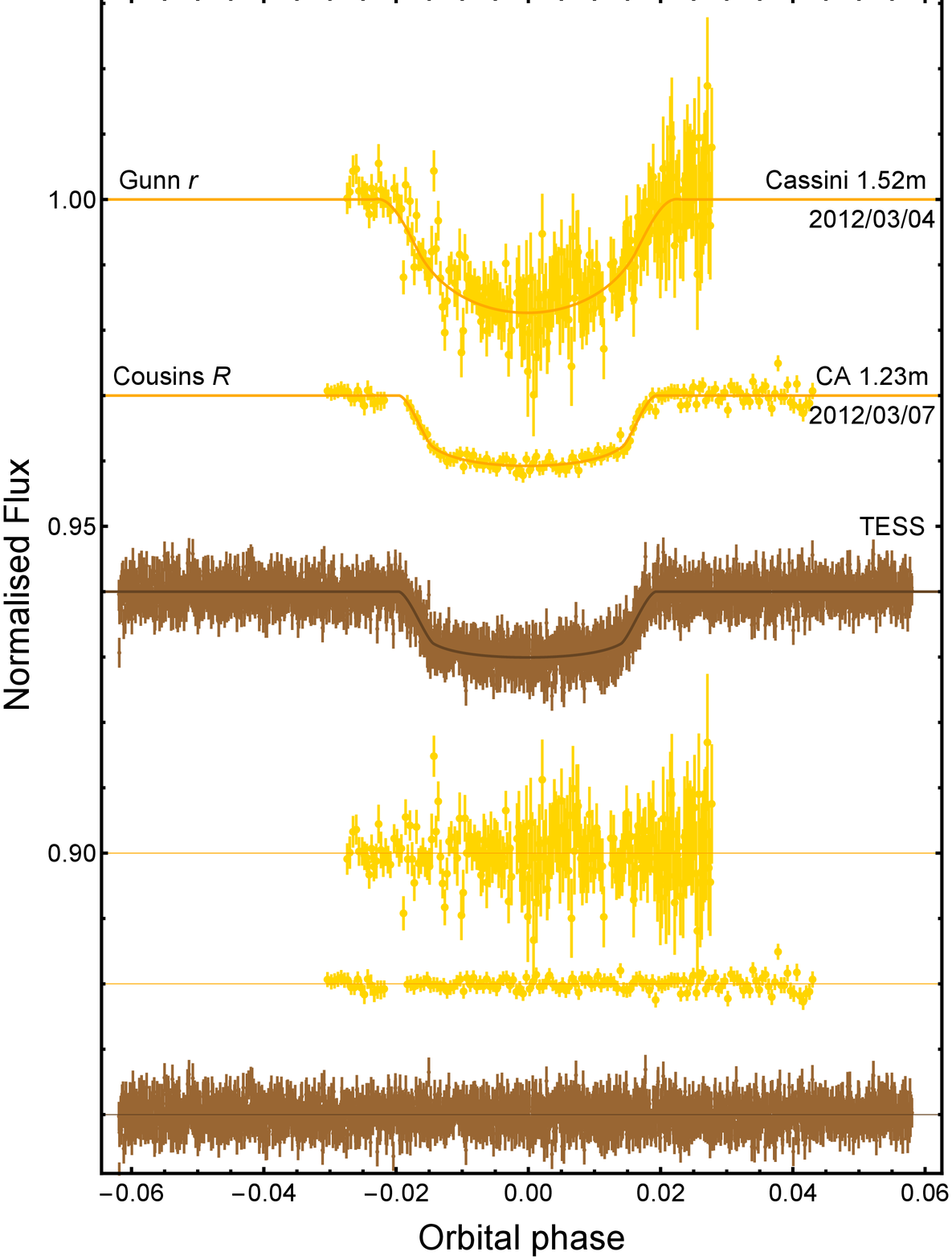}
\caption{Phased light curves of HAT-P-21\,b transits presented in this work. Two transits were observed with ground-based telescopes, and five were obtained by TESS. The light curves are compared with the best {\sc jktebop} fits. The dates, telescopes, and filters related to the observation of each transit event are indicated. Residuals from the fits are plotted at the bottom of the figure.}
\label{fig:hatp21_lc}
\end{figure}
\begin{figure}
\centering
\includegraphics[width=\hsize]{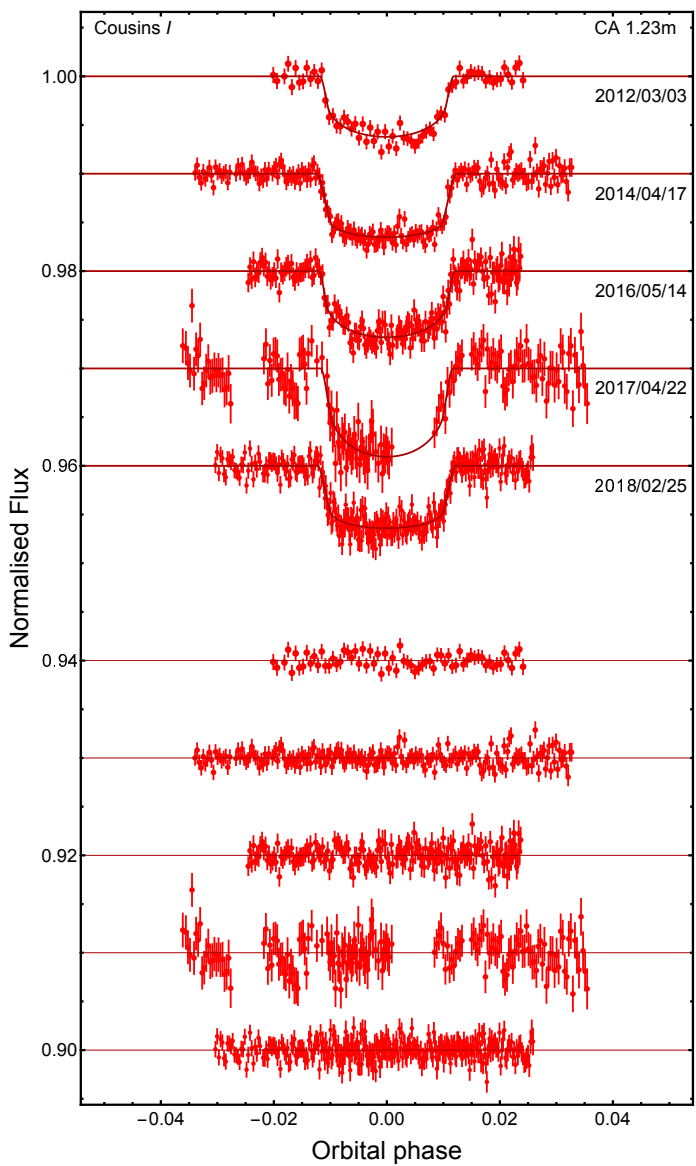}
\caption{Phased light curves of HAT-P-26\,b transits presented in this work. Five transits were observed with the CA\,1.23\,m telescope. These phased light curves are compared with the best {\sc jktebop} fits. The dates, telescopes, and filters related to the observation of each transit event are indicated. Residuals from the fits are plotted at the bottom of the figure.}
\label{fig:hatp26_lc}
\end{figure}
\begin{figure}
\centering
\includegraphics[width=\hsize]{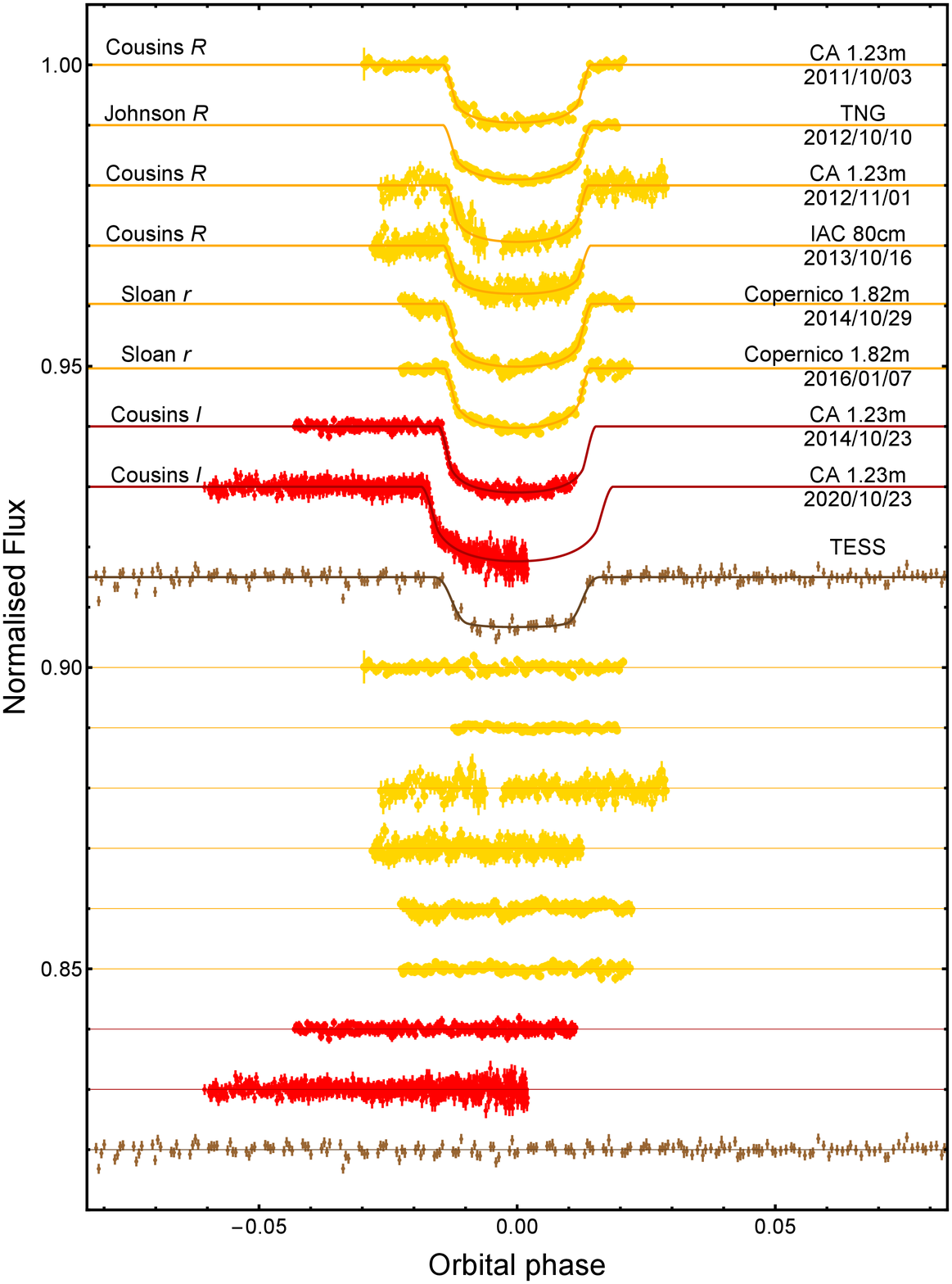}
\caption{Phased light curves of HAT-P-29\,b transits presented in this work. Two transits were observed with ground-based telescopes. Other four with TESS; they are plotted phased. The light curves obtained with the Copernico and the TNG telescopes have been also binned. All the light curves are compared with the best {\sc jktebop} fits. The dates, telescopes, and filters related to the observation of each transit event are indicated. Residuals from the fits are plotted at the bottom of the figure.}
\label{fig:hatp29_lc}
\end{figure}

\section{Times of mid-transit} 
\label{Times_of_mid-transit}
The tables in this appendix report the value of the mid-transit times analysed for reviewing the orbital ephemeris of the HAT-P-17, HAT-P-21, HAT-P-26, and HAT-P-29 planetary systems. The figures in this appendix contain the corresponding plots of the residuals of the timings of mid-transit versus a linear ephemeris.
\begin{table} 
\centering
\caption{Times of mid-transit for HAT-P-17\,b and their residuals for a constant period
($P_{\rm orb} = 10.33853781 \pm 0.00000060$).}
\label{tab:transit_times_hatp17}
\resizebox{\hsize}{!}{
\begin{tabular}{l r r c} 
\hline  \\[-6pt]
~~~Time of minimum      & Cycle~ & ${\rm O}-{\rm C}$~~~ & Reference \\
BJD(TDB)$-2\,400\,000$  & no.~   & (day)~~~             & \\
\hline \\ [-6pt]
$54\,801.16943 \pm 0.00020$ & $   0$ & $ -0.000002 $ & \citet{howard:2012} \\ [2pt]
$56\,124.50251 \pm 0.00055$ & $ 128$ & $  0.000239 $ & Cassini (this work) \\ [2pt]
$56\,858.53815 \pm 0.00055$ & $ 199$ & $ -0.000306 $ & CA (this work)      \\ [2pt]
$58\,719.47555 \pm 0.00031$ & $ 379$ & $  0.000289 $ & TESS (this work)    \\ [2pt]
$58\,729.81360 \pm 0.00027$ & $ 380$ & $ -0.000199 $ & TESS (this work)    \\ [2pt]
\hline
\end{tabular}
}
\end{table}
\begin{table} 
\centering
\caption{Times of mid-transit for HAT-P-21\,b and their residuals for a constant period ($P_{\rm orb} = 4.12449009 \pm 0.00000090$).}
\label{tab:transit_times_hatp21}
\resizebox{\hsize}{!}{
\begin{tabular}{l r r c} 
\hline  \\[-6pt]
~~~Time of minimum      & Cycle~ & ${\rm O}-{\rm C}$~~~ & Reference \\
BJD(TDB)$-2\,400\,000$  & no.~   & (day)~~~             & \\
\hline \\ [-6pt]
$54\,996.41312 \pm 0.00069$ & $   0$ & $  0.000685 $ & \citet{bakos:2011} \\ [2pt]
$55\,994.53827 \pm 0.00063$ & $ 242$ & $ -0.000767 $ & CA (this work)     \\ [2pt] 
$58\,902.30440 \pm 0.00108$ & $ 947$ & $ -0.000152 $ & TESS (this work)   \\ [2pt] 
$58\,906.43060 \pm 0.00120$ & $ 948$ & $  0.001558 $ & TESS (this work)   \\ [2pt] 
$58\,910.55289 \pm 0.00093$ & $ 949$ & $ -0.000642 $ & TESS (this work)   \\ [2pt] 
$58\,918.80358 \pm 0.00105$ & $ 951$ & $  0.001067 $ & TESS (this work)   \\ [2pt]
$58\,922.92610 \pm 0.00115$ & $ 952$ & $ -0.000903 $ & TESS (this work)   \\ [2pt]
\hline
\end{tabular}
}
\end{table}
\begin{table} 
\centering
\caption{Times of mid-transit for HAT-P-26\,b and their residuals for a constant period ($P_{\rm orb} = 4.23450213 \pm 0.00000076$). With the exception of the last four values, the others were taken from the compilation made by \citet{vonEssen:2019}.}
\label{tab:transit_times_hatp26}
\resizebox{\hsize}{!}{
\begin{tabular}{l r r c} 
\hline  \\[-6pt]
~~~Time of minimum      & Cycle~ & ${\rm O}-{\rm C}$~~~ & Reference \\
BJD(TDB)$-2\,400\,000$  & no.~   & (day)~~~             & \\
\hline \\ [-6pt]
$54\,860.02786 \pm 0.00147$ & $-105.0$ & $-0.00179$ & \citet{hartman:2011}   \\ %
$55\,342.76262 \pm 0.00041$ & $   9.0$ & $-0.00016$ & \citet{hartman:2011}   \\ %
$55\,304.65218 \pm 0.00025$ & $   0.0$ & $-0.00009$ & \citet{stevenson:2016} \\ %
$56\,405.62370 \pm 0.00090$ & $ 260.0$ & $ 0.00113$ & \citet{stevenson:2016} \\ %
$56\,545.36220 \pm 0.00030$ & $ 293.0$ & $ 0.00110$ & \citet{stevenson:2016} \\ %
$57\,129.72248 \pm 0.00017$ & $ 431.0$ & $ 0.00022$ & \citet{stevenson:2016} \\ %
$57\,413.43284 \pm 0.00017$ & $ 498.0$ & $-0.00100$ & \citet{wakeford:2017}  \\ %
$57\,460.01327 \pm 0.00016$ & $ 509.0$ & $-0.00008$ & \citet{wakeford:2017}  \\ %
$57\,510.82710 \pm 0.00016$ & $ 521.0$ & $-0.00026$ & \citet{wakeford:2017}  \\ %
$57\,616.69010 \pm 0.00011$ & $ 546.0$ & $ 0.00021$ & \citet{wakeford:2017}  \\ %
$57\,112.78503 \pm 0.00058$ & $ 427.0$ & $ 0.00077$ & \citet{vonEssen:2019}  \\ %
$57\,125.48930 \pm 0.00063$ & $ 430.0$ & $ 0.00154$ & \citet{vonEssen:2019}  \\ %
$57\,129.72283 \pm 0.00063$ & $ 431.0$ & $ 0.00057$ & \citet{vonEssen:2019}  \\ %
$57\,163.59738 \pm 0.00040$ & $ 439.0$ & $-0.00089$ & \citet{vonEssen:2019}  \\ %
$57\,180.53670 \pm 0.00057$ & $ 443.0$ & $ 0.00042$ & \citet{vonEssen:2019}  \\ %
$57\,197.47376 \pm 0.00046$ & $ 447.0$ & $-0.00052$ & \citet{vonEssen:2019}  \\ %
$57\,523.53041 \pm 0.00072$ & $ 524.0$ & $-0.00046$ & \citet{vonEssen:2019}  \\ %
$57\,887.69984 \pm 0.00089$ & $ 610.0$ & $ 0.00187$ & \citet{vonEssen:2019}  \\ %
$57\,904.63796 \pm 0.00066$ & $ 614.0$ & $ 0.00199$ & \citet{vonEssen:2019}  \\ %
$57\,921.57698 \pm 0.00078$ & $ 618.0$ & $ 0.00301$ & \citet{vonEssen:2019}  \\ %
$58\,226.45772 \pm 0.00093$ & $ 690.0$ & $-0.00034$ & \citet{vonEssen:2019}  \\ %
$55\,990.64139 \pm 0.00078$ & $ 162.0$ & $-0.00007$ & CA (this work) \\   %
$56\,765.55640 \pm 0.00086$ & $ 345.0$ & $ 0.00124$ & CA (this work) \\   %
$57\,523.53031 \pm 0.00055$ & $ 524.0$ & $-0.00056$ & CA (this work) \\   %
$58\,175.64477 \pm 0.00081$ & $ 678.0$ & $ 0.00073$ & CA (this work) \\   %
\hline
\end{tabular}
}
\end{table}
\begin{table} 
\centering
\caption{Times of mid-transit for HAT-P-29\,b and their residuals for a constant period
($P_{\rm orb} = 5.7233746  \pm 0.0000034$).}
\label{tab:transit_times_hatp29}
\resizebox{\hsize}{!}{
\begin{tabular}{l r r c} 
\hline  \\[-6pt]
~~~Time of minimum      & Cycle~ & ${\rm O}-{\rm C}$~~~ & Reference \\
BJD(TDB)$-2\,400\,000$  & no.~   & (day)~~~             & \\
\hline \\ [-6pt]
$55\,197.57540 \pm 0.00181$ & $ -112$ & $-0.001260$ & \citet{buchhave:2011} \\ [2pt]
$55\,563.87156 \pm 0.00065$ & $ -48 $ & $-0.001075$ & \citet{wang:2018}     \\ [2pt]
$55\,586.76257 \pm 0.00061$ & $ -44 $ & $-0.003564$ & \citet{wang:2018}     \\ [2pt]
$55\,838.59570 \pm 0.00050$ & $  0  $ & $ 0.001084$ & CA (this work)        \\ [2pt]
$56\,210.61502 \pm 0.00056$ & $ 65  $ & $ 0.001054$ & TNG (this work)       \\ [2pt]
$56\,233.51031 \pm 0.00064$ & $ 69  $ & $ 0.002846$ & CA (this work)        \\ [2pt]
$56\,582.63483 \pm 0.00120$ & $ 130 $ & $ 0.001514$ & IAC\, 80\,cm (this work) \\ [2pt]
$56\,611.25060 \pm 0.00139$ & $ 135 $ & $ 0.000411$ & \citet{wang:2018}     \\ [2pt]
$56\,634.13560 \pm 0.00293$ & $ 139 $ & $-0.008087$ & \citet{wang:2018}     \\ [2pt]
$56\,657.03630 \pm 0.00109$ & $ 143 $ & $-0.000885$ & \citet{wang:2018}     \\ [2pt]
$56\,697.10060 \pm 0.00166$ & $ 150 $ & $-0.000208$ & \citet{wang:2018}     \\ [2pt]
$56\,954.65758 \pm 0.00131$ & $ 195 $ & $ 0.004915$ & CA (this work)        \\ [2pt]
$56\,960.37608 \pm 0.00036$ & $ 196 $ & $ 0.000040$ & Copernico (this work) \\ [2pt]
$56\,983.26700 \pm 0.00427$ & $ 200 $ & $-0.002538$ & \citet{wang:2018}     \\ [2pt]
$57\,029.05882 \pm 0.00068$ & $ 208 $ & $ 0.002285$ & \citet{wang:2018}     \\ [2pt]
$57\,395.35157 \pm 0.00033$ & $ 272 $ & $-0.000940$ & Copernico (this work) \\ [2pt]
$58\,797.57696 \pm 0.00122$ & $ 517 $ & $-0.002329$ & TESS (this work)      \\ [2pt]
$59\,146.70788 \pm 0.00217$ & $ 578 $ & $ 0.002740$ & CA (this work)        \\ [2pt]
\hline
\end{tabular}
}
\end{table}
\begin{figure*}
\centering
\includegraphics[width=\hsize]{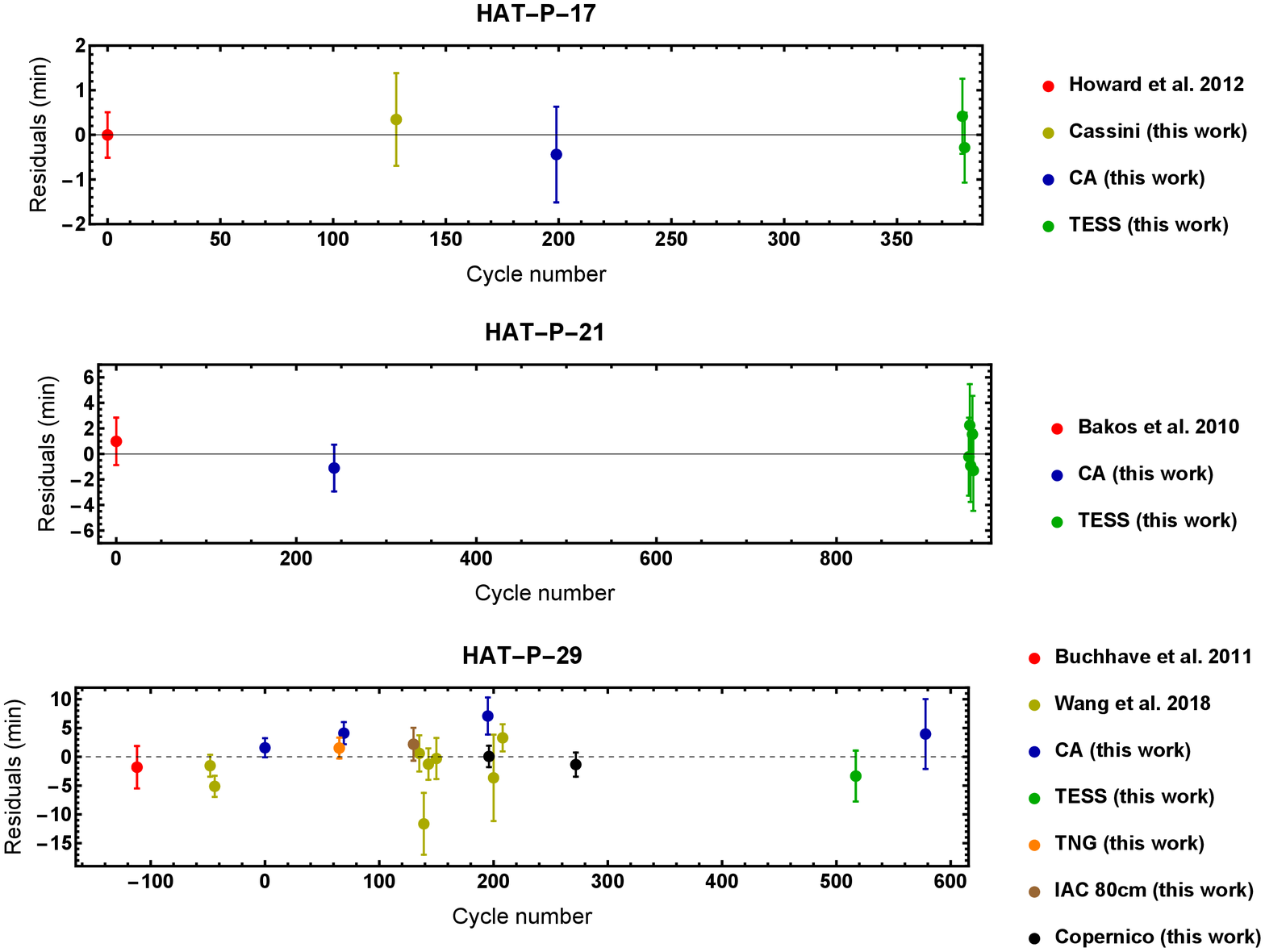}
\caption{Plot of the residuals of the timings of mid-transit of HAT-P-17\,b, HAT-P-21\,b, and HAT-P-29\,b versus a linear ephemeris.}
\label{fig:OC_plots}
\end{figure*}

\section{Revised physical parameters of the planetary systems} %
\label{appendix_Revised_physical_parameters}
The tables in this appendix report the final values that we obtained for the main physical parameters of the planetary systems under study. The values
obtained in this work (Sect.~\ref{sec:Final_Physical_parameters}) are compared with those taken from the literature. Where two error bars are given, the first refers to the statistical uncertainties and the second to the systematic errors.
%
\begin{table*}
\tiny
\centering %
\caption{Physical parameters of the planetary system HAT-P-15 derived in this work.} %
\label{tab:hatp15_final_parameters} %
\begin{tabular}{l c c c c | c }
\hline %
\hline  \\[-8pt]
Parameter & Nomen. & Unit & This Work & Source & \citet{kovacs:2010} \\
\hline  \\[-6pt]
\multicolumn{1}{l}{\textbf{Stellar parameters}} \\ [2pt] %
Spectra class \dotfill & & & & &  G5 \\
Effective temperature \dotfill & $T_{\mathrm{eff}}$ & K & $\begin{tabular}{c}
$5620\pm20$ \\
$5620\pm20$ \\
\end{tabular}$ &
$\begin{tabular}{c}
{\sc EXOFASTv2} \\
{\sc HARPS-N} \\
\end{tabular}$
& $5568 \pm 90$ \\ [8pt] %
Metallicity \dotfill & [Fe/H] & & 
$\begin{tabular}{c}
$+0.237^{+0.093}_{-0.098}$ \\
$~+0.24 \pm 0.10$ \\
\end{tabular}$ &
$\begin{tabular}{c}
{\sc EXOFASTv2} \\
{\sc HARPS-N} \\
\end{tabular}$
& $0.22 \pm 0.08$ \\ [8pt] %
Projected rotational velocity$^{(a)}$ \dotfill & $v\,\sin{i_{\star}}$ & km\,s$^{-1}$ & 
$\begin{tabular}{c}
$1.53 \pm 0.25$ \\
$2.3 \pm 0.5$ \\
\end{tabular}$ &
$\begin{tabular}{c}
RM fit \\
{\sc HARPS-N} \\
\end{tabular}$
& $2.0\pm0.5$ \\ [8pt] %
Convective blue-shift velocity$^{(a)}$ \dotfill & CBV & km\,s$^{-1}$ & $>-1.73\pm0.85$ & RM fit & --  \\ [2pt] %
Luminosity  \dotfill & $L_{\star}$ & $L_{\sun}$ & $1.072^{+0.100}_{-0.097}$ & {\sc EXOFASTv2} & $1.00 \pm 0.11$ \\ [2pt] %
Mass    \dotfill & $M_{\star}$ & $M_{\sun}$ & $1.020_{-0.060}^{+0.066}$ & {\sc EXOFASTv2} & $1.013 \pm 0.043$ \\ [2pt]%
Radius  \dotfill & $R_{\star}$ & $R_{\sun}$ & $1.092_{-0.050}^{+0.049}$ & {\sc EXOFASTv2} & $1.080 \pm 0.039$ \\ [2pt] %
Mean density \dotfill & $\rho_{\star}$ & $\rho_{\sun}$ & $1.11_{-0.15}^{+0.18}$  & {\sc EXOFASTv2} & -- \\ [2pt] %
Logarithmic surface gravity \dotfill & $\log{g_{\star}}$ & cgs & $4.371_{-0.045}^{+0.047}$ & {\sc EXOFASTv2} & $4.38\pm0.03$ \\ [2pt]%
Equal Evolutionary Phase \dotfill & $EEP$ & & $389^{+26}_{-42}$ & {\sc EXOFASTv2} & -- \\ [2pt]  %
$V$-band extinction \dotfill & $A_{\rm V}$ & mag & $1.04^{+0.12}_{-0.13}$ & {\sc EXOFASTv2} & -- \\ [2pt]  %
SED photometry error scaling \dotfill & $\sigma_{SED}$ & & $7.3^{+2.0}_{-1.3}$ & {\sc EXOFASTv2} & --\\ [2pt]  %
Parallax \dotfill & $\varpi$ & mas & $5.185^{+0.016}_{-0.016}$ & {\sc EXOFASTv2} & -- \\ [2pt]  %
Distance \dotfill & $d$ & pc & $192.85^{+0.61}_{-0.60}$ & {\sc EXOFASTv2} & $190 \pm 8$ \\ [2pt]  %
Age \dotfill & & Gyr & $6.6_{-3.4}^{+3.7} $ & {\sc EXOFASTv2} & $6.8^{+2.5}_{-1.6}$ \\ [2pt]  %
\hline \\[-6pt]%
\multicolumn{1}{l}{\textbf{Orbital parameters}} \\ [2pt] %
RV-curve semi-amplitude \dotfill & $K_{\star}$ &  m\,s$^{-1}$ & -- & -- & $180.6 \pm 3.5$  \\ [2pt] %
Barycentric RV  \dotfill & $\gamma$ & km\,s$^{-1}$ & $31.7622 \pm 0.0009$ &  RM fit & $31.7616 \pm0.0014$\\  [2pt] %
Projected spin-orbit angle  \dotfill & $\lambda$ & degree  & $13 \pm 6$ & RM fit & --  \\ [2pt] %
\hline %
\end{tabular}
\end{table*}
%
\begin{table*}
\tiny
\centering %
\caption{Physical parameters of the planetary system HAT-P-17 derived in this work.} %
\label{tab:hatp17_final_parameters} %
\resizebox{\hsize}{!}{
\begin{tabular}{l c c c c | c c }
\hline %
\hline  \\[-8pt]
Parameter & Nomen. & Unit & {\bf This Work} & Source & \citet{howard:2012} & \citet{fulton:2013}  \\
\hline  \\[-6pt]
\multicolumn{1}{l}{\textbf{Stellar parameters}} \\ [2pt] %
Spectra class \dotfill & & & & &  G0 \\
Effective temperature \dotfill & $T_{\mathrm{eff}}$ & K & $\begin{tabular}{c}
$5351^{+21}_{-20}$ \\
$~~~5350 \pm 20$ \\
\end{tabular}$ &
$\begin{tabular}{c}
{\sc EXOFASTv2} \\
{\sc HARPS-N} \\
\end{tabular}$
& $5246 \pm 80$ & -- \\ [8pt] %
Metallicity \dotfill & [Fe/H] & & 
$\begin{tabular}{c}
$+0.012^{+0.082}_{-0.080}$ \\
$~+0.02 \pm 0.09$ \\
\end{tabular}$ &
$\begin{tabular}{c}
{\sc EXOFASTv2} \\
{\sc HARPS-N} \\
\end{tabular}$
& $0.00 \pm 0.08$ & -- \\ [8pt] %
Projected rotational velocity$^{(a)}$ \dotfill & $v\,\sin{i_{\star}}$ & km\,s$^{-1}$ &
$\begin{tabular}{c}
$0.84 \pm 0.07$ \\
$0.5 \pm 0.5$ \\
\end{tabular}$ &
$\begin{tabular}{c}
RM fit \\
{\sc HARPS-N} \\
\end{tabular}$
& $0.3 \pm 0.5$ & $0.54 \pm 0.15$ \\ [6pt] %
Convective blue-shift velocity \dotfill & CBV & km\,s$^{-1}$ & $-0.46\pm0.11$ & RM fit & -- & $-0.65 \pm 0.23$  \\ [2pt] %
Luminosity  \dotfill & $L_{\star}$ & $L_{\sun}$ & $0.540^{+0.037}_{-0.034}$ & {\sc EXOFASTv2} & $0.48 \pm 0.04$ & -- \\ [2pt] %
Mass    \dotfill & $M_{\star}$ & $M_{\sun}$ & $0.886 \pm 0.030 \pm 0.024$ & ABSDIM & $0.857 \pm 0.039$ & -- \\ [2pt]%
Radius  \dotfill & $R_{\star}$ & $R_{\sun}$ & $0.860 \pm 0.023 \pm 0.008$ & ABSDIM & $0.838 \pm 0.021$ & -- \\ [2pt] %
Mean density \dotfill & $\rho_{\star}$ & $\rho_{\sun}$ & $1.39 \pm 0.12$  & ABSDIM & -- & --\\ [2pt] %
Logarithmic surface gravity \dotfill & $\log{g_{\star}}$ & cgs & $4.516 \pm 0.027 \pm 0.004$ & ABSDIM & $4.52 \pm 0.02$ & --\\ [2pt]%
Equal Evolutionary Phase \dotfill & $EEP$ & & $348^{+29}_{-26}$ & {\sc EXOFASTv2} & -- & --\\ [2pt]  %
$V$-band extinction \dotfill & $A_{\rm V}$ & mag & $0.250^{+0.089}_{-0.087}$ & {\sc EXOFASTv2} & -- & -- \\ [2pt]  %
SED photometry error scaling \dotfill & $\sigma_{SED}$ & & $23.0^{+5.6}_{-4.0}$ & {\sc EXOFASTv2} & --& -- \\ [2pt]  %
Parallax \dotfill & $\varpi$ & mas & $10.820^{+0.018}_{-0.019}$ & {\sc EXOFASTv2} & -- & -- \\ [2pt]  %
Distance \dotfill & $d$ & pc & $92.42^{+0.16}_{-0.16}$ & {\sc EXOFASTv2} & $90 \pm 3$ & -- \\ [2pt]  %
Age \dotfill & & Gyr & $6.5_{-2.7\,-2.2}^{+3.4\,+1.6} $ & ABSDIM & $7.8 \pm 3.3$ & -- \\ [2pt]  %
\hline \\[-6pt]%
\multicolumn{1}{l}{\textbf{Transit parameters}} \\ [2pt] %
Sum of the fractional radii \dotfill & $r_{\rm p}+r_{\star}$ &  & $0.0503 \pm 0.0015$  & JKTEBOP & -- & --\\ [2pt] %
Ratio of the fractional radii  \dotfill & $r_{\rm p}/r_{\star}$ &  & $0.1212 \pm  0.0012$  & JKTEBOP & $0.1238 \pm 0.0010$ & --\\ [2pt] %
Impact parameter   \dotfill & $b$ &  & $0.27 \pm 0.11$  & JKTEBOP & $0.311^{+0.045}_{-0.067}$ & --\\ [2pt] %
\hline \\[-6pt]%
\multicolumn{1}{l}{\textbf{Planetary parameters}} \\ [2pt] %
Mass \dotfill   & $M_{\mathrm{p}}$ & $M_{\mathrm{Jup}}$ & $0.558 \pm 0.015 \pm 0.010$ & ABSDIM & $0.534 \pm 0.018$ & $0.532^{+0.018}_{-0.017}$\\ [2pt] %
Radius \dotfill & $R_{\mathrm{p}}$ & $R_{\mathrm{Jup}}$ & $1.015 \pm 0.040 \pm 0.009$ & ABSDIM & $1.010 \pm 0.029$ & --\\ [2pt] %
Mean density \dotfill&$\rho_{\mathrm{p}}$&$\rho_{\mathrm{Jup}}$ & $0.499\pm 0.058\pm 0.004$ & ABSDIM & $0.518 \pm 0.047$ & --\\[2pt]%
Surface gravity  \dotfill & $g_{\mathrm{p}}$ & m\,s$^{-2}$ & $13.4 \pm 1.0$ & ABSDIM & $12.9 \pm 0.6$ & --\\ [2pt] %
Equilibrium temperature \dotfill & $T_{\mathrm{eq}}$ & K & $800 \pm 11$ & ABSDIM & $792 \pm 15$ & --\\  [2pt] %
Safronov number \dotfill & $\Theta$ & & $0.1105 \pm 0.0046 \pm 0.0010$ & ABSDIM & $0.109 \pm 0.004$ & --\\ [2pt]%
\hline \\[-6pt]%
\multicolumn{1}{l}{\textbf{Orbital parameters}} \\ [2pt] %
Reference epoch of mid-transit \dotfill & $T_{0}$ & BJD\,(TDB)& 2\,454\,801.16943\,(15) & Timing fit & $2\,454\,801.16943\,(20)$ & --\\ [2pt] %
Period  \dotfill & $P_{\mathrm{orb}}$ & days    & $10.33853781\,(60)$ & Timing fit & $10.338523\,(9)$ & \\  [2pt]%
Inclination \dotfill & $i$ & degree & $89.29 \pm 0.30$ & JKTEBOP & $89.2^{+0.2}_{-0.1}$ & --\\ [2pt] %
Semi-major axis \dotfill & $a$ & au & $0.0892 \pm 0.0010 \pm 0.0008$ & ABSDIM & $0.0882 \pm 0.0014$ & --\\ [2pt]%
RV-curve semi-amplitude \dotfill & $K_{\star}$ &  m\,s$^{-1}$ & -- & -- & $58.8 \pm 0.9$ & $58.58^{+0.69}_{-0.68}$ \\ [2pt] %
Barycentric RV  \dotfill & $\gamma$ & km\,s$^{-1}$ & $20.3141 \pm 0.0021$ & RM fit & $20.13 \pm 0.21$ & $20^{+27}_{-16}$\\  [2pt] %
Projected spin-orbit angle\,$^{(a)}$  \dotfill & $\lambda$ & degree  & $-27.5 \pm 6.7$ & RM fit & -- & $19^{+14}_{-16}$ \\ [2pt] %
\hline %
\end{tabular}
}
\tablefoot{
\tablefoottext{a}{Note that \citet{fulton:2013} used a convention with the opposite sign for $\lambda$ than ours. Therefore, the two results are fully compatible.}}
\end{table*}

\begin{table*}
\tiny
\centering %
\caption{Physical parameters of the planetary system HAT-P-21 derived in this work.} %
\label{tab:hatp21_final_parameters} %
\begin{tabular}{l c c c c | c}
\hline %
\hline  \\[-8pt]
Parameter & Nomen. & Unit & This Work & Source & \citet{bakos:2011} \\
\hline  \\[-6pt]
\multicolumn{1}{l}{\textbf{Stellar parameters}} \\ [2pt] %
Spectra class \dotfill & & & & &  G3 \\
Effective temperature \dotfill & $T_{\mathrm{eff}}$   & K & $\begin{tabular}{c}
$5699\pm44$ \\
$5695 \pm 45$ \\
\end{tabular}$ &
$\begin{tabular}{c}
{\sc EXOFASTv2} \\
{\sc HARPS-N} \\
\end{tabular}$
& $5588\pm80$ \\ [8pt] %
Metallicity \dotfill & [Fe/H] & & 
$\begin{tabular}{c}
$+0.046^{+0.085}_{-0.084}$ \\
$~+0.04 \pm 0.09$ \\
\end{tabular}$ &
$\begin{tabular}{c}
{\sc EXOFASTv2} \\
{\sc HARPS-N} \\
\end{tabular}$
& $0.01\pm0.08$ \\ [8pt] %
Projected rotational velocity \dotfill & $v\,\sin{i_{\star}}$ & km\,s$^{-1}$ & $3.9 \pm 0.9 $ & RM fit & $3.5\pm0.5$ \\ [2pt] %
Luminosity  \dotfill & $L_{\star}$ & $L_{\sun}$ & $1.52^{+0.11}_{-0.10}$ & {\sc EXOFASTv2} & $1.06^{+0.20}_{-0.16}$\\ [2pt] %
Mass    \dotfill & $M_{\star}$ & $M_{\sun}$ & $0.998 \pm 0.035 \pm 0.020$ & ABSDIM & $0.947 \pm 0.042$ \\ [2pt]%
Radius  \dotfill & $R_{\star}$ & $R_{\sun}$ & $1.248 \pm 0.049 \pm 0.009$ & ABSDIM & $1.105 \pm 0.083$ \\ [2pt] %
Mean density \dotfill & $\rho_{\star}$ & $\rho_{\sun}$ & $0.513 \pm 0.058$  & ABSDIM & -- \\ [2pt] %
Logarithmic surface gravity \dotfill & $\log{g_{\star}}$ & cgs & $4.245 \pm 0.033 \pm 0.003$ & ABSDIM & $4.33 \pm 0.06$ \\ [2pt]%
Equal Evolutionary Phase \dotfill & $EEP$ & & $428^{+10}_{-18}$ & {\sc EXOFASTv2} & -- \\ [2pt]  %
$V$-band extinction \dotfill & $A_{\rm V}$ & mag & $0.25\pm0.12$ & {\sc EXOFASTv2} & -- \\ [2pt]  %
SED photometry error scaling \dotfill & $\sigma_{SED}$ & & $5.6^{+1.5}_{-1.0}$ & {\sc EXOFASTv2} & -- \\ [2pt]  %
Parallax \dotfill & $\varpi$ & mas & $3.519\pm0.023$ & {\sc EXOFASTv2} & -- \\ [2pt]  %
Distance \dotfill & $d$ & pc & $284.2\pm1.8$ & {\sc EXOFASTv2} & $254 \pm 19$ \\ [2pt]  %
Age \dotfill & & Gyr & $7.8_{-1.1\,-2.6}^{+2.6\,+4.4} $ & ABSDIM & $10.2 \pm 2.5$ \\ [2pt]  %
Rotational period \dotfill & $P_{\rm rot}$ & day & $15.88 \pm 0.02$ & TFA & -- \\ [2pt]  %
\hline \\[-6pt]%
\multicolumn{1}{l}{\textbf{Transit parameters}} \\ [2pt] %
Sum of the fractional radii \dotfill & $r_{\rm p}+r_{\star}$ &  & $0.1258 \pm 0.0064$  & JKTEBOP & -- \\ [2pt] %
Ratio of the fractional radii  \dotfill & $r_{\rm p}/r_{\star}$ &  & $0.0931 \pm 0.0012$  & JKTEBOP & $0.0950 \pm 0.0022$ \\ [2pt] %
Impact parameter   \dotfill & $b$ &  & $0.615 \pm 0.056$  & JKTEBOP & $0.631^{+0.025}_{-0.028}$ \\ [2pt] %
\hline \\[-6pt]%
\multicolumn{1}{l}{\textbf{Planetary parameters}} \\ [2pt] %
Mass \dotfill   & $M_{\mathrm{p}}$ & $M_{\mathrm{Jup}}$ & $4.23 \pm 0.15 \pm  0.06$ & ABSDIM & $4.063 \pm 0.161$ \\ [2pt] %
Radius \dotfill & $R_{\mathrm{p}}$ & $R_{\mathrm{Jup}}$ & $1.130 \pm 0.056 \pm  0.008$ & ABSDIM & $1.024 \pm 0.092$ \\ [2pt] %
Mean density \dotfill&$\rho_{\mathrm{p}}$&$\rho_{\mathrm{Jup}}$ & $2.74 \pm 0.40 \pm  0.02$ & ABSDIM & $3.77^{+1.28}_{-0.80}$  \\[2pt]%
Surface gravity  \dotfill & $g_{\mathrm{p}}$ & m\,s$^{-2}$ & $82.1 \pm 8.1$ & ABSDIM & $95 \pm 18$ \\ [2pt] %
Equilibrium temperature \dotfill & $T_{\mathrm{eq}}$ & K & $1367 \pm 27$ & ABSDIM & $1283 \pm 50$ \\  [2pt] %
Safronov number \dotfill & $\Theta$ & & $0.377 \pm 0.021 \pm 0.003$ & ABSDIM & $0.413 \pm 0.038$ \\ [2pt]%
\hline \\[-6pt]%
\multicolumn{1}{l}{\textbf{Orbital parameters}} \\ [2pt] %
Reference epoch of mid-transit \dotfill & $T_{0}$& BJD(TDB)&$2\,454\,996.41243\,(60)$ & JKTEBOP & $2\,454\,996.41312\,(69)$ \\ [2pt] %
Period      \dotfill & $P_{\mathrm{orb}}$ & days & $4.12449009\,(90)$  & JKTEBOP & $4.124481\,(07)$ \\  [2pt]%
Inclination \dotfill & $i$ & degree & $86.47 \pm 0.50 $ & JKTEBOP & $87.2 \pm 0.7$ \\  [2pt]%
Semi-major axis \dotfill & $a$ & au & $0.05037 \pm 0.00058 \pm 0.00034$ & ABSDIM & $0.0494 \pm 0.0007$ \\ [2pt]%
RV-curve semi-amplitude \dotfill & $K_{\rm A}$ &  m\,s$^{-1}$ & -- & -- & $548.3 \pm 14.2$ \\ [2pt] %
Barycentric RV  \dotfill & $\gamma$ & km\,s$^{-1}$ & $-53.018 \pm 0.005$ & RM fit & $-53.190 \pm 0.090$ \\  [2pt] %
Projected spin-orbit angle  \dotfill & $\lambda$ & degree  & $-0.7 \pm 12.5$ & RM fit & -- \\ [2pt] %
\hline %
\end{tabular}
\end{table*}

\begin{table*}
\tiny
\centering %
\caption{Physical parameters of the planetary system HAT-P-26 derived in this work.} %
\label{tab:hatp26_final_parameters} %
\begin{tabular}{l c c c c | c}
\hline %
\hline  \\[-8pt]
Parameter & Nomen. & Unit & This Work & Source & \citet{hartman:2011} \\
\hline  \\[-6pt]
\multicolumn{1}{l}{\textbf{Stellar parameters}} \\ [2pt] %
Spectra class \dotfill & & & & &  K1 \\
Effective temperature \dotfill & $T_{\mathrm{eff}}$   & K & $\begin{tabular}{c}
$5102^{+20}_{-19}$ \\
$~~~5100 \pm 20$ \\
\end{tabular}$ &
$\begin{tabular}{c}
{\sc EXOFASTv2} \\
{\sc HARPS-N} \\
\end{tabular}$
& $5079\pm88$ \\ [8pt] %
Metallicity \dotfill & [Fe/H] & & 
$\begin{tabular}{c}
$+0.005^{+0.091}_{-0.088}$ \\
$~~~+0.05 \pm 0.10$ \\
\end{tabular}$ &
$\begin{tabular}{c}
{\sc EXOFASTv2} \\
{\sc HARPS-N} \\
\end{tabular}$
& $-0.04\pm0.08$ \\ [8pt] %
Projected rotational velocity \dotfill & $v\,\sin{i_{\star}}$ & km\,s$^{-1}$ & $1.9 \pm 0.4$ & {\sc HARPS-N} & $1.8\pm0.5$ \\ [2pt] %
Luminosity  \dotfill & $L_{\star}$ & $L_{\sun}$ & $0.357^{+0.021}_{-0.012}$ & {\sc EXOFASTv2} & $0.38^{+0.16}_{-0.06}$\\ [2pt] %
Mass    \dotfill & $M_{\star}$ & $M_{\sun}$ & $0.796 \pm 0.015 \pm 0.026$ & ABSDIM & $0.816 \pm 0.033$ \\ [2pt]%
Radius  \dotfill & $R_{\star}$ & $R_{\sun}$ & $0.916 \pm 0.062 \pm 0.010$ & ABSDIM & $0.788^{+0.098}_{-0.043}$ \\ [2pt] %
Mean density \dotfill & $\rho_{\star}$ & $\rho_{\sun}$ & $1.04 \pm 0.22$  & ABSDIM & -- \\ [2pt] %
Logarithmic surface gravity \dotfill & $\log{g_{\star}}$ & cgs & $4.416 \pm 0.063 \pm 0.005$ & ABSDIM & $4.56\pm0.06$ \\ [2pt]%
Equal Evolutionary Phase \dotfill & $EEP$ & & $338^{+15}_{-29}$ & {\sc EXOFASTv2} & -- \\ [2pt]  %
$V$-band extinction \dotfill & $A_{\rm V}$ & mag & $0.062^{+0.082}_{-0.044}$ & {\sc EXOFASTv2} & -- \\ [2pt]  %
SED photometry error scaling \dotfill & $\sigma_{SED}$ & & $14.8^{+3.9}_{-2.6}$ & {\sc EXOFASTv2} & -- \\ [2pt]  %
Parallax \dotfill & $\varpi$ & mas & $6.999\pm0.020$ & {\sc EXOFASTv2} & -- \\ [2pt]  %
Distance \dotfill & $d$ & pc & $142.88^{+0.42}_{-0.40}$ & {\sc EXOFASTv2} & $134^{+18}_{-8}$ \\ [2pt]  %
Age \dotfill & & Gyr & $5.9_{-3.8}^{+4.8}$ & {\sc EXOFASTv2} & $10.2 \pm 2.5$ \\ [2pt]  %
\hline \\[-6pt]%
\multicolumn{1}{l}{\textbf{Transit parameters}} \\ [2pt] %
Sum of the fractional radii \dotfill & $r_{\rm p}+r_{\star}$ &  & $0.0962312^{+0.0070}_{-0.0068}$  & JKTEBOP & -- \\ [2pt] %
Ratio of the fractional radii  \dotfill & $r_{\rm p}/r_{\star}$ &  & $0.0732 \pm 0.0011$  & JKTEBOP & $0.0737\pm0.0012$ \\ [2pt] %
Impact parameter   \dotfill & $b$ &  & $0.55^{+0.17}_{-0.13}$  & JKTEBOP & $0.30^{+0.11}_{-0.12}$ \\ [2pt] %
\hline \\[-6pt]%
\multicolumn{1}{l}{\textbf{Planetary parameters}} \\ [2pt] %
Mass \dotfill   & $M_{\mathrm{p}}$ & $M_{\mathrm{Jup}}$ & $0.0577 \pm 0.0069 \pm 0.0013$ & ABSDIM & $0.059 \pm 0.007$ \\ [2pt] %
Radius \dotfill & $R_{\mathrm{p}}$ & $R_{\mathrm{Jup}}$ & $0.652 \pm 0.055 \pm 0.007$ & ABSDIM & $0.565^{+0.072}_{-0.032}$ \\ [2pt] %
Mean density \dotfill&$\rho_{\mathrm{p}}$&$\rho_{\mathrm{Jup}}$ & $0.195 \pm 0.055 \pm 0.002$ & ABSDIM & $0.32 \pm 0.08$  \\[2pt]%
Surface gravity  \dotfill & $g_{\mathrm{p}}$ & m\,s$^{-2}$ & $3.37 \pm 0.69$ & ABSDIM & $4.47^{+0.90}_{-0.92}$ \\ [2pt] %
Equilibrium temperature \dotfill & $T_{\mathrm{eq}}$ & K & $1080 \pm 39$ & ABSDIM & $1001^{+66}_{-37}$ \\  [2pt] %
Safronov number \dotfill & $\Theta$ & & $0.0106 \pm 0.0015 \pm 0.0001$ & ABSDIM & $0.012 \pm 0.002$ \\ [2pt]%
\hline \\[-6pt]%
\multicolumn{1}{l}{\textbf{Orbital parameters}} \\ [2pt] %
Reference epoch of mid-transit \dotfill & $T_{0}$& BJD(TDB)&$2\,455\,304.65234\,(35)$ & Timing fit & 2\,455\,304.65122\,(35) \\ [2pt] %
Period      \dotfill & $P_{\mathrm{orb}}$ & days & 4.23450213\,(76) & Timing fit & 4.234516\,(15) \\  [2pt]%
Inclination \dotfill & $i$ & degree & $87.20\pm0.86$ & JKTEBOP & $88.6^{+0.5}_{-0.9}$ \\  [2pt]%
Semi-major axis \dotfill & $a$ & au & $0.04748\pm0.00030\pm0.00052 $ & ABSDIM & $0.0479\pm0.0006$ \\ [2pt]%
RV-curve semi-amplitude \dotfill & $K_{\rm A}$ &  m\,s$^{-1}$ & -- & -- & $8.5\pm1.0$ \\ [2pt] %
Barycentric RV  \dotfill & $\gamma$ & km\,s$^{-1}$ & $Unconstrained$ & RM fit & $14.72\pm0.10$ \\  [2pt] %
Projected spin-orbit angle  \dotfill & $\lambda$ & degree  & $Unconstrained$ & RM fit & -- \\ [2pt] %
\hline %
\end{tabular}
\end{table*}

\begin{table*}
\tiny
\centering %
\caption{Physical parameters of the planetary system HAT-P-29 derived in this work.} %
\label{tab:hatp29_final_parameters} %
\begin{tabular}{l c c c c | c}
\hline %
\hline  \\[-8pt]
Parameter & Nomen. & Unit & This Work & Source & \citet{buchhave:2011} \\
\hline  \\[-6pt]
\multicolumn{1}{l}{\textbf{Stellar parameters}} \\ [2pt] %
Spectra class \dotfill & & & & &  G \\
Effective temperature \dotfill & $T_{\mathrm{eff}}$ & K &
$\begin{tabular}{c}
$6140^{+29}_{-30}$ \\
$~~~6140 \pm 30$ \\
\end{tabular}$ &
$\begin{tabular}{c}
{\sc EXOFASTv2} \\
{\sc HARPS-N} \\
\end{tabular}$ & $6087\pm88$ \\ [8pt] %
Metallicity \dotfill & [Fe/H] & & 
$\begin{tabular}{c}
$+0.240^{+0.078}_{-0.080}$ \\
$~+0.25 \pm 0.08$ \\
\end{tabular}$ & 
$\begin{tabular}{c}
{\sc EXOFASTv2} \\
{\sc HARPS-N} \\
\end{tabular}$ &
$0.21\pm0.08$ \\ [8pt] %
Projected rotational velocity \dotfill & $v\,\sin{i_{\star}}$ & km\,s$^{-1}$ & 
$\begin{tabular}{c}
$5.2 \pm 0.7$ \\
$4.5 \pm 0.8$ \\
\end{tabular}$ &
$\begin{tabular}{c}
RM fit \\
{\sc HARPS-N} \\
\end{tabular}$ & $3.9\pm0.5$ \\ [8pt] %
Convective blue-shift velocity \dotfill & CBV & km\,s$^{-1}$ & $>-0.58$ & RM fit & --  \\ [2pt] %
Luminosity  \dotfill & $L_{\star}$ & $L_{\sun}$ & $2.05^{+0.13}_{-0.12}$ & {\sc EXOFASTv2} & $1.84^{+0.47}_{-0.26}$\\ [2pt] %
Mass    \dotfill & $M_{\star}$ & $M_{\sun}$ & $1.206_{-0.052\,-0.016}^{+0.048\,+0.011}$ & ABSDIM & $1.207\pm0.046$ \\ [2pt]%
Radius  \dotfill & $R_{\star}$ & $R_{\sun}$ & $1.272_{-0.042\,-0.006}^{+0.032\,+0.004}$ & ABSDIM & $1.224^{+0.133}_{-0.075}$ \\ [2pt] %
Mean density \dotfill & $\rho_{\star}$ & $\rho_{\sun}$ & $0.586_{-0.039}^{+0.049}$  & ABSDIM & -- \\ [2pt] %
Logarithmic surface gravity \dotfill & $\log{g_{\star}}$ & cgs & $4.311_{-0.020\,-0.002}^{+0.026\,+0.001}$ & ABSDIM & $4.34\pm0.06$ \\ [2pt]%
Equal Evolutionary Phase \dotfill & $EEP$ & & $347^{+41}_{-28}$ & {\sc EXOFASTv2} & -- \\ [2pt]  %
$V$-band extinction \dotfill & $A_{\rm V}$ & mag & $0.58\pm0.11$ & {\sc EXOFASTv2} & -- \\ [2pt]  %
SED photometry error scaling \dotfill & $\sigma_{SED}$ & & $4.87^{+1.40}_{-0.88}$ & {\sc EXOFASTv2} & -- \\ [2pt]  %
Parallax \dotfill & $\varpi$ & mas & $3.135^{+0.020}_{-0.019}$ & {\sc EXOFASTv2} & -- \\ [2pt]  %
Distance \dotfill & $d$ & pc & $318.9^{+1.9}_{-2.0}$ & {\sc EXOFASTv2} & $322^{+35}_{-21}$ \\ [2pt]  %
Age \dotfill & & Gyr & $2.5_{-1.7\,-0.4}^{+1.2\,+0.4}$ & ABSDIM & $2.2\pm1.0$ \\ [2pt]  %
\hline \\[-6pt]%
\multicolumn{1}{l}{\textbf{Transit parameters}} \\ [2pt] %
Sum of the fractional radii \dotfill & $r_{\rm p}+r_{\star}$ &  & $0.0964^{+0.0030}_{-0.0020}$  & JKTEBOP & -- \\ [2pt] %
Ratio of the fractional radii  \dotfill & $r_{\rm p}/r_{\star}$ &  & $0.08777\pm0.00063$  & JKTEBOP & $0.0927\pm0.0028$ \\ [2pt] %
Impact parameter   \dotfill & $b$ &  & $0.37\pm0.12$  & JKTEBOP & $0.591^{+0.062}_{-0.094}$ \\ [2pt] 
\hline \\[-6pt]%
\multicolumn{1}{l}{\textbf{Planetary parameters}} \\ [2pt] %
Mass \dotfill   & $M_{\mathrm{p}}$ & $M_{\mathrm{Jup}}$ & $0.773_{-0.050\,-0.007}^{+0.050\,+0.005}$ & ABSDIM & $0.778^{+0.076}_{-0.040}$ \\ [2pt] %
Radius \dotfill & $R_{\mathrm{p}}$ & $R_{\mathrm{Jup}}$ & $1.105_{-0.037\,-0.005}^{+0.031\,+0.003}$ & ABSDIM & $1.107^{+0.136}_{-0.082}$ \\ [2pt] %
Mean density \dotfill&$\rho_{\mathrm{p}}$&$\rho_{\mathrm{Jup}}$ & $0.536_{-0.053\,-0.002}^{+0.056\,+0.002}$ & ABSDIM & $0.54 \pm 0.14$  \\[2pt]%
Surface gravity  \dotfill & $g_{\mathrm{p}}$ & m\,s$^{-2}$ & $15.7_{-1.2}^{+1.3}$ & ABSDIM & $15.84 \pm 2.55$ \\ [2pt] %
Equilibrium temperature \dotfill & $T_{\mathrm{eq}}$ & K & $1281_{-26}^{+23}$ & ABSDIM & $1260^{+64}_{-45}$ \\  [2pt] %
Safronov number \dotfill & $\Theta$ & & $0.0772_{-0.0050\,-0.0002}^{+0.0052\,+0.0003}$ & ABSDIM & $0.077 \pm 0.007$ \\ [2pt]%
\hline \\[-6pt]%
\multicolumn{1}{l}{\textbf{Orbital parameters}} \\ [2pt] %
Reference epoch of mid-transit \dotfill & $T_{0}$& BJD(TDB)&$2\,455\,838.59462\,(65)$ & Timing fit & $2\,455\,197.5754\,(18)$ \\ [2pt] %
Period      \dotfill & $P_{\mathrm{orb}}$ & days & 5.7233746\,(34) & Timing fit & 5.723186\,(49) \\  [2pt]%
Inclination \dotfill & $i$ & degree & $88.82 \pm 0.50$ & JKTEBOP & $87.1^{+0.5}_{-0.7}$ \\  [2pt]%
Semi-major axis \dotfill & $a$ & au & $0.06667_{-0.00096\,-0.00029}^{+0.00089\,+0.00021}$ & ABSDIM & $0.0667\pm0.0008$ \\ [2pt]%
RV-curve semi-amplitude \dotfill & $K_{\rm A}$ &  m\,s$^{-1}$ & -- & -- & $78.3 \pm 5.9$ \\ [2pt] %
Barycentric RV  \dotfill & $\gamma$ & km\,s$^{-1}$ & $-21.651\pm0.002$ & RM fit & $-21.670\pm0.08$ \\  [2pt] %
Projected spin-orbit angle  \dotfill & $\lambda$ & degree  & $-26 \pm 16$ & RM fit & -- \\ [2pt] %
\hline %
\end{tabular}
\end{table*}

\end{appendix}

\end{document}